\documentclass[journal,10pt]{IEEEtran}
\renewcommand{\baselinestretch}{0.99}
\usepackage{hyperref}
\usepackage{cite}
\usepackage[T1]{fontenc}
\usepackage[utf8]{inputenc}
\DeclareUnicodeCharacter{010D}{\v{c}}
\usepackage{amsmath,amsthm,amssymb,amsfonts,steinmetz}
\usepackage{multirow} 
\usepackage{graphicx}
\usepackage{algorithm}
\usepackage{algpseudocode,flushend}
\theoremstyle{remark}
\newtheorem{remark}{Remark}
\usepackage{booktabs}

\def\BibTeX{{\rm B\kern-.05em{\sc i\kern-.025em b}\kern-.08em
    T\kern-.1667em\lower.7ex\hbox{E}\kern-.125emX}}

\usepackage{tikz}
\usetikzlibrary{calc}
\makeatletter
\newcommand{\gettikzxy}[3]{%
  \tikz@scan@one@point\pgfutil@firstofone#1\relax
  \edef#2{\the\pgf@x}%
  \edef#3{\the\pgf@y}%
}
\makeatother

\usepackage{comment}
\usepackage{subcaption}

\DeclareMathAlphabet\mathbfcal{OMS}{cmsy}{b}{n}

\begin{document}
%\title{2D Waveguide-Fed Metasurface Antennas: Physically Consistent Modeling, Validation, and Optimization}
\title{2D Waveguide-Fed Metasurfaces: Physically Consistent Modeling, Validation, and Optimization}
%\title{2D Waveguide-Fed Metasurface Antennas: Physics-Based~Modeling,~Validation,~and~Optimization}
\author{Panagiotis Gavriilidis,~\IEEEmembership{Graduate~Student~Member,~IEEE}, and George C. Alexandropoulos,~\IEEEmembership{Senior~Member,~IEEE}
\thanks{This work has been supported by the SNS JU project TERRAMETA under the EU's Horizon Europe research and innovation programme under Grant Agreement number 101097101, including also top-up funding by UKRI under the UK government's Horizon Europe funding guarantee. P. Gavriilidis also acknowledges the support of the Fulbright Program and the Fulbright Foundation in Greece through a Fulbright Fellowship.}
\thanks{The authors are with the Department of Informatics and Telecommunications, National and Kapodistrian University of Athens, 16122 Athens, Greece. G. C. Alexandropoulos is also with the Department of Electrical and Computer Engineering, University of Illinois Chicago, IL 60601, USA (e-mails: \{pangavr, alexandg\}@di.uoa.gr).}
}

\maketitle

\begin{abstract}
Antenna array architectures based on programmable metasurfaces are emerging as a promising solution for scalable implementations of the eXtremely Large Multiple-Input Multiple-Output (XL-MIMO) systems paradigm, envisioned for $6$-th Generation (6G), and beyond, wireless networks. However, their accurate modeling, quantifying the role of key structural features, such as strong mutual coupling and guided-wave excitation, remains challenging, amplifyisng the need for physically consistent representations of the constituent metamaterial elements. In this paper, capitalizing on the coupled dipole formulation, we develop a comprehensive electromagnetics-compliant framework for $2$-Dimensional (2D) waveguide-fed metasurface antennas. The proposed model extends relevant existing modeling approaches by incorporating both electric and magnetic dipoles' responses, accounting for multiple excitation feeds, and enabling accurate characterization in both the near- and far-field regimes. Radiation-reaction corrections based on passivity constraints are derived and shown to ensure the passivity of the overall dipole system. In addition, we present a novel input impedance model for the considered architecture enabling explicit computation of the accepted power, and facilitating efficient beamforming design under realistic power constraints. All modeling components developed in this paper are validated against full-wave electromagnetic simulations. Furthermore, the analytical structure of the proposed model enables the formulation of a differentiable beamforming design optimization problem over both the considered metasurface geometry and its feed excitations. The presented numerical results demonstrate the effectiveness of the proposed modeling framework in achieving both directive beamforming and sector-wide coverage. Overall, the proposed framework provides a physically grounded and computationally efficient tool for the analysis and optimization of 2D waveguide-fed metasurface antennas.
\end{abstract}

\begin{IEEEkeywords}
Metasurface, coupled dipole method, XL MIMO, 2D waveguide, 
mutual coupling, physically consistent modeling, polarizability. %, full-wave validation.
\end{IEEEkeywords}

\section{Introduction}
% XL-MIMO motivation, hybrid MIMO and metasurface-based architectures

To meet the stringent throughput demands of $6$-th Generation (6G), and beyond, wireless networks, novel antenna array architectures beyond fully digital Multiple-Input Multiple-Output (MIMO) arrays are being actively investigated~\cite{8264743}. In this direction, hybrid Analog and Digital (A/D) beamforming architectures have emerged as a promising solution to reduce hardware complexity and power consumption~\cite{8030501}. In those systems, digital precoding/combining is carried in baseband feeding, in the sequel, a limited number of Radio-Frequency Chains (RFCs), followed by analog beamforming/combining implemented through networks of power splitters and phase shifters connected at the RFCs outputs~\cite{9933358}. This feature enables improved beamforming/combining capabilities without increasing the number of RFCs, which are the most power-intensive components of those antenna array architectures. However, power splitter networks and phase shifter circuits introduce significant hardware complexity, non-negligible power consumption, and performance degradation over wide bandwidths, hindering their scalability potential and, thus, their applicability to eXtremely Large (XL) MIMO systems.

The technology of programmable metasurfaces, which was initially considered primarily for reflectarrays and Reconfigurable Intelligent Surfaces (RISs)~\cite{AlexandropoulosRIS}, constitutes a natural enabler for highly scalable multi-element antenna arrays, which can be capable of realizing energy-efficient analog beamforming/combining through reconfigurable metamaterial elements excited via waveguide structures~\cite{10352433,10124713,Receiving_RISs}. Dynamic Metasurface Antennas (DMAs) represent a recent prominent implementation of this paradigm \cite{Shlezinger2021Dynamic}, consisting of, possibly XL large numbers and subwavelength-spaced~\cite{Gong2024HMIMO}, response-tunable metamaterial elements connected with limited numbers of RFCs. This emerging antenna array architecture has lately attracted significant research interest, both in theoretical studies and experimental investigations, including communication system design~ \cite{Shlezinger2021Dynamic,Shlezinger2019_DMA_uplink_MIMO,Li2023_DMA_energy_efficiency,10505154,DMA1bit_comms,10938788}, localization and sensing \cite{guohao2020_DMAsensingMag,9500663,DMA_CRB,DMA1bit_Loc,Gavras2025DMA_bistatic_Optimization,10403512,11161718,Rezvani2024_DMA_channelestimation,10694467}, as well as computational imaging \cite{Gollub2017LargeMetasurfaceImaging,Boyarsky2018SingleFrequency3D,Diebold2018PhaselessGhostImaging,9769592,11362106}. Furthermore, such architectures can naturally support the envisioned tri-hybrid MIMO configurations when combined with feeding networks based on phase shifters~\cite{Heath_2026_trihybrid}.

% Metasurfaces require dedicated electromagnetic modeling
Conventional hybrid A/D and emerging metasurface-based transceivers have some fundamental differences, a fact that necessitates distinct tailored ElectroMagnetic (EM) modeling frameworks. In contrast to conventional arrays, metasurface unit elements exhibit strong mutual coupling driven by both the guided-wave excitation mechanism and the radiation in free space, features that need to be modeled in a fully coupled manner \cite{pulidomancera2018,williams2023EM_DMA,gavras20262D_DMA}. In addition, the excitation of the elements is governed by the underlying waveguide or microstrip structure, which necessitates a physics-consistent representation of the guided fields. Furthermore, the metamaterial elements themselves must be described through physically consistent models that capture their frequency-dependent behavior while satisfying passivity constraints. Moreover, in metasurface-based transceiver architectures, power is injected only at the digital/RFC level, however, the waveguide and the metamaterial elements constitute a passive structure that must properly account for radiation and dielectric losses~\cite{davidsmith2017,gavriilidis2025microstrip}. Existing modeling approaches often do not explicitly enforce passivity \cite{Shlezinger2019_DMA_uplink_MIMO,williams2023EM_DMA,Carlson2026WidebandDMABeamforming}, which may lead to non-physical artifacts, such as predicted radiated power exceeding the injected power. These considerations highlight that, despite their potential as an enabling technology for XL-MIMO systems, metasurface-based antenna arrays still lack comprehensive, physically consistent modeling frameworks, a fact that hinders the establishment of reliable performance baselines and fair comparisons with conventional hybrid A/D MIMO transceiver architectures.

% Conventional DMA architectures and motivation for PPW-based designs
Up to date DMA designs typically rely on multiple stacked $1$-Dimensional (1D) waveguides, also termed as microstrip lines, each fed through a dedicated excitation port, with metamaterial elements etched on top~~\cite{Shlezinger2021Dynamic,Shlezinger2019_DMA_uplink_MIMO,Li2023_DMA_energy_efficiency,10505154,DMA1bit_comms,10938788,guohao2020_DMAsensingMag,9500663,DMA_CRB,DMA1bit_Loc,Gavras2025DMA_bistatic_Optimization,10403512,11161718,Rezvani2024_DMA_channelestimation,10694467,Gollub2017LargeMetasurfaceImaging,Boyarsky2018SingleFrequency3D,Diebold2018PhaselessGhostImaging,9769592,11362106}. In those architectures, scalability is inherently limited by the number of microstrips and their associated feeding circuitry. In contrast, the utilization of $2$-Dimensional (2D) waveguides, such as Parallel-Plate Waveguides (PPWs), enables excitation of the entire aperture with a small number of feeds, or even a single one. This leads to highly energy-efficient and naturally scalable designs, where the aperture size can be increased without increasing the active hardware components~\cite{tretyakov2020ppwdoublepenetrable,hosseini2023ppw_huygens,pulidomancera2018}. Moreover, PPW-fed architectures offer favorable bandwidth characteristics, as they support Transverse Electromagnetic Mode (TEM) propagation without a low cut-off frequency. In contrast, 1D waveguides exhibit both lower and upper frequency limits for their fundamental mode, due to boundary-induced effects of the metallic side walls~\cite{balanis2012advanced}. On the other hand, PPW-fed metasurfaces cannot be accurately described using simple transmission line models (e.g.,~\cite{davidsmith2017}), since the field distribution is governed by radial guided-wave propagation and strong mutual coupling between unit elements through the waveguide and free space. 

% Discrete dipole approximation 
A powerful tool for addressing the modeling challenges of such waveguide-fed metasurface-based antenna array architectures is the Coupled Dipole (CD) method \cite{CDF_tacit,rashidfaqiri2023physfad}, which is conceptually related to the discrete dipole approximation used for modeling EM scattering from continuous objects~\cite{Landy2014DiscreteDipoleMetamaterial,YURKIN2007DiscreteDipoleApproximation}. In classical discrete dipole approximation methods, dipoles arise from discretizing a continuous medium, in contrast to this, the dipoles in metasurface-based antenna arrays directly represent the physical response of their constituent metamaterials. Owing to their subwavelength size, those unit elements can be accurately modeled as polarizable dipoles, whose responses are governed by their electric and/or magnetic polarizabilities, while their mutual interactions are efficiently captured through analytical Green’s function expressions. Building on this framework, the works in \cite{williams2023EM_DMA} and \cite{pulidomancera2018} developed dipole-based models for stacked 1D and 2D waveguide-fed metasurface architectures, respectively, contributing toward their physically consistent understanding. However, those models do not explicitly enforce passivity constraints, as they do not incorporate detailed modeling of the metamaterial responses and their physically admissible behavior. Moreover, they are restricted to magnetic dipole representations, neglecting electric dipole contributions, and do not extend to the Near-Field (NF) regime, which becomes increasingly important for XL antenna apertures \cite{liu2023near,Gong2024HMIMO}.

In this paper, we study 2D waveguide-fed metasurface antenna arrays, and compared to~\cite{pulidomancera2018}, which focuses on magnetic-dipole-based modeling under single-feed excitation, we develop a physically consistent and more general framework that incorporates both electric and magnetic dipole responses, accounts for multiple excitation feeds, and derives the physically admissible set of metamaterial responses, enabling seamless integration with conventional (XL) MIMO system models. In particular, in Section~\ref{sec: Modeling Magnetic Dipoles Only}, we extend the CD method to multi-feed excitation, establish the corresponding power conservation properties for the considered coupled system, and derive NF compatible expressions for the radiated field. In Section~\ref{sec: Both electric and magnetic dipoles}, we present a generalization of our model that includes both electric and magnetic dipoles, deriving all associated coupling mechanisms and rigorously characterizing the power injected in the PPW-fed antenna array structure. Section~\ref{sec: Polarizability Retrieval} presents the retrieval and validation of the polarizability models, while Section~\ref{sec: Model Validation} provides extensive model validation in both the Far-Field (FF) and NF regimes, as well as at the feed level. Capitalizing on the proposed physically consistent modeling framework, in Section~\ref{sec:sector_optimization}, we formulate and solve a sector-wide beamforming optimization problem for the fabrication design of the considered PPW-fed metasurface antennas. In Section~\ref{sec: Numerical results and discussion}, the validity and effectiveness of the proposed modeling and design optimization approaches is corroborated through numerical investigations, while Section~\ref{sec: Conclusion} includes the paper's concluding remarks. Overall, the proposed framework bridges EM modeling and conventional (XL) MIMO models, providing a fully analytical and physically grounded tool for the analysis and optimization of 2D waveguide-fed metasurface antenna arrays. The contributions of the paper are summarized as follows.
\begin{itemize}
    \item An electromagnetically consistent modeling framework for PPW-fed metasurface antenna arrays is presented, which extends prior magnetic-dipole-based formulations to account for multiple excitation feeds and enables the accurate characterization of the array in both NF and FF regimes. The proposed model is validated through Full-Wave EM (FW) simulations. 
    \item We derive a unified CD model that incorporates both electric and magnetic dipoles' responses, capturing all relevant propagation and coupling mechanisms. Comparisons with FW simulations and the magnetic-only model show that neglecting electric dipole effects leads to degraded modeling accuracy, particularly at high elevation angles where their impact becomes more pronounced.    
    \item We derive the necessary passivity conditions for the radiating metamaterials of the considered PPW-fed metasurface architecture for both electric and magnetic dipoles. Then, we rigorously establish the passivity of the resulting system, and validate the derived passivity conditions with FW simulations of elliptic iris elements.
    \item We develop the input impedance model for the PPW-fed metasurface, which is validated by means of FW simulations. This enables an explicit computation of the accepted power, providing physically consistent power modeling. Most importantly, this allows the formulation of beamforming  design problems under realistic power constraints, and highlights the role of the metamaterial response to the power consumption of the overall array.
    \item We formulate and solve a joint optimization problem for the metasurface fabrication design parameters and feed excitations for sector-wide beamforming. The use of closed-form polarizability expressions ensures full differentiability, and we derive analytical gradients with respect to a generic design parameter vector, enabling efficient gradient-based optimization. This formulation readily extends to reconfigurable metasurface elements, e.g., varactor-loaded metamaterials, by mapping circuit models to equivalent polarizability representations. The effectiveness of the proposed design framework in enabling both sector-wide coverage and directive beamforming is validated through numerical evaluations, highlighting scalability with aperture size, the trade-off between coverage and peak gain, as well as the capability of PPW-fed metasurfaces to efficiently realize scalable XL MIMO systems with fixed numbers of RFCs.
 %   \item Through extensive numerical evaluations, we demonstrate the effectiveness of the proposed framework in enabling both sector-wide coverage and directive beamforming. The results highlight the scalability with aperture size, the trade-off between coverage and peak gain, and the ability of the architecture to realize scalable hybrid MIMO systems with a fixed number of RF chains.
\end{itemize}

\textit{Notations:} Vectors and matrices are represented by boldface lowercase and boldface capital letters, respectively. $\mathbf{I}_{n}$, and \(\mathbf{0}_n\) are the $n\times n$ identity matrix, and the \(n \times n\) zeros matrix, respectively. $[\mathbf{A}]_{i,j}$ is the $(i,j)$-th element of $\mathbf{A}$, $[\mathbf{a}]_i$ and $||\mathbf{a}||_2$ denote $\mathbf{a}$'s $i$-th element and Euclidean norm, \(\mathbf{A}^{\rm T}\) and \(\mathbf{A}^{\rm H}\) denote the transpose and the Hermitian transpose of \(\mathbf{A}\), $\mathbf{x}\sim\mathcal{CN}(\mathbf{a},\mathbf{A})$ indicates a complex Gaussian random vector with mean $\mathbf{a}$ and covariance matrix $\mathbf{A}$, \(E[\cdot]\) denotes the expectation operator, and \(\mathbf B \otimes \mathbf A\) denotes the Kronecker product between two matrices. \(\mathbf{A} = \operatorname{diag}(\mathbf{a})\) denotes a diagonal matrix whose diagonal entries are given by the elements of \(\mathbf{a}\). More generally, \(\operatorname{diag}(\mathbf{a}_1,\ldots,\mathbf{a}_k)\) denotes a diagonal matrix formed by concatenating the elements of \(\mathbf{a}_1,\ldots,\mathbf{a}_k\) along its diagonal. When matrix arguments are used, \(\operatorname{diag}(\mathbf{A}_1,\ldots,\mathbf{A}_k)\) denotes the block-diagonal matrix with blocks \(\mathbf{A}_1,\ldots,\mathbf{A}_k\) on its main diagonal. Furthermore, $\jmath\triangleq\sqrt{-1}$ is the imaginary unit, while \(\rm Re\{\cdot\}\) and \(\rm Im\{\cdot\}\) are the real and imaginary operators. Additionally, $\mathrm{acos}(\cdot)$ denotes the arc-cosine function and $\mathrm{atan}(y/x)$ denotes the two-argument arctangent function, which returns the angle whose tangent is $y/x$ while preserving the correct quadrant of $(x,y)$. Moreover, \(\mathbf a \times \mathbf b\) and \(\mathbf a \cdot \mathbf b\) denote the cross and inner products between two vectors. Finally, \(\nabla_{\mathbf{z}} \mathbf{A}\) denotes the gradient with respect to (w.r.t.) the vector \(\mathbf{z}\), while, for spatial vectors in Cartesian coordinates such as \(\mathbf{f}\triangleq f_x\hat{\mathbf{x}} + f_y\hat{\mathbf{y}} + f_z\hat{\mathbf{z}}\), we use the notations \(\nabla\cdot \mathbf{f} \triangleq \partial f_x/\partial x \,\hat{\mathbf{x}} + \partial f_y/\partial y \,\hat{\mathbf{y}} + \partial f_z/\partial z \,\hat{\mathbf{z}} \), and \(\nabla\mathbf{f} \triangleq\partial f_x/\partial x + \partial f_y/\partial y + \partial f_z/\partial z  \), where \(\hat{\mathbf{x}}=[1,0,0]^{\rm T},\hat{\mathbf{y}}=[0,1,0]^{\rm T},\) and \(\hat{\mathbf{z}}=[0,0,1]^{\rm T}\) are the 3D Cartesian unit vectors.

%\section{Modeling of 2D Waveguide-Fed Metasurface}\label{sec: Modeling Magnetic Dipoles Only}
\section{PPW-Fed Metasurfaces:\\Modeling as Magnetic Dipoles}\label{sec: Modeling Magnetic Dipoles Only}

\begin{figure}
    \centering
    \includegraphics[width=\linewidth]{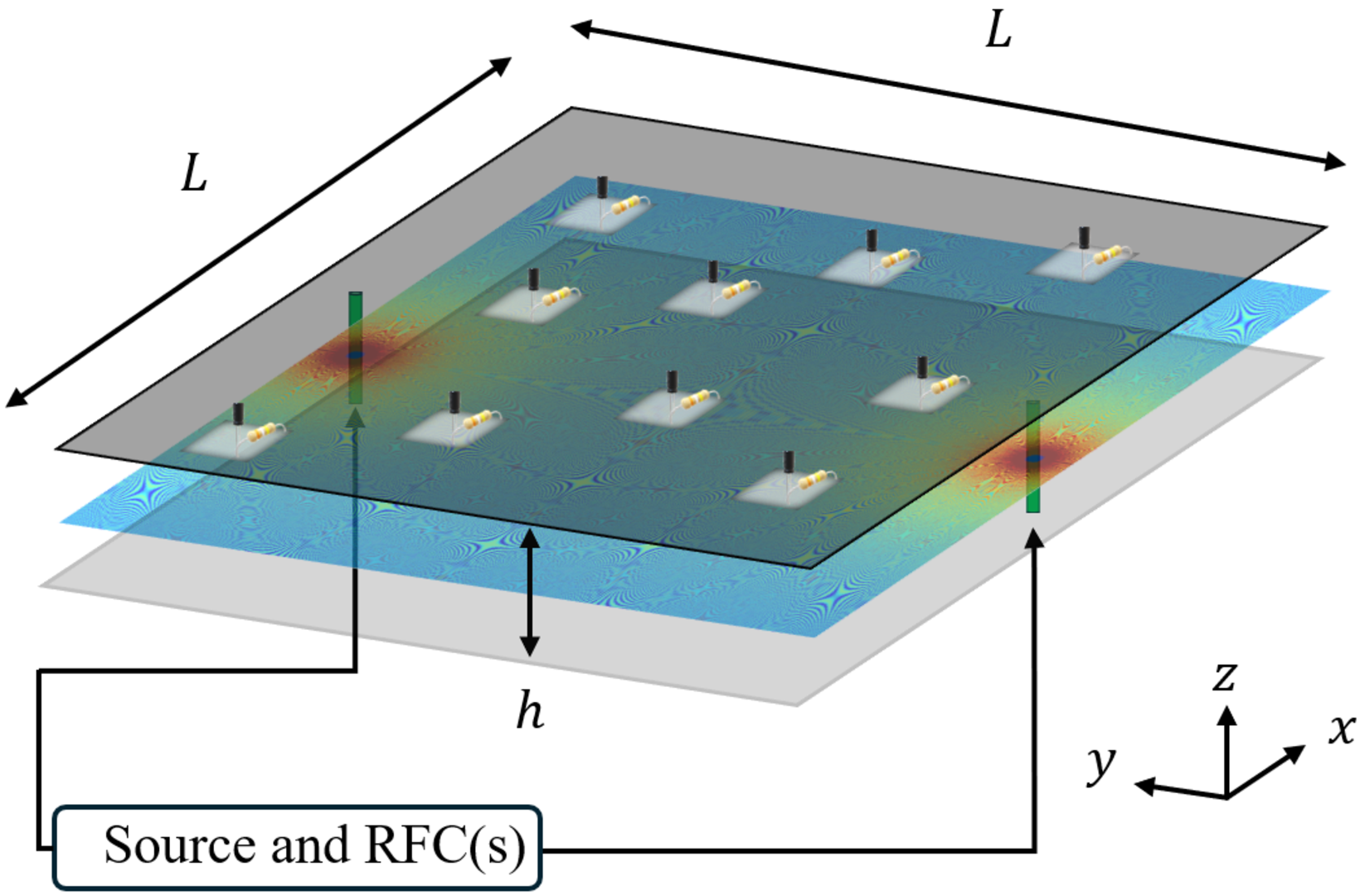}
    \caption{The considered 2D PPW-fed metasurface antenna array comprising randomly distributed reconfigurable elements, for the example case of two current source feeds. The constituent elements are illustrated as subwavelength gaps in the top metallic layer; in one possible physical implementation, each of them can be loaded with a tunable capacitor and a resistor. In the analysis presented in this paper, each metamaterial element is modeled more generally via its effective polarizability, whereas each feed is modeled as a thin-wire cylinder excited by a transmit RFC, emulating coaxial probe excitation. The elements are etched on the top layer (an $L\times L$ plate), and the bottom layer is a metallic plane (again of size $L\times L$) which, together with the top layer, forms the waveguide of height $h$. The drawing also includes as a virtual intermediate layer a heatmap with the magnitude of the guided electric field inside the PPW for the case of an in-phase and equal amplitude excitation of the feeds.}
    \label{fig:metasurface reconfig}
\end{figure}

We consider a PPW-fed metasurface antenna array, as shown in Fig.~\ref{fig:metasurface reconfig}, comprising multiple metamaterials etched on the top plate of the waveguide, acting as the antenna elements, and multiple thin wires in between the two plates, each excited by a digital source along with a transmission RFC. This array structure, considered in this paper as an XL multi-antenna Transmitter (TX), exhibits strong mutual coupling since its constituent elements interact both through the guided mode inside the waveguide and via radiation in free space.
%TX implemented as a 2D waveguide-fed metasurface of height \(h\), filled with air (Fig.~\ref{fig:metasurface fig}). This configuration exhibits strong mutual coupling, as the metamaterials etched on the top plate of the waveguide interact both through the guided mode inside the waveguide and via radiation in free space.
To accurately capture these interactions, we adopt the CD formalism~\cite{pulidomancera2018,CDF_tacit}. Following relevant studies in waveguide-fed metasurfaces \cite{williams2023EM_DMA,pulidomancera2018,davidsmith2017}, a simplified model, in which each metamaterial element is represented solely as a magnetic dipole, is first considered, serving as a baseline model. Then, in Section~\ref{sec: Both electric and magnetic dipoles}, the model will be extended to incorporate both electric and magnetic dipoles, leading to a more general and physically complete description of the PPW-fed metasurface antenna array architecture under investigation.

Under the magnetic dipole model, each metamaterial element is generally associated with a $3$-Dimensional (3D) polarizability matrix relating the local magnetic field to the induced dipole moments. However, in the PPW configuration, the magnetic field component normal to the surface vanishes due to the boundary conditions imposed by the conducting plates. Consequently, only the in-plane components are excited, leading to a \(2 \times 2\) magnetic polarizability matrix. Without loss of generality, and following \cite{pulidomancera2018}, we consider the metasurface to lie in the $x$–$y$ plane at $z=0$ (see Fig.~\ref{fig:metasurface reconfig}), thus, the polarizability matrix for each $n$-th element ($n=1,\ldots,N$, with $N$ denoting the total number of metamaterials) is characterized as follows: 
\begin{equation}\label{eq: Polarizability Matrix}
    \mathbf{A}_n \triangleq
\begin{bmatrix}
\alpha_n^{xx} & \alpha_n^{xy} \\
\alpha_n^{yx} & \alpha_n^{yy}
\end{bmatrix},
\end{equation}
where \(\alpha^{xx}_n\) and \(\alpha^{yy}_n\) denote the self-polarization terms, while \(\alpha^{xy}_n\) and \(\alpha^{yx}_n\) correspond to the cross-polarization terms. It is noted that, the polarizability is, in general, frequency dependent, however, for notational simplicity, the explicit dependence of $\mathbf{A}_n$, and its entries, on the frequency, \(\omega\), is omitted in the remainder of the paper. For reference, in single-polarized elements with polarizability \(\alpha_n\), the frequency dependence of the polarizability is typically modeled using a Lorentzian-type resonance~\cite{Novotny_Hecht_2006}:
\begin{equation}
    \alpha_n(\omega) \triangleq \frac{F_n \omega^2}{\omega_{0,n}^2 - \omega^2 + \jmath \Gamma_n \omega},
    \label{eq: Lorentzian}
\end{equation}
where \(F_n\) is a real constant related to the resonance strength, \(\omega_{0,n}\) is the resonance frequency of the \(n\)-th element, and \(\Gamma_n\) accounts for radiation damping and absorption losses. It is noted that, due to reciprocity in linear, time-invariant media, the polarizability matrix is symmetric, i.e., \(\mathbf{A}_n = \mathbf{A}^{\rm T}_n\). Furthermore, if each element is symmetric with respect to both axes of the waveguide, \(\mathbf{A}_n\) becomes diagonal, implying that inter-polarization coupling is eliminated \cite{bohren2008absorption}. %It is noted that polarizability is, in general, frequency-dependent, i.e., \(\mathbf{A}_n = \mathbf{A}_n(\omega)\). For notational simplicity, the explicit dependence on \(\omega\) is omitted in the remainder of the paper. 
% In the book FIELD THEORY OF GUIDED WAVES table 12.1 there are polarizability matrices for examples of passive elements. 
Finally, let $\mathbf{m}_n\triangleq[m_n^x,m_n^y] \in \mathbb{C}^{2 \times 1}$ denote the magnetic dipole moment of each $n$-th element, and $\mathbf{h}_{\text{loc},n}\triangleq[h^x_{\text{loc},n},h^y_{\text{loc},n}]\in \mathbb{C}^{2 \times 1}$ represent the local magnetic field at this element's position. These two features are related through $\mathbf{A}_n$ as follows:
\begin{equation}
    \mathbf{m}_n = \mathbf{A}_n \mathbf{h}_{\text{loc},n}.
\end{equation}
% UNITS: dipole moment m has units A*m^2, polarizabilities in m^3 units, the magnetic field in A/m units. In this paper, we define Green functions with m^-3 units so [Green * dipole moment] is A/m that corresponds to the units of magnetic field.

The local field acting on each $n$-th element is given by the sum of the guided excitation $\mathbf{h}_{0,n}\triangleq [h^{x}_{0,n},h^y_{0,n}]\in \mathbb{C}^{2 \times 1}$ from the feed and the scattered fields generated by all other dipoles:
\begin{equation}
    \mathbf{h}_{\text{loc},n} = \mathbf{h}_{0,n} + %\sum_{\substack{j=1 \\ j\neq n}}^{N} 
    \sum_{j=1, j\neq n}^{N}\mathbf{G}_{n,j}\, \mathbf{m}_j,
    \label{eq:Hloc_expanded_final}
\end{equation}
where $\mathbf{G}_{n,j}\in \mathbb{C}^{2 \times 2}$ models the interaction between the $j$-th and $n$-th elements, while the self-term is excluded. Defining $\mathbf{r}_n \triangleq [r_{x_n},r_{y_n},0]$ as the position vector of the $n$-th element, where we imply \(r_{z_{n}}=0\,\forall n\), the interactions are modeled as:
\begin{equation}\label{eq: Total Green Function}
\mathbf{G}_{n,j} \triangleq 
\begin{cases}
\mathbf{G}_{\text{WG}}\left(\mathbf{r}_n,\mathbf{r}_j\right)+ \mathbf{G}_{\text{FS}}\left(\mathbf{r}_n,\mathbf{r}_j\right), & n\neq j\\
0, & n=j
\end{cases},
\end{equation}
with $\mathbf{G}_{\text{WG}}(\cdot,\cdot)$ describing the coupling through the waveguide and $\mathbf{G}_{\text{FS}}(\cdot,\cdot)$ the coupling through free space for \(x\)- and \(y\)-polarized fields. For the considered PPW of height \(h\), the waveguide contribution is obtained by following the methodology of \cite{pulidomancera2018}. For completeness, the corresponding expressions are also independently derived in Appendix~\ref{app: Magnetic and electric dipoles propagation}. The resulting Green's function components are given by:
\begin{align}\label{eq: Waveguide Green}
    \!\!\!{G}^{xx}_{\text{WG}}\left(\mathbf{r}_n,\mathbf{r}_j\right)
    \!= & \frac{\text{-}\jmath k^2}{8h}\!\left(
        H^{(2)}_{0}\!\left(k\rho_{n,j}\right)\!+\! \cos\left(2\psi_{n,j}\right)H^{(2)}_{2}\!\left(k\rho_{n,j}\right)\right)\!,\nonumber\\
\!\!\!{G}^{xy}_{\text{WG}}\left(\mathbf{r}_n,\mathbf{r}_j\right)\!
    = & \frac{\text{-}\jmath k^2}{8h} \!\sin\!\left(2\psi_{n,j}\right)\!H^{(2)}_{2}\!\left(k\rho_{n,j}\right),{G}^{yx}_{\text{WG}}
    \!=\!{G}^{xy}_{\text{WG}},\\
\!\!\!{G}^{yy}_{\text{WG}}\left(\mathbf{r}_n,\mathbf{r}_j\right)
    \!=\! & \frac{\text{-}\jmath k^2}{8h}\!\left(
        H^{(2)}_{0}\!\left(k\rho_{n,j}\right)\!-\! \cos\left(2\psi_{n,j}\right)H^{(2)}_{2}\!\left(k\rho_{n,j}\right)\right)\!.\nonumber
\end{align}
The dyadic Green's function, relating the magnetic moment to magnetic field in free space, is derived from~\cite[eq.~(8.52)]{Novotny_Hecht_2006} by: 1) using the duality between electric and magnetic properties from \cite[Table~7.2]{balanis2012advanced}; 2) keeping only the terms associated with the \(x\) and \(y\) polarizations; and 3) accounting for the image of the dipole created due to the metallic surfaces that it is placed on, yielding the following expression for $\mathbf{G}_{\text{FS}}(\cdot,\cdot)$:
\begin{align}
    & \mathbf{G}_{\text{FS}}(\mathbf{r}_n,\mathbf{r}_j) = \left(
    \left(\frac{3}{k^2 \rho_{n,j}^2}+\frac{3\jmath}{k\rho_{n,j}}-1\right)\mathbf{R}\right.\label{eq: Free Space Green}\\
    &\left. \quad \qquad \qquad\qquad+\left(1-\frac{\jmath}{k\rho_{n,j}}-\frac{1}{k^2 \rho_{n,j}^2}\right)\mathbf{I}_2
    \right)\frac{k^2e^{-\jmath k\rho_{n,j}}}{2\pi\rho_{n,j}},\nonumber\\
     &\mathbf{R}\triangleq \begin{bmatrix}
\cos^2(\psi_{n,j}) & \cos(\psi_{n,j})\sin(\psi_{n,j}) \\
\cos(\psi_{n,j})\sin(\psi_{n,j}) & \sin^2(\psi_{n,j})
\end{bmatrix}.\label{eq: Projection matrix}
\end{align}
In expresssions~\eqref{eq: Waveguide Green}--\eqref{eq: Projection matrix}, $H^{(2)}_{\nu}(\cdot)$ denotes the Hankel function of the second kind and order $\nu$\cite[eqs.~(V-14) and (V-15)]{balanis2016antenna}, $k$ is the propagation constant of the guided mode, $\rho_{n,j}\triangleq\lvert\mathbf{r}_n-\mathbf{r}_j\rvert=\sqrt{(r_{x_n}-r_{x_j})^2+(r_{y_n}-r_{y_j})^2}$, \(\mathbf{R}\) is the projection matrix, and $\psi_{n,j}\triangleq\mathrm{atan}\left((r_{y_n}-r_{y_j})/(r_{x_n}-r_{x_j})\right)$. 

We henceforth extend the single-feed-based formulation of~\cite{pulidomancera2018} to the general case of \(N_{\rm f}\geq1\) excitation feeds. Those components are modeled as thin-wire current sources whose excitation currents are prescribed and independently controlled. This corresponds to a current-driven excitation model that enforces the desired currents at the feed terminals. Under this assumption, mutual coupling does not modify the imposed currents, but instead affects the terminal voltages required to sustain them, and consequently the injected power. 
Owing to the linearity of Maxwell’s equations, the magnetic field incident on each \(n\)-th metasurface element due to the existence of \(N_{\rm f}\) feeds is given by the superposition of the fields radiated by each feed, yielding the expressions:
\begin{equation}\label{eq: Excitation Field}
\begin{split}
    h^{x}_{0,n}\!\!\triangleq& \frac{\jmath k}{4}\!\sum_{i=1}^{N_{\rm f}}\!I_i \, H^{(2)}_{1}\!\!\left(k|\mathbf{r}_n-\mathbf{b}_i|\right)\!\sin\!\!\left(\!\!\mathrm{atan}\!\left( \frac{r_{y_n}-b_{y_i}}{r_{x_n}-b_{x_i} }\right)\!\!\right),\\
    h^{y}_{0,n}\!\!\triangleq& \frac{\text{-}\jmath k}{4}\!\sum_{i=1}^{N_{\rm f}}\!I_i \, H^{(2)}_{1}\!\!\left(k|\mathbf{r}_n-\mathbf{b}_i|\right)\!\cos\!\!\left(\!\!\mathrm{atan}\!\left(\! \frac{r_{y_n}-b_{y_i} }{r_{x_n}-b_{x_i} }\!\right)\!\!\right)\!,
\end{split}
\end{equation}
where \(\mathbf{b}_i \triangleq [b_{x_i}, b_{y_i}, 0]\) denotes the position of the \(i\)-th feed ($i=1,\ldots,N_{\rm f}$), and \(I_i\) represents the externally applied current at that feed. The considered feeds effectively generate a single-polarized excitation, with the magnetic field oriented along the azimuthal direction in cylindrical coordinates. However, since the metasurface elements are characterized by distinct \(x\)- and \(y\)-directed polarizabilities, it is convenient to express the induced magnetic field in Cartesian components. This representation highlights that the metasurface can independently control the responses along the two in-plane directions through its polarizability matrix. As a result, although the excitation is single-polarized, the aperture can produce different responses for the \(x\)- and \(y\)-components, enabling, interestingly, polarization-dependent beamforming.

For compactness, the concatenated dipole moment vector is defined as $\mathbf{m} \triangleq [\mathbf{m}_1,\ldots,\mathbf{m}_N]  \in \mathbb{C}^{2N\times 1}$. Analogously, the polarizability matrix containing the polarizabilities of all elements is defined as $\bar{\mathbf{A}} \triangleq {\rm diag} [\mathbf{A}_1,\ldots,\mathbf{A}_N] \in \mathbb{C}^{2N\times 2N}$, i.e., \([\bar{\mathbf{A}}]_{2n-1:2n,2n-1:2n}=\mathbf{A}_n\), and similarly, we define the interaction array \(\bar{\mathbf{G}}\in \mathbb{C}^{2 N \times 2 N}\) for which \([\bar{\mathbf{G}}]_{2n-1:2n,2j-1:2j}=\mathbf{G}_{n,j}\) \(\forall \,n,j=1,\ldots,N\). Additionally, the excitation-field vector is expressed as $\mathbf{h}_0 \triangleq [\mathbf{h}_{0,1},\ldots,\mathbf{h}_{0,N}] \in \mathbb{C}^{2N\times 1}$. Using the latter definitions, \eqref{eq:Hloc_expanded_final} holds for all $N$ metamaterial elements simultaneously and, by solving collectively for the stacked dipole moment vector, the following expression is deduced:
\begin{equation}\label{eq: Dipole Moment}
    \mathbf{m} = \left(\bar{\mathbf{A}}^{-1} - \bar{\mathbf{G}}\right)^{-1}\mathbf{h}_0.
\end{equation}
In fact, in this expression, we can include the explicit contribution of the source currents \(\{I_i\}_{i=1}^{N_{\rm f}}\), as follows. Let us introduce the matrix \(\mathbf{H}_{\rm f}\in\mathbb{C}^{2N \times N_{\rm f}}\), where \([\mathbf{H}_{\rm f}]_{2n-1,i}\triangleq\frac{\jmath k}{4}H^{(2)}_{1}\!\left(k|\mathbf{r}_n-\mathbf{b}_i|\right)\sin\!\left(\mathrm{atan}\!\left( \frac{ r_{y_n}- p_{y_i}}{ r_{x_n}-p_{x_i}}\right)\right)\) and \([\mathbf{H}_{\rm f}]_{2n,i}\triangleq\frac{-\jmath k}{4}H^{(2)}_{1}\!\left(k|\mathbf{r}_n-\mathbf{b}_i|\right)\cos\!\left(\mathrm{atan}\!\left( \frac{r_{y_n}-p_{y_i}}{ r_{x_n}-p_{x_i}}\right)\right)\). Also, by introducing the definition of the currents' vector \(\mathbf{i}\triangleq [I_1,\ldots,I_{N_{\rm f}}]\in \mathbb{C}^{N_{\rm f}\times 1}\), $\mathbf{h}_0$ in~\eqref{eq: Dipole Moment} can be expressed as follows:
\begin{equation}\label{eq: Excitation Field in Matrix Form}
    \mathbf{h}_0 = \mathbf{H}_{\rm f} \mathbf{i}.
\end{equation}
Expressions~\eqref{eq: Dipole Moment} and~\eqref{eq: Excitation Field in Matrix Form} provide a compact characterization of the underlying coupled-dipole-based physics for the considerd PPW-fed metasurface antenna arrays. The final expression linking electric currents in the sources to dipole moments is:
\begin{equation}\label{eq: Dipole Moment from currents}
    \mathbf{m} = \left(\bar{\mathbf{A}}^{-1} - \bar{\mathbf{G}}\right)^{-1}\mathbf{H}_{\rm f} \mathbf{i}.
\end{equation}

\subsection{Power Conservation for a Single Dipole}\label{subsec: Power Conservation Magnetic Dipoles}
For the considered metamaterial elements lacking any signal amplification capability, the magnetic polarizability of each dipole needs to satisfy a passivity constraint following the conservation of energy principle: the power supplied to each $n$-th dipole, $P_{\mathrm{sup},n}$, cannot be smaller than its radiated power, $P_{\mathrm{rad},n}$. While this principle has been established for electric dipoles in free space \cite{Tretyakov2000,bohren2008absorption}, it is missing for magnetic dipoles in a 2D waveguide, despite their common use in modeling metamaterial elements~\cite{pulidomancera2018,davidsmith2017,williams2023EM_DMA}.

Let us assume time-harmonic fields with the convention $\exp(\jmath\omega t)$. For this convention and considering passive elements with positive radiation damping, the associated Lorentz-oscillator model implies that the imaginary part of the polarizability is \emph{negative} \cite{bohren2008absorption} (see \eqref{eq: Lorentzian}).  A negative imaginary part signifies that the dipole absorbs power from the local field rather than supplying power to it, which is the defining characteristic of a passive element. Note that, when matrices are in place instead of scalar polarizability values, this requirement generalizes to the imaginary part of the polarizability matrix in~\eqref{eq: Polarizability Matrix}, \(\rm Im\{\mathbf{A}_n\}\), being a negative definite matrix. The physics behind the latter condition will become apparent in the sequel, where it will be shown that this needs to serve as a necessary condition so that the supplied power in the dipole is positive.

It follows from the Poynting theorem~\cite[eq.~(1--76b)]{balanis2012advanced} that the power absorbed by the $n$-th dipole due to the environment, i.e., other elements and the feeds, can be derived by the following integration over its effective area $A_n$:
\begin{equation}\label{eq:Poynting_Thm}
P_{\mathrm{sup},n}
\triangleq \tfrac{1}{2}{\rm Re}\!\left\{\int_{A_n} \mathbf{j}^{\rm T}_n(\mathbf r)\mathbf{h}_{\mathrm{loc},n}^{*}(\mathbf r)d\mathbf r\right\},
\end{equation}
where $\mathbf{j}_n(\mathbf r)\triangleq\jmath\omega\mu_0 (m_n^x\delta(x-r_{x_n})\hat{\mathbf{x}}_{2{\rm D}} +m_n^y\delta(y-r_{y_n})\hat{\mathbf{y}}_{2{\rm D}} )$ is the magnetic current density \cite[eq.~(8.49)]{Novotny_Hecht_2006}, with \(\hat{\mathbf{x}}_{2{\rm D}}\triangleq[1 ,0]^{\rm T}\) and \(\hat{\mathbf{y}}_{2{\rm D}}\triangleq[0,1]^{\rm T}\) being the 2D unit vectors of the \(x\)- and \(y\)-axis, respectively.  
Due to the delta function in $\mathbf{j}_n(\mathbf r)$, the integration collapses to computing a single term in the dipoles position, hence, \eqref{eq:Poynting_Thm} becomes as follows:
\begin{equation}\label{eq:Psup_final}
P_{\mathrm{sup},n}
= -\frac{\omega\mu_0}{2}\,{\rm Im}\left\{\mathbf{h}^{\rm H}_{\text{loc},n}\mathbf{A}_n\mathbf{h}_{\text{loc},n}\right\},
\end{equation}
which is equivalent to inserting the imaginary operator only on \(\mathbf{A}_n\) as: \( \mathbf{h}^{\rm H}_{\text{loc},n}{\rm Im}\{\mathbf{A}_n\}\mathbf{h}_{\text{loc},n}\), since \(\mathbf{A}_n\) is symmetric. Hence, for the supplied power to be positive, the imaginary part of matrix \(\mathbf{A}_n\) needs to be negative, i.e., ${\rm Im}\{\mathbf{A}_n\}<\mathbf{0}_2$.

On the other hand, to compute the radiated power, the Poynting theorem can be again used, but the induced magnetic field needs to be substituted by the scattered field of the $n$-th element. To this end, the sign needs to change, since now the power flowing out of the volume is computed, yielding \(P_{{\rm rad},n}\triangleq\text{-}0.5{\rm Re}\!\left\{\int_{A_n} \mathbf{j}^{\rm T}_n(\mathbf r)\mathbf{h}_{\mathrm{sc},n}^{*}(\mathbf r)d\mathbf r\right\}\). In this expression, the scattered field can be expressed as $\mathbf{h}_{\mathrm{sc},n}(\mathbf r)\triangleq \mathbf{G}(\mathbf r,\mathbf r_n)\mathbf{m}_n$, with $\mathbf{G}(\cdot)\triangleq\mathbf{G}_{\mathrm{FS}}(\cdot)+\mathbf{G}_{\mathrm{WG}}(\cdot)$,
which yields:
\begin{equation}\label{eq:Prad_1}
P_{\mathrm{rad},n}
= -\tfrac{\omega\mu_0}{2}
\mathbf{h}^{\rm H}_{\text{loc},n}\mathbf{A}_n^{\rm H}{\rm Im}\!\left\{\mathbf{G}(0)\right\}\mathbf{A}_n\mathbf{h}_{\text{loc},n}.
\end{equation}
This radiated power is positive since \({\rm Im}\{\mathbf{G}(0)\}\) is a negative definite matrix. 
Evaluating the singularity terms in the four entries of the self-Green's function, yields ${\rm Im}\{\mathbf{G}_{\mathrm{FS}}(0)\}=-k^3/(3\pi)\mathbf{I}_2$ \cite{tretyakov2020magneticdipoles,williams2023EM_DMA}. This can be derived by taking the limit of \eqref{eq: Free Space Green} at $\mathbf r_n\to\mathbf r_j$ and using the Taylor expansion of $\exp(-j k\rho)$ near $\rho=0$ \cite[eq.~(4)]{tretyakov2020magneticdipoles}. For the waveguide term, the singular parts of the Hankel functions are computed via \cite[eqs.~(V-11) and (V-12)]{balanis2016antenna}, yielding ${\rm Im}\{\mathbf{G}_{\mathrm{WG}}(0)\}=-k^2/(8h)\mathbf{I}_2$; recall that $h$ denotes the waveguide height. Hence, combining the above derivations, the imaginary part of the self-term of the 2D Green's matrix is given as follows:
\begin{equation}\label{eq: Self-term Green}
    {\rm Im}\{\mathbf{G}(0)\} = -\left(k^3/(3\pi)+k^2/(8h)\right)\mathbf{I}_2.
\end{equation}

When imposing energy conservation for a passive element $n$, i.e.,
\(
P_{\mathrm{sup},n} \geq P_{\mathrm{rad},n},
\)
by substituting \eqref{eq:Psup_final} and \eqref{eq:Prad_1}, yields:
\begin{equation}
\mathbf{A}_n^{\rm H}{\rm Im}\left\{\mathbf{G}(0)\right\}\mathbf{A}_n - {\rm Im}\left\{\mathbf{A}_n\right\} \geq \mathbf{0}_2.
\end{equation}
Using the matrix identity ${\rm Im}\{\mathbf{A}_n\}=-\mathbf{A}_n{\rm Im}\{\mathbf{A}^{-1}_n\}\mathbf{A}_n^{\rm H}$ \cite{petersen2008matrix}, the latter passivity constraint can be rewritten as follows:
\begin{equation}\label{eq: passivity_constraint}
{\rm Im}\{ \mathbf{A}_n^{-1} \}+{\rm Im}\{\mathbf{G}(0)\}\geq \mathbf{0}_2.
\end{equation}
The latter condition can be interpreted as: the imaginary part of the matrix \(\mathbf{A}^{-1}+\mathbf{G}(0)\) must be semi-positive definite. This condition is general referring also to other architectures (i.e., not only restricted to the 2D waveguide considered in this paper), but the exact values of \(\mathbf{G}(0)\) need to be computed for a specific setup; for example, \eqref{eq: Self-term Green} reveals the necessary passivity condition for the present case.

It is noted that, instead of enforcing~\eqref{eq: passivity_constraint} as an explicit inequality, one can build it directly into the definition of the polarizability via the standard Radiation--Reaction (RR) correction \cite{tretyakov2003analytical}. For this, we distinguish between an intrinsic polarizability $\mathbf{A}_{n}^{\rm int}$, which is obtained from a quasi-static model (such as \cite[eq.~(27)]{mancera2017polarizability}) and does \emph{not} include the dipole’s self-field\footnote{In the intrinsic polarizability definition, the constitutive relations of the local magnetic field would include, in the summation of~\eqref{eq:Hloc_expanded_final}, the term related to the self-field, as also shown in~\cite[eq.~(3)]{mancera2017polarizability}.}, and the effective polarizability $\mathbf{A}_n$, which includes it. Essentially, the intrinsic polarizability is the property of each $n$-th element due to its geometry, without accounting for the radiation environment, while the effective polarizability accounts for the environment (the 2D waveguide in our case). For a dipole $n$ embedded in the PPW, the self-interaction is represented by the Green’s function evaluated at the dipole location, hence, the RR-corrected polarizability is written as: 
\begin{equation}\label{eq: rr_correction}
\mathbf{A}_n
= \mathbf{A}_{n}^{\rm int} \left(\mathbf{I}_2 - \jmath {\rm Im}\{\mathbf{G}(0)\}\mathbf{A}_{n}^{\rm int}\right)^{-1},
\end{equation}
which is the matrix equivalent of \cite[eq.~(1)]{mancera2017polarizability}.
Taking the imaginary parts in both sides of this expression and after performing some algebraic manipulations, yields:  
\begin{equation}
{\rm Im}\left\{\mathbf{A}^{-1}_n\right\}={\rm Im}\left\{(\mathbf{A}_{n}^{\rm int})^{-1}\right\} - {\rm Im}\{\mathbf{G}(0)\}.
\end{equation}
Since, for a passive element $n$, it holds ${\rm Im}\{(\mathbf{A}_{n}^{\rm int})^{-1}\}> \mathbf{0}_2$, then, it directly applies that:  
\(
{\rm Im}\left\{\mathbf{A}_n^{-1}\right\} + {\rm Im}\{\mathbf{G}(0)\}
\;\geq\;
\mathbf{0}_2,
\)
which is exactly the passivity condition in~\eqref{eq: passivity_constraint}. In the lossless limit, ${\rm Im}\{\mathbf{A}_n^{\rm int}\}=0$, indicating that the corrected polarizability \eqref{eq: rr_correction} saturates the bound, i.e., all dissipation is purely radiative.

\subsection{Power Conservation for a CD System}\label{subsec: Power Conservation Coupled Dipole System}

%Though the previous derivations are sufficient for single element passivity, 
In environments with multiple dipoles and increased coupling among them, the radiated power of a single dipole cannot be computed only via its self-field. This is due to the fact that, since the same dipole radiates to other nearby elements which, in turn, radiate back to the original element, the radiated power inevitably includes mutual coupling terms. To establish power conservation for a system with $N$ dipoles, as our PPW-fed metasurface antenna array, we compare the power supplied to the dipoles by the external excitation with the power they radiate. We distinguish between the feeding and metamaterial systems, where the supplied power corresponds to the impact of the incident (guided) field generated by the feeds on the dipoles, and the radiated power corresponds to the power emitted by the dipoles into the surrounding environment. The total supplied power to all dipoles from the feeds is given as:
\begin{align}\label{eq: total supplied power magnetic}
    P_{\rm sup}\triangleq &\sum_{n=1}^N 0.5{\rm Re}\left\{ \jmath \omega \mathbf{m}_n^{\rm T}\mathbf{h}_{0,n}^{*}\right\}\\
    =& 0.5\omega \mu_0{\rm Im}\left\{\mathbf{m}^{\rm H}(\bar{\mathbf{A}}^{-1}-\bar{\mathbf{G}})\mathbf{m}\right\}.\nonumber
\end{align}
On the other hand, for the total radiated power from the dipoles, we use both the self-term and the mutual interaction terms, i.e., \(\mathbf{h}_{{\rm sc},n}\triangleq (\mathbf{G}(0) +\sum_{j=1,i\neq n}^N \mathbf{G}_{j,n} )\mathbf{m}_n\). Hence, by defining \(\bar{\mathbf{G}}_{\rm full} \triangleq \jmath {\rm Im}\{\mathbf{G}(0)\}\otimes \mathbf{I}_{N}+\bar{\mathbf{G}}\) and letting \(P_{\rm rad}\) denote the total radiated power of all $N$ metamaterials, it holds:
\begin{equation}\label{eq: total radiated power magnetic}
    P_{\rm rad} = -0.5 \omega \mu_0{\rm Im}\left\{\mathbf{m}^{\rm H}\mathbf{G}_{\rm full}\mathbf{m}\right\}.
\end{equation}
It is shown in Appendix~\ref{app: power magnetic} that \(P_{\rm sup}\geq P_{\rm rad}\) holds automatically when the RR correction from \eqref{eq: rr_correction} is applied.

\subsection{Integration into MIMO Systems}
\subsubsection{MIMO Channel}\label{MIMO_channel_magnetic}
By solving the radiation problem in the domain outside of the waveguide, the electric field radiated from each \(n\)-th metamaterial, \(\mathbf{e}_{{\rm sc},n} \in \mathbb{C}^{2 \times 1}\), can be expressed as a function of its magnetic moment \(\mathbf{m}_n\). In~\cite{pulidomancera2018}, using the FF approximation, this electric field was derived for the free space. In this paper, we do not make the complete FF approximation, but, as per the radiative NF approximation~\cite{liu2023near}, we only eliminate terms divided by the square distance, \(R^2_n\), between the \(n\)-th element and the observation point. To this end, we compute the electric field in an arbitrary point \(\mathbf{s}_{\ell}\in \mathbb{R}^{3\times 1}\) in free space as \(\mathbf{e}_{{\rm sc},n}(\mathbf{s}_{\ell})=\jmath \omega \mu_0 (\nabla \!\times \!\tilde{\mathbf{G}}_{\rm FS}(\mathbf{s}_{\ell},\mathbf{r}_n))\mathbf{m}_n\), where \(\tilde{\mathbf{G}}_{\rm FS}\in \mathbb{C}^{3 \times 3}\) represents the dyadic Green function in free space, as given in \cite[eq.~(8.61)]{Novotny_Hecht_2006}, but doubled due to the dipole image created by the top wall of the waveguide\footnote{To circumvent the size incompatibility issue in the \((\nabla\!\times\! \tilde{\mathbf{G}}_{\rm FS}(\mathbf{s}\text{-}\mathbf{r}_n))\mathbf{m}_n\) multiplication, a zero needs to be padded to \(\mathbf{m}_n\).}. Then, since the use of spherical coordinates can simplify the analysis due to the fact that the longitudinal component of both the electric and magnetic fields (i.e., the component parallel to propagation) is zero, the electric field can be transformed from Cartesian to spherical coordinates. To this end, only the azimuth, \(\phi\), and elevation, \(\theta\), components are of interest. Consequently, the electric field vector \(\mathbf{e}_{{\rm sc},n}(\mathbf{s}_{\ell})\) at an observation point \(\mathbf{s}_{\ell}\), due to the existence of the $n$-th dipole located at \(\mathbf{r}_n\), consists of the following two components:
\begin{align}\label{eq: electric_field_due_to_nth_dipole}
\!\!\! \! e_{{\rm sc},n}^{\theta}(\mathbf{s}_{\ell})\!
&\triangleq\!
\frac{\eta k^{2}e^{-\jmath kR_{\ell,n}}}{2\pi  R_{\ell,n}}\!
\big(m_{n}^{x}\sin(\phi_{\ell,n}) \!-\! m_{n}^{y}\cos(\phi_{\ell,n})\big),\\[2mm]
\!\!\!\! e_{{\rm sc},n}^{\phi}(\mathbf{s}_{\ell})
\!&\triangleq\!
\frac{\eta k^{2}e^{-\jmath kR_{\ell,n}}}{2\pi R_{\ell,n}}
\!\big(\!m_{n}^{x}\!\cos(\phi_{\ell,n}) \!\!+ \!\!m_{n}^{y}\sin(\phi_{\ell,n})\!\big)\!\!\cos(\theta_{\ell,n}),\nonumber
\end{align}
where \(\eta\) represents the free-space impedance, \(R_{\ell,n}\) is the radial distance between \(\mathbf{s}_{\ell}\) and \(\mathbf{r}_n\), and \(\theta_n \triangleq{\rm acos}\left({s_{\ell_{z}}}/{R_{\ell,n}}\right)\) as well as \(\phi_{\ell,n} \triangleq {\rm atan}\left( ({s_{\ell_{y}}- r_{y_n}})/({s_{\ell_{x}}- r_{x_n}})\right)\) are respectively the elevation and azimuth angles between them. 
In analogy with the focusing vector in conventional MIMO systems,  
the electric field components scattered toward the observation point 
$\mathbf{s}_{\ell}$ can be written in a linear form w.r.t. the stacked
magnetic moments $\mathbf{m}\in\mathbb{C}^{2N\times 1}$.  
However, due to the radiative NF formulation, the elevation and azimuth angles between each \(n\)-th dipole and the observation point, i.e.,
$\theta_{\ell,n}$ and $\phi_{\ell,n}$, respectively, depend on the specific dipole location
$\mathbf{r}_n$, and, thus, the field components
$e_{{\rm sc},n}^{\theta}(\mathbf{s}_{\ell})$ and
$e_{{\rm sc},n}^{\phi}(\mathbf{s}_{\ell})$ in
\eqref{eq: electric_field_due_to_nth_dipole} are expressed in
dipole-dependent local spherical bases. As a result, these components do not correspond to a common polarization basis and cannot be directly summed.

To enable coherent superposition, each dipole contribution can be projected onto a common spherical basis associated with the observation point $\mathbf{s}_\ell$, and defined w.r.t. the TX center. Let $(\theta_\ell,\phi_\ell)$ denote the elevation and azimuth
angles of $\mathbf{s}_\ell$ w.r.t. that TX center, and let
$\hat{\boldsymbol{\theta}}_\ell$ and $\hat{\boldsymbol{\phi}}_\ell$
denote the corresponding spherical-to-Cartesian projection vectors,
defined as
$\hat{\boldsymbol{\theta}}_\ell
\triangleq
[\cos\theta_\ell\cos\phi_\ell,\,
 \cos\theta_\ell\sin\phi_\ell,\,
 -\sin\theta_\ell]^{\mathrm T}$
and
$\hat{\boldsymbol{\phi}}_\ell
\triangleq
[-\sin\phi_\ell,\,
 \cos\phi_\ell,\,
 0]^{\mathrm T}$.
Similarly, $\hat{\boldsymbol{\theta}}_{\ell,n}$ and
$\hat{\boldsymbol{\phi}}_{\ell,n}$ denote the spherical-to-Cartesian
projection vectors associated with the direction
$\mathbf{s}_\ell-\mathbf{r}_n$, obtained by replacing
$(\theta_\ell,\phi_\ell)$ with $(\theta_{\ell,n},\phi_{\ell,n})$ in the
above definitions. Then, we can define the following $2\times2$ projection matrix for each $n$-th dipole:
\begin{equation}\label{eq:Tnell_def}
\mathbf T_{n\rightarrow \ell}
\triangleq
\begin{bmatrix}
\hat{\boldsymbol{\theta}}_\ell^{\mathrm T}\hat{\boldsymbol{\theta}}_{\ell,n} &
\hat{\boldsymbol{\theta}}_\ell^{\mathrm T}\hat{\boldsymbol{\phi}}_{\ell,n}\\
\hat{\boldsymbol{\phi}}_\ell^{\mathrm T}\hat{\boldsymbol{\theta}}_{\ell,n} &
\hat{\boldsymbol{\phi}}_\ell^{\mathrm T}\hat{\boldsymbol{\phi}}_{\ell,n}
\end{bmatrix},
\end{equation}
which maps the transverse components of the $n$-th dipole, \(e_{{\rm sc},n}^{\phi}(\mathbf{s}_{\ell})\) and \(e_{{\rm sc},n}^{\theta}(\mathbf{s}_{\ell})\), from their local basis into the common basis. %, by projecting through $\mathbf T_{n\rightarrow \ell}\,\forall n$. 
To express all dipole contributions in the common spherical basis, we first introduce the local focusing vectors $\mathbf a^{\theta}(\mathbf s_\ell)$ and
$\mathbf a^{\phi}(\mathbf s_\ell)$, which collect the dipole-dependent phase and
angular responses prior to basis unification; these vector are defined element-wise $\forall n=1,\ldots,N$ as follows:
\begin{equation}\label{eq:focusing_vector_definitions}
\begin{alignedat}{2}
    &\big[\mathbf{a}^{\theta}(\mathbf{s}_{\ell})\big]_{2n-1} &&\triangleq {1}/({2\pi R_{\ell,n}})\sin\phi_{\ell,n}\,e^{-jkR_{\ell,n}}, \\
    &\big[\mathbf{a}^{\theta}(\mathbf{s}_{\ell})\big]_{2n}   &&\triangleq {-1}/({2\pi R_{\ell,n}})\cos\phi_{\ell,n}\,e^{-jkR_{\ell,n}}, \\
    &\big[\mathbf{a}^{\phi}(\mathbf{s}_{\ell})\big]_{2n-1}  &&\triangleq {1}/({2\pi R_{\ell,n}})\cos\phi_{\ell,n}\,\cos\theta_{\ell,n}\,e^{-jkR_{\ell,n}}, \\
    &\big[\mathbf{a}^{\phi}(\mathbf{s}_{\ell})\big]_{2n}    &&\triangleq {1}/({2\pi R_{\ell,n}})\sin\phi_{\ell,n}\,\cos\theta_{\ell,n}\,e^{-jkR_{\ell,n}}.
\end{alignedat}
\end{equation}
% \begin{align}\label{eq:focusing_vector_definitions}
%   \left[[\mathbf a^{\theta}(\mathbf{s}_{\ell})]_{2n-1},
% [\mathbf a^{\theta}(\mathbf{s}_{\ell})]_{2n}\right]^{\rm T}
% & =
% \frac{e^{-\jmath k R_{\ell,n}}}{2 \pi R_{\ell,n}}
% [\sin(\phi_{\ell,n}),
% -\cos(\phi_{\ell,n})]^{\rm T},\\  
% \left[[\mathbf a^{\phi}(\mathbf{s}_{\ell})]_{2n-1},
% [\mathbf a^{\phi}(\mathbf{s}_{\ell})]_{2n}
% \right]^{\rm T}
% &=\frac{e^{-\jmath k R_{\ell,n}}\cos(\theta_{\ell,n})}{2 \pi R_{\ell,n}}
% [\cos(\phi_{\ell,n}),
% \sin(\phi_{\ell,n})
% ]^{\rm T}.\nonumber
% \end{align}
%
Consequently, to account for the basis mismatch, the focusing vectors are transformed
element-wise through $\mathbf T_{n\rightarrow\ell}$'s, yielding the
projected (common basis) focusing vectors
$\tilde{\mathbf a}^{\theta}(\mathbf s_\ell)\in\mathbb C^{2N\times1}$ and
$\tilde{\mathbf a}^{\phi}(\mathbf s_\ell)\in\mathbb C^{2N\times1}$, defined $\forall n=1,\ldots,N$ as:
\begin{equation}\label{eq:projected_focusing_vectors}
\begin{alignedat}{2}
&\begin{bmatrix}
    \big[\tilde{\mathbf a}^{\theta}(\mathbf s_\ell)\big]_{2n-1}\\
 \big[\tilde{\mathbf a}^{\phi}(\mathbf s_\ell)\big]_{2n-1}   
\end{bmatrix}
&&\triangleq\mathbf{T}_{n\rightarrow \ell}\begin{bmatrix}
\big[\mathbf a^{\theta}(\mathbf s_\ell)\big]_{2n-1}\\
\big[\mathbf a^{\phi}(\mathbf s_\ell)\big]_{2n-1}
\end{bmatrix},\\
&\begin{bmatrix}
    \big[\tilde{\mathbf a}^{\theta}(\mathbf s_\ell)\big]_{2n}\\
 \big[\tilde{\mathbf a}^{\phi}(\mathbf s_\ell)\big]_{2n}   
\end{bmatrix}
&&\triangleq\mathbf{T}_{n\rightarrow \ell}\begin{bmatrix}
\big[\mathbf a^{\theta}(\mathbf s_\ell)\big]_{2n}\\
\big[\mathbf a^{\phi}(\mathbf s_\ell)\big]_{2n}
\end{bmatrix}.
\end{alignedat}
\end{equation}
Then, the scattered field components at the observation point $\mathbf{s}_\ell$ can be expressed as follows:
\begin{equation}\label{eq: total electric field_projected}
\begin{aligned}
e_{\rm sc}^{\theta}(\mathbf{s}_{\ell})
&=
{\eta k^{2}}\,
\tilde{\mathbf a}^{\theta}(\mathbf s_\ell)^{\mathrm T}\mathbf m,\\
e_{\rm sc}^{\phi}(\mathbf{s}_{\ell})
&=
{\eta k^{2}}\,
\tilde{\mathbf a}^{\phi}(\mathbf s_\ell)^{\mathrm T}\mathbf m .
\end{aligned}
\end{equation}
This expression can be further used to define the dual-polarized channel matrix between the PPW-fed metasurface TX and the \(L\) observation points as
\(\mathbf{H} \in \mathbb{C}^{2L\times 2N}\),
with the rows associated with \(\mathbf{s}_{\ell}\) \(\forall \ell = 1,\ldots,L\) given by:
\begin{equation}\label{eq: channel_matrix_definition}
\begin{alignedat}{2}
&[\mathbf H]_{2\ell-1,:}
&&\triangleq
{\eta k^{2}}\,
\tilde{\mathbf a}^{\theta}(\mathbf s_\ell)^{\mathrm T},\\
&[\mathbf H]_{2\ell,:}
&&\triangleq
{\eta k^{2}}\,
\tilde{\mathbf a}^{\phi}(\mathbf s_\ell)^{\mathrm T}.
\end{alignedat}
\end{equation}
In this expression, each \(2\times2\) submatrix \(\mathbf{H}_{2\ell-1:2\ell,2n-1:2n}\) associates the dual-polarized field at the \(\ell\)-th point \(\mathbf{s}_{\ell}\) with the \(x\) and \(y\) magnetic dipole moments of the \(n\)-th element.
%With this definition, each \((\ell,n)\)-th pair is defined by a \(2\times2\) submatrix of \(\mathbf{H}\) that associates the dual-polarized field at the \(\ell\)-th point \(\mathbf{s}_{\ell}\) with the \(x\) and \(y\) magnetic dipole moments of the \(n\)-th element.

\subsubsection{Received Signal Model} The $2L$-element vector containing the electric field observed at the \(L\) points can be expressed in baseband form as \(\mathbf{y} \triangleq \mathbf{H}\mathbf{m} + \mathbf{n}\), where \(\mathbf{n}\) denotes additive white Gaussian noise, modeling thermal effects, and is distributed as \(\mathcal{CN}(\mathbf{0}_{2L}, \sigma^2 \mathbf{I}_{2L})\). To reformulate this vector w.r.t. the TX's reconfigurable parameters, we substitute \(\mathbf{m}\) for the general $N_{\rm f}$-fed PPW case in~\eqref{eq: Dipole Moment from currents}, yielding the baseband input-output MIMO expression: 
\begin{equation}\label{eq: Received Signal}
    \mathbf{y} = \mathbf{H} \left(\bar{\mathbf{A}}^{-1} - \bar{\mathbf{G}}\right)^{-1}\mathbf{H}_{\rm f} \mathbf{i} + \mathbf{n}.
\end{equation}
 Clearly, since vector \(\mathbf{i}\in\mathbb{C}^{N_{\rm f}\times1}\) bears both the digital precoding and the transmitted data stream, it can be composed as \(\mathbf{i}=\mathbf{Fs}\)~\cite{7037416,Nossek2010CircuitTheoryCommunication}, with \(\mathbf{F}\in\mathbb{C}^{N_{\rm f}\times N_b}\) representing the beamforming matrix precoding the complex-valued column vector $\mathbf{s}$, consisting of the $N_b$ independent data symbols which are usually chosen from a discrete modulation set. %\(\mathbf{s}\sim\mathcal{CN}(\mathbf{0},\mathbf{I}_{N_{b}})\) the information symbol.

% Regarding line source's self-impedance if needed: We adopt the standard regularization strategy (see [DMA paper]) whereby the divergent real part of the self-Green’s function is subtracted, as it is an artifact of the infinitesimal line-source model and does not correspond to any physical loss mechanism. The resulting finite imaginary part \(h/4\) is retained. Any actual ohmic loss of the feed can be incorporated separately via an added real impedance term if required, but is neglected here for simplicity. For accurate computation of the impedance of the probe see Metamaterial-Enabled Transformation Optics page 142-147.

\subsubsection{The FF Channel as a Special Case}\label{subsec: Far field magnetic}
In the FF regime, the distance between each $n$-th dipole and each
observation point $\mathbf{s}_\ell$ can be approximated as
$R_{\ell,n}\approx R_\ell-\hat{\mathbf u}_\ell^{\mathrm T}\mathbf r_n$,
where $\hat{\mathbf u}_\ell\triangleq
[\sin\theta_\ell\cos\phi_\ell,\,
 \sin\theta_\ell\sin\phi_\ell,\,
 \cos\theta_\ell]^{\mathrm T}$
denotes the propagation direction associated with $\mathbf{s}_\ell$, while \(R_{\ell}\) denotes the distance from the TX's center to \(\mathbf{s}_{\ell}\). Under this approximation, the angles
$\theta_{\ell,n}$ and $\phi_{\ell,n}$ become independent of the dipole
index $n$, since
$\theta_{\ell,n}\approx\theta_\ell$ and
$\phi_{\ell,n}\approx\phi_\ell$ $\forall n$.
Consequently, all $N$ dipole contributions can be expressed in the same spherical basis, and each NF projection matrix
$\mathbf T_{n\rightarrow\ell}$ reduces to an identity matrix, i.e.,
$\mathbf T_{n\rightarrow\ell}=\mathbf I_2$ $\forall n$, yielding \(\tilde{\mathbf{a}}^{\theta}(\cdot)=\mathbf{a}^{\theta}(\cdot)\) as well as \(\tilde{\mathbf{a}}^{\phi}(\cdot)=\mathbf{a}^{\phi}(\cdot)\). Accordingly, the focusing vectors reduce to their FF forms, by replacing all angles \(\phi_{\ell,n}\)'s and \(\theta_{\ell,n}\)'s in~\eqref{eq:focusing_vector_definitions} with \(\phi_{\ell}\) and \(\theta_{\ell}\), respectively, as well as substituting each radial distance \(R_{\ell,n}\) with its approximation \(R_\ell-\sin(\theta_{\ell})\cos(\phi_{\ell})r_{x_n}-\sin(\theta_{\ell})\sin(\phi_{\ell})r_{y_n}\). Finally, the corresponding FF dual-polarized channel matrix %, $\mathbf H_{\rm FF}\in\mathbb{C}^{2L\times2N}$, 
can be obtained from~\eqref{eq: channel_matrix_definition} by replacing the NF focusing vectors therein with their latter FF counterparts, and omitting the projection matrices. Hence, the proposed NF MIMO channel model in~\eqref{eq: channel_matrix_definition} naturally reduces to the conventional steering-vector-based formulation (i.e., FF case)
as a special case.

\section{Extension to Electric and Magnetic Dipoles}\label{sec: Both electric and magnetic dipoles}

In the previous section, the metamaterial elements were modeled exclusively as magnetic dipoles, neglecting any electric polarizability. However, in the considered PPW, the EM field at the aperture includes a non-zero electric component along the \(z\)-direction, in addition to the in-plane magnetic field components. As a result, an electric dipole moment oriented along the \(z\)-axis can also be excited. To account for this effect, we extend the CD model of Section~\ref{sec: Modeling Magnetic Dipoles Only} by introducing an electric polarizability \(\alpha^e_n\) for each \(n\)-th element. Then, the induced electric dipole moment of each $n$-th element, $p_n$, is related to the local electric field, \(E_{{\rm loc},n}\), via the relationship \(p_n = \varepsilon_0 \alpha^e_n E_{{\rm loc},n}\). where \(\varepsilon_0\) denotes the permittivity of free space \cite{balanis2012advanced}. Furthermore, we assume that the elements are not bianisotropic; this is typically the case for symmetric metasurface designs \cite{mancera2017polarizability,yang2019surface}. Under this assumption, electric and magnetic responses are decoupled, i.e., the magnetic field does not induce electric dipole moments, and vice versa. Therefore, the CD formalism relations become:
\begin{equation}
    \mathbf{m}_n \!=\! \mathbf{A}_n\!\underbrace{\left(\mathbf{h}_{0,n} +\!\!\!\!\sum_{j=1, j\neq n}^{N}\!\!\! \mathbf{G}_{n,j}\, \mathbf{m}_j +\!\!\!\! \sum_{j=1, j\neq n}^{N}\!\!\!  \mathbf{g}^{me}_{n,j}\, {p}_j\right) }_{=\mathbf{h}_{{\rm loc},n}}, \label{eq: Coupled Dipole Equations both Electric and Magnetic: Magnetic}
\end{equation}
\begin{equation}
    p_n \!=\! \varepsilon_0\alpha^{e}_n\!\underbrace{\left( \!\!E_{0,n} +\!\!\!\! \sum_{j=1, j\neq n}^{N}\!\!\! {G}^{ee,zz}_{n,j}\, {p}_j +\!\!\!\! \sum_{j=1, j\neq n}^{N}\!\!\! (\mathbf{g}^{em}_{n,j})^{\rm T} \mathbf{m}_j \!\!\right)}_{=E_{{\rm loc},n}}\!.\label{eq: Coupled Dipole Equations both Electric and Magnetic:Electric}
\end{equation}
Note that an electric dipole induces a magnetic field, and vice versa, hence, these CD equations for electric (expression~\eqref{eq: Coupled Dipole Equations both Electric and Magnetic: Magnetic}) and magnetic (expression~\eqref{eq: Coupled Dipole Equations both Electric and Magnetic:Electric}) responses are inherently coupled. In particular, the vectors \(\mathbf{g}^{me}_{n,j},\mathbf{g}^{em}_{n,j}\in \mathbb{C}^{2 \times 1}\) model the coupling between the electric dipole moment at the \(j\)-th element and the magnetic field at the \(n\)-th element, and vice versa, respectively, while \(E_{0,n}\) denotes the electric field induced by the feeds. Moreover, \(G^{ee,zz}_{n,j}\) models the coupling between the electric dipole moment at the \(j\)-th element and the electric field at the \(n\)-th element. It is noted that all the aforementioned Green's functions are defined to be zero for \(n=j\). Hence, \(\forall n,j=1,\ldots,N\) except \(n=j\), it holds that:
\begin{align}\label{eq:gemnj_vector}
    \mathbf{g}^{em}_{n,j} \triangleq& \left[G^{em,zx}_{\rm FS}\left(\mathbf{r}_{n} ,\mathbf{r}_{j}\right) +G^{em,zx}_{\rm WG}\left(\mathbf{r}_{n} , \mathbf{r}_{j}\right);\right.\\
    & \left.G^{em,zy}_{\rm FS}\left(\mathbf{r}_{n} ,\mathbf{r}_{j}\right) +G^{em,zy}_{\rm WG}\left(\mathbf{r}_{n} ,\mathbf{r}_{j}\right)\right]\in\mathbb{C}^{2\times1},\nonumber
\end{align}
where the superscript ``\(em\)'' indicates that we transfer from magnetic moment to electric field. The definition of \(\mathbf{g}^{me}_{n,j}\in\mathbb{C}^{2\times1}\) is identical by transposing \(e\) with \(m\) as well as \(z\) with \(x\) or \(y\), respectively. Using Appendix~\ref{app: Magnetic and electric dipoles propagation}, we have that:
\begin{align}
    G^{em,zx}_{\rm WG}\left(\mathbf{r}_n,\mathbf{r}_j\right) =& -\frac{ k^2 \eta }{4h} H^{(2)}_{1}\!\left(k\rho_{n,j}\right)\sin\left(\psi_{n,j}\right),\label{eq: Waveguide Green Magnetic to electric zx} \\
    G^{em,zy}_{\rm WG}\left(\mathbf{r}_n,\mathbf{r}_j\right) = & \frac{ k^2 \eta }{4h} H^{(2)}_{1}\!\left(k\rho_{n,j}\right)\cos\left(\psi_{n,j}\right),\label{eq: Waveguide Green Magnetic to electric zy}
\end{align}
while, for the free-space interactions, the following holds:
\begin{align}
        \!\!G^{em,zx}_{\rm FS}\left(\mathbf{r}_n,\mathbf{r}_j\right)  & =  \!\frac{\text{-}\eta k^2e^{-\jmath k\rho_{n,j}}}{2\pi\rho_{n,j}} \sin(\psi_{n,j})\left(1-\frac{\jmath}{k\rho_{n,j}}\right)\!,\!\label{eq: Free Space Green Magnetic to Electric zx}\\
        \!\!G^{em,zy}_{\rm FS}\left(\mathbf{r}_n,\mathbf{r}_j\right) &= \frac{\eta k^2e^{-\jmath k\rho_{n,j}}}{2\pi\rho_{n,j}} \cos(\psi_{n,j})\left(1-\frac{\jmath}{k\rho_{n,j}}\right)\!.\!\label{eq: Free Space Green Magnetic to Electric zy}
\end{align}
The latter two expressions were derived from \cite[eq.~(8.53)]{Novotny_Hecht_2006} using the electric-magnetic duality in~\cite[Table~(7.2)]{balanis2012advanced}. It is, thus, concluded that expressions \eqref{eq: Waveguide Green Magnetic to electric zx}--\eqref{eq: Free Space Green Magnetic to Electric zy} summarize the definition of the \(\mathbf{g}^{em}_{n,j}\) vector in~\eqref{eq:gemnj_vector} appearing in~\eqref{eq: Coupled Dipole Equations both Electric and Magnetic:Electric}. 

It is shown in Appendix~\ref{app: Magnetic and electric dipoles propagation} that \(G^{me,xz}_{\rm WG}(\cdot)= 1/(\eta^2\varepsilon_0) G^{em,zx}_{\rm WG}(\cdot)\)  and \(G^{me,yz}_{\rm WG}(\cdot)= 1/(\eta^2\varepsilon_0) G^{em,zy}_{\rm WG}(\cdot)\). The same relation can be found between the free-space coupling interactions, using once again the electric-magnetic duality, yielding \(\mathbf{g}^{me}_{n,j}=1/(\eta^2\varepsilon_0)\mathbf{g}^{em}_{n,j}\) (appearing in~\eqref{eq: Coupled Dipole Equations both Electric and Magnetic: Magnetic}), where \(\mathbf{g}^{me}_{n,j}\triangleq[G^{me,xz}_{\rm FS}\left(\mathbf{r}_{j} ,\mathbf{r}_{n}\right) +G^{me,xz}_{\rm WG}\left(\mathbf{r}_{j} , \mathbf{r}_{n}\right); \left.G^{me,yz}_{\rm FS}\left(\mathbf{r}_{j},\mathbf{r}_{n}\right) +G^{me,yz}_{\rm WG}\left(\mathbf{r}_{j} ,\mathbf{r}_{n}\right)\right]\in\mathbb{C}^{2\times1}\). Regarding the free-space and waveguide coupling between an electric dipole and the associated electric field, i.e., \(G^{ee,zz}(\cdot)=G^{ee,zz}_{\rm FS}(\cdot)+G^{ee,zz}_{\rm WG}(\cdot) \), the following holds:
\begin{align}
        G^{ee,zz}_{\rm FS}\left(\mathbf{r}_n,\mathbf{r}_j\right)=& \left(1-\frac{\jmath}{k\rho_{n,j}}-\frac{1}{k^2 \rho_{n,j}^2}\right)\frac{k^2e^{-\jmath k\rho_{n,j}}}{2\varepsilon_0\pi\rho_{n,j}}, \label{eq: Electric to electric interactions: free space}\\
    G^{ee,zz}_{\rm WG}\left(\mathbf{r}_n,\mathbf{r}_j\right) = &\frac{k^2}{4\jmath \varepsilon_0 \,h}\,
H_0^{(2)}\!\left(k\,\rho_{n,j}\right),\label{eq: Electric to electric interactions: waveguide}
\end{align}
where the latter expression is derived in Appendix~\ref{app: Magnetic and electric dipoles propagation}, while the former is taken from \cite[eq.~(8.52)]{Novotny_Hecht_2006}. Having computed \(G^{ee,zz}_{\rm WG}\), the formula relating the induced electric field due to the feeds w.r.t. the source currents can be interpreted. Specifically, each \(i\)-th line source can be viewed as an electric dipole with a scalar dipole moment \(p^{\rm f}_i\) in the \(z\)-axis. The relation between the source current \(I_i\) and the dipole moment \(p^{\rm f}_i\) is given via \(I_i = \jmath \omega p^{\rm f}_i/h\). Hence, using the definition \(E_{0,n}\triangleq\sum_{i=1}^{N_{\rm f}}G^{ee,zz}_{\rm WG}(\mathbf{r}_n,\mathbf{b}_i)p^{\rm f}_i\) and the latter current--dipole moment relation, \(E_{0,n}\) can be re-expressed as follows:
\begin{equation}\label{eq: Excitation field electric dipoles}
    E_{0,n}=\frac{\text{-}k\eta }{4}\sum_{i=1}^{N_{\rm f}} I_i H^{(2)}_{0}\!\left(k|\mathbf{r}_n-\mathbf{b}_i|\right).
\end{equation}
This expression concludes all the necessary definitions to compute the CD formalism expressions~\eqref{eq: Coupled Dipole Equations both Electric and Magnetic: Magnetic} and~\eqref{eq: Coupled Dipole Equations both Electric and Magnetic:Electric}.

%Specifically,  the formula \(E_{z,n} = [{G}^{em,zx}_{\rm WG},{G}^{em,zy}_{\rm WG}]\mathbf{m}_n\) due to duality can be written as \(H_{y,n}=\) changes to \(1/\eta\) and the magnetic moment \(\mu_0 m\) changes to electric dipole moment \(p\). Finally, \(G^{ee,zz}_{n,j} = G^{ee,zz}_{\rm FS}(\mathbf{r}_n-\mathbf{r}_j) + G^{ee,zz}_{\rm WG}(\mathbf{r}_n-\mathbf{r}_j)\).
%At this point, we note the following duality between  we establish the following duality between electric and magnetic equations, consider as \(E^{em}_{n,j}\) the electric field created from the \(j\)-th magnetic dipole to the \(n\)-th electric dipole: \(E^{em}_{n,j} = (\mathbf{g}^{em}_{n,})^{\rm T}\mathbf{m}_{j}\) a Green function 

\subsection{Computation of the Dipole Moments}
Let us define the stacked electric dipole vectors as \(\mathbf{p}\triangleq[p_1,\ldots,p_N]\in \mathbb{C}^{N \times 1}\) and the stacked electric excitation fields as \(\mathbf{e}_0\triangleq[E_{0,1},\ldots,E_{0,N}]\in \mathbb{C}^{N \times 1}\). The electric polarizability vector is defined as \({\boldsymbol{\alpha}}^{e} \triangleq 
\varepsilon_0\,[\alpha_1^{e},\ldots,\alpha_N^{e}]
\in \mathbb{C}^{N\times 1}\) and \(\mathbf{A}^e\triangleq {\rm diag}(\boldsymbol{\alpha}^e)\). Furthermore, we amend the notation of the previously defined in~\eqref{eq: Dipole Moment}: magnetic--magnetic interaction array from \(\bar{\mathbf{G}}\) to \(\bar{\mathbf{G}}^{mm}\) and the block diagonal matrix containing the magnetic polarizabilities from \(\bar{\mathbf{A}}\) to \(\mathbf{A}^m\). to highlight the electric and magnetic correspondancies. In addition, the electric--electric interaction array is defined as
\(\bar{\mathbf{G}}^{ee} \in \mathbb{C}^{N\times N}\) with
\([\bar{\mathbf{G}}^{ee}]_{n,j} \triangleq G^{ee,zz}_{n,j}\) \(\forall n,j=1,\ldots,N\). The magnetic field generated by electric dipoles is described by \(\bar{\mathbf{G}}^{me} \in \mathbb{C}^{2N\times N}\) with \(
[\bar{\mathbf{G}}^{me}]_{2n-1:2n,\,j} \triangleq \mathbf{g}^{me}_{n,j}\), and conversely, the electric field generated by magnetic dipoles is described by
\(\bar{\mathbf{G}}^{em} \in \mathbb{C}^{N\times 2N}\) with \([\bar{\mathbf{G}}^{em}]_{n,\,2j-1:2j} \triangleq (\mathbf{g}^{em}_{n,j})^{\rm T}\), and it holds that \(\bar{\mathbf{G}}^{me} = -1/(\eta^2 \varepsilon_0)(\bar{\mathbf{G}}^{em})^{\rm T}\). 
Using all above definitions, the coupled system for all $N$ metamaterials can be compactly written as:
\begin{equation}\label{eq: coupled_block_system}
\begin{bmatrix}
\mathbf{I}_{2N} -{\mathbf{A}}^m\bar{\mathbf{G}}^{mm} 
& \text{-}\,{\mathbf{A}}^m\bar{\mathbf{G}}^{me} \\[1mm]
-\mathbf{A}^e\bar{\mathbf{G}}^{em}
& \mathbf{I}_{N} - \mathbf{A}^e\bar{\mathbf{G}}^{ee}
\end{bmatrix}
\begin{bmatrix}
\mathbf{m} \\
\mathbf{p}
\end{bmatrix}
=
\begin{bmatrix}
{\mathbf{A}}^m\mathbf{h}_0 \\
\mathbf{A}^e\mathbf{e}_0
\end{bmatrix}.
\end{equation}
By left-multiplying this expression with the inverse of 
\(
{\rm diag}({\mathbf{A}}^m,\,\mathbf{A}^e)
\)
and solving for the dipole moments, we obtain:
\begin{equation}\label{eq: compact_inverse_solution}
\begin{bmatrix}
\mathbf{m} \\[1mm]
\mathbf{p}
\end{bmatrix}
\!\!=\!\!
\underbrace{
\left(
\begin{bmatrix}
({\mathbf{A}}^{m})^{-1} - \bar{\mathbf{G}}^{mm} & -\,\bar{\mathbf{G}}^{me} \\[1mm]
-\,\bar{\mathbf{G}}^{em} & (\mathbf{A}^{e})^{-1} - \bar{\mathbf{G}}^{ee}
\end{bmatrix}
\right)^{-1}
}_{\triangleq\,\mathbf{K}^{-1}}
\begin{bmatrix}
\mathbf{h}_0 \\[1mm]
\mathbf{e}_0
\end{bmatrix}\!\!.\!\!
\end{equation}

We now extend the previously
defined magnetic excitation matrix $\mathbf{H}_{\rm f} \in \mathbb{C}^{2N\times N_{\rm f}}$ in~\eqref{eq: Excitation Field in Matrix Form} to incorporate the contribution of the electric dipoles. To this end, we introduce the matrix $\mathbf{H}_{\rm f}^{e} \in \mathbb{C}^{N\times N_{\rm f}}$ whose 
entries describe the electric field generated at each $n$-th electric dipole location by each $i$-th source current. This matrix is expressed $\forall n=1,\ldots,N$ and $\forall i=1,\ldots,N_{\rm f}$ as \([\mathbf{H}_{\rm f}^{e}]_{n,i}
\triangleq\frac{-k\eta}{4}
H^{(2)}_{0}\left(k|\mathbf{r}_n-\mathbf{b}_i|\right)\).
By vertically stacking the magnetic and electric excitation matrices, the overall excitation matrix is composed as \(\bar{\mathbf{H}}_{\rm f} \triangleq
[\mathbf{H}_{\rm f} ;\mathbf{H}_{\rm f}^{e}]\in \mathbb{C}^{3N\times N_{\rm f}}\),
which now maps the source currents to the combined magnetic and electric incident
fields. This implies that~\eqref{eq: Excitation Field in Matrix Form} now extends to:
\begin{equation}\label{eq: stacked_excitation_matrix_form}
\begin{bmatrix}
\mathbf{h}_0 \\[1mm]
\mathbf{e}_0
\end{bmatrix}
=
\bar{\mathbf{H}}_{\rm f} \mathbf{i}.
\end{equation}
Substituting this expression into
\eqref{eq: compact_inverse_solution}, the dipole moments due to the
source currents can be compactly expressed as follows:
\begin{equation}\label{eq: dipoles_from_currents_coupled}
\begin{bmatrix}
\mathbf{m} \\[1mm]
\mathbf{p}
\end{bmatrix}
=
\mathbf{K}^{-1}\bar{\mathbf{H}}_{\rm f} \mathbf{i}.
\end{equation}
Note that this expression extends~\eqref{eq: Dipole Moment from currents} (which is valid only for magnetic dipoles) to the case where each metamaterial element of the considered PPW-fed metasurface antenna array architecture is represented as a dipole with both electric and magnetic responses.

\subsection{Integration into MIMO Systems}
We commence by computing the total electric field in free space. To this end, we only need to compute the electric field scattered due to the electric dipole, since the electric field due to the magnetic dipoles remains the same as in~\eqref{eq: electric_field_due_to_nth_dipole}. Hence, the scattered electric field in free space due to each \(n\)-th metamaterial element, considering both electric and magnetic polarizabilities, is given as follows: 
\begin{align}\label{eq: Elevation Electric Field: electric and magnetic}
    &e_{{\rm sc},n}^{\theta}(\mathbf{s_{\ell}})= 
\frac{\eta k^{2}e^{-\jmath kR_{\ell,n}}}{2\pi  R_{\ell,n}}
\\ &\times\left(m_{n}^{x}\sin(\phi_{\ell,n}) - m_{n}^{y}\cos(\phi_{\ell,n})-\frac{p_n}{\eta \varepsilon_0}\sin(\theta_{\ell,n})\right),\nonumber
\end{align}
while $e_{{\rm sc},n}^{\phi}(\mathbf{s})$ remains identical to
\eqref{eq: electric_field_due_to_nth_dipole}; the electric dipole has no contribution in the azimuthal \(\phi\)-polarization.
Similar to the magnetic case, the angles $\theta_{\ell,n}$'s and $\phi_{\ell,n}$'s depend on
the dipole location and, thus, the above field components can be expressed in dipole-dependent local spherical bases.

To enable coherent superposition across all $N$ metamaterials, the electric-dipole contribution must also be projected onto the common
spherical basis associated with each observation point $\mathbf{s}_\ell$.
Following the same procedure as in the magnetic case (Section~\ref{MIMO_channel_magnetic}), we introduce a
local electric focusing vector
$\mathbf{a}^{e,\theta}(\mathbf{s}_\ell)\in\mathbb{C}^{N\times 1}$,
whose elements capture the NF response of the electric
dipoles in their respective local bases as:
\begin{equation}
[\mathbf{a}^{e,\theta}(\mathbf{s}_\ell)]_n
\triangleq
-1/(2\pi R_{\ell,n})\sin(\theta_{\ell,n})\,e^{-jkR_{\ell,n}}. 
\end{equation}
Then, we transform the local focusing vector \(\mathbf{a}^{e,\theta}(\mathbf{s}_{\ell})\) to the common basis, by introducing the vectors \(\tilde{\mathbf{a}}^{e,\theta}(\mathbf{s}_{\ell})\in \mathbb{C}^{N\times 1}\) and \(\tilde{\mathbf{a}}^{e,\phi}(\mathbf{s}_{\ell})\in \mathbb{C}^{N\times 1}\), which are given element-wise as:
\begin{equation}
\begin{bmatrix}
    [\tilde{\mathbf{a}}^{e,\theta}(\mathbf{s}_{\ell})]_n\\
    [\tilde{\mathbf{a}}^{e,\phi}(\mathbf{s}_{\ell})]_n
\end{bmatrix}    \triangleq [\mathbf{T}_{n\rightarrow \ell}]_{:,1}[\mathbf{a}^{e,\theta}(\mathbf{s}_{\ell})]_n.
\end{equation}
It is noted that, if the local and common basis angles do not differ substantially, then, \(\hat{\boldsymbol{\phi}}^{\rm T}_{\ell}\hat{\boldsymbol{\theta}}_{\ell,n}\approx 0\) holds, implying that each electric dipole still contributes mostly at the \(\theta\)-component of the free-space electric field.

By stacking the projected magnetic and electric focusing vectors, the total dual-polarized electric field at $\mathbf{s}_\ell$ can be expressed as follows (extending~\eqref{eq: total electric field_projected}):
\begin{equation}\label{eq: local_field_vector_form_mp}
\mathbf{e}_{\rm sc}(\mathbf{s}_\ell)
=
{k^2}
\begin{bmatrix}
\eta\,\tilde{\mathbf a}^{\theta}(\mathbf{s}_\ell)^{\mathrm T} &
\varepsilon_0^{-1}\tilde{\mathbf a}^{e,\theta}(\mathbf{s}_\ell)^{\mathrm T}\\
\eta\,\tilde{\mathbf a}^{\phi}(\mathbf{s}_\ell)^{\mathrm T} &
\varepsilon_0^{-1}\tilde{\mathbf a}^{e,\phi}(\mathbf{s}_\ell)^{\mathrm T}
\end{bmatrix}\!
\begin{bmatrix}
\mathbf m\\
\mathbf p
\end{bmatrix}.
\end{equation}

\subsubsection{Received Signal Model}
By stacking the $\theta$- and $\phi$-polarized scattered field components
at all $L$ observation points $\mathbf{s}_\ell$'s, the received electric signal can be
written in the classical MIMO baseband respresentation, as (extending~\eqref{eq: Received Signal}):
\begin{equation}\label{eq: received_signal_mp}
    \mathbf{y}
    =
    \mathbf{H}_{mp}\mathbf{K}^{-1}\bar{\mathbf{H}}_{\rm f}\,\mathbf{i}
    + \mathbf{n},
\end{equation}
where $\mathbf{H}_{mp}\in\mathbb{C}^{2L\times 3N}$ represents the
dual-polarized channel matrix between the PPW-fed metasurface
TX and the $L$ observation points, accounting for both magnetic and
electric dipole radiation. Its rows are given $\forall \ell=1,\ldots,L$ as follows:
\begin{equation}\label{eq: channel_matrix_definition_mp}
\begin{alignedat}{2}
    &[\mathbf{H}_{mp}]_{2\ell-1,:}
    &&=
        {k^{2}}
    \begin{bmatrix}
        \eta\,\tilde{\mathbf{a}}^{\theta}(\mathbf{s}_{\ell})^{\mathrm{T}} &
        \varepsilon_0^{-1}\tilde{\mathbf{a}}^{e,\theta}(\mathbf{s}_{\ell})^{\mathrm{T}}
    \end{bmatrix}\!,\!\\
    &[\mathbf{H}_{mp}]_{2\ell,:}
    &&=
    {k^{2}}
    \begin{bmatrix}
        \eta\,\tilde{\mathbf{a}}^{\phi}(\mathbf{s}_{\ell})^{\mathrm{T}} &
        \varepsilon_0^{-1}\tilde{\mathbf{a}}^{e,\phi}(\mathbf{s}_{\ell})^{\mathrm{T}}
    \end{bmatrix}\!,\!
\end{alignedat}
\end{equation}
which extends the derivations in~\eqref{eq: channel_matrix_definition} that are valid only for the magnetic dipole case.

\subsubsection{The FF Channel as a Special Case}\label{subsec:FF}
The FF channel for the case of dipoles with both electric and
magnetic responses can be easily obtained by substituting \(\phi_{\ell,n}\)'s, \(\theta_{\ell,n}\)'s, and \(R_{\ell,n}\)'s $\forall \ell,n$ with their FF formulas, as given in Section~\ref{subsec: Far field magnetic}, as well as by omitting the projection matrices setting \(\mathbf{T}_{n\rightarrow\ell}=\mathbf{I}_2\)~\(\forall n\).

\subsection{Power Conservation for an Electric Dipole}
According to our assumption in the beginning of Section~\ref{sec: Both electric and magnetic dipoles}, the considered metamaterial elements are not magnetoelectric (i.e., they do not exhibit bianisotropic polarizability). This means that the electric and magnetic dipole moments depend solely on their corresponding local fields. Specifically, the magnetic dipole moment depends only on the local magnetic field, while the electric dipole moment depends only on the local electric field. Under this assumption, the power conservation constraint derived in Section~\ref{subsec: Power Conservation Magnetic Dipoles} remains valid, and can be applied independently to the electric and magnetic polarizabilities~\cite{yatsenko2003plane}. To this end, we herein repeat the same reasoning for an electric dipole $n$ characterized by a scalar polarizability \(\alpha^{e}_n\) and an electric dipole moment \(p_n\). Consequently, we apply the Poynting theorem~\cite[eq.~(1--76b)]{balanis2012advanced} twice: one to compute the supplied power due to \(E_{{\rm loc},n}\) (represented by $P^{e}_{{\rm sup},n}$); and another to compute the radiated power due to the \(z\)-component of the scattered electric field \(E_{{\rm sc},n}(\mathbf{r}) \triangleq G^{ee,zz}(\mathbf{r},\mathbf{r}_n)p_n\) (denoted by $P^{e}_{{\rm rad},n}$), yielding the following expressions and relationship:
\begin{align}\label{eq: Supplied and Radiated Power Electric Dipole}
    P^{e}_{{\rm sup},n} & = -\frac{\omega \varepsilon_0}{2}|E_{{\rm loc},n}|^2{\Im}\{\alpha^{e}_n\},\\
    P^{e}_{{\rm rad},n} & = -\frac{\omega \varepsilon^2_0}{2} |E_{{\rm loc},n}|^2|\alpha^e_n|^2{\rm Im}\{G^{ee,zz}(0)\},
\end{align}
\vspace{-0.3 cm}
\begin{equation}\label{eq: Passivity constraint electric dipole}
        P^{e}_{{\rm rad},n} \leq P^e_{{\rm sup},n} \!\Rightarrow \!{\rm Im}\{(\alpha^e_n)^{\text{-}1}\} \geq - \varepsilon_0 {\rm Im}\{G^{ee,zz}(0)\}.
\end{equation}
Similar to Section~\ref{subsec: Power Conservation Magnetic Dipoles}, the singularities of the Green function at \(\mathbf{r}\to \mathbf{r}_n\) can be computed with the help of the Taylor expansion near zero, yielding \({\rm Im}\{G^{ee,zz}(0)\} = -(k^3/(3 \varepsilon_0\pi)+ k^2/(4 \varepsilon_0h))\). In addition, given an intrinsic polarizability \(\bar{\alpha}_n^e\), the RR correction can be used to ensure that the effective polarizability \(\alpha^e_n \) adheres to the passivity constraint, i.e.:
\begin{equation}\label{eq: Radiation-reaction electric dipole}
    \alpha^e_n = \frac{{\alpha}_n^{{\rm int}.e}}{1+\jmath {\alpha}_n^{{\rm int}.e} \left(k^3/(3 \pi)+ k^2/(4 h)\right)}.
\end{equation}

\subsection{Power Conservation for a CD System}
Similar to Section~\ref{subsec: Power Conservation Coupled Dipole System}, we distinguish between the total supplied power from the $N_{\rm f}$ feeds to the overall CD system (i.e., to all $N$ metamaterials), \(P_{\rm sup}\), and the total radiated power of the CD system, \(P_{\rm rad}\). The relationship between these two quantities should be \(P_{\rm sup}\geq P_{\rm rad}\). In the case of dipoles with both electric and magnetic responses, these quantities are given by the contribution of both magnetic and electric dipoles, i.e.:
\begin{align}
    P_{\rm sup}=& 0.5 \omega \left(\mu_0{\rm Im}\left\{\mathbf{m}^{\rm H}\mathbf{h}_0\right\} +{\rm Im}\left\{\mathbf{p}^{\rm H}\mathbf{e}_0\right\}\right)\nonumber\\
    =& 0.5 \omega{\rm Im}\left\{\mathbf{x}^{\rm H}\mathbf{S}_{\mu_0}\mathbf{K}\mathbf{x}\right\},\label{eq: total supplied power magnetic and electric dipoles} \\
    P_{\rm rad}=&-0.5 \omega {\rm Im}\left\{\mathbf{x}^{\rm H}\mathbf{S}_{\mu_0}\mathbf{G}_{\rm full}\mathbf{x}\right\} \label{eq: total radiated power electric and magnetic dipoles},
\end{align}
where we have used the definitions \(\mathbf{x}\triangleq[\mathbf{m};\mathbf{p}]\), \(\mathbf{S}_{\mu_0}\triangleq{\rm diag}(\mu_0\mathbf{1}_{2N\times 1}; \mathbf{1}_{N \times 1})\), and \(\mathbf{G_{\rm full}} \triangleq \mathbf{G}_{\rm self} + \mathbf{G}_{\rm mut}\), with \(\mathbf{G}_{\rm mut}\triangleq[\bar{\mathbf{G}}^{mm},\bar{\mathbf{G}}^{me};\bar{\mathbf{G}}^{em},\bar{\mathbf{G}}^{ee}]\) and \(\mathbf{G}_{\rm self} \in \mathbb{C}^{3N\times 3N}\) being a diagonal purely imaginary matrix that contains only the imaginary parts of the Green's function singularity. In the Appendix~\ref{app: power magnetic and electric}, it is shown that, when the RR corrections are applied, the condition \(P_{\rm sup}\geq P_{\rm rad}\) holds true.

\subsection{Power Injected to the PPW-Fed Metasurface}\label{subsec: Power Delivered to PPW}

We now compute the power delivered from the $N_{\rm f}$ thin-wire current sources, connected to the respective feeds, to the PPW-fed metasurface antenna. Specifically, the circuit form of the Poynting theorem~\cite[eq.~(1-77)]{balanis2012advanced} is used, according to which:
\begin{equation}\label{eq:Ptot_port}
P_{\rm tot}\triangleq\frac{1}{2}\Re\!\left\{\mathbf{i}^{\rm H}\mathbf{v}\right\},
\end{equation}
where $\mathbf{v}\in\mathbb{C}^{N_{\rm f}\times 1}$ denotes the vector of voltages at the feeds.

\paragraph{Voltage at the Feeds}
Each $i$-th feed, positioned at the 3D point $\mathbf{b}_i$, is modeled as a thin vertical wire of height $h$, hence, the voltage across its edges is defined as the line integral of the total electric field, \(E_{z}^{\rm tot}(\mathbf{b}_i)\), along the wire, i.e.:
\begin{equation}\label{eq:voltage_def}
[\mathbf{v}]_i \triangleq -\int_{-h}^{0} E_{z}^{\rm tot}(\mathbf{b}_i){\rm d}z .
\end{equation}
By assuming that the electric field along the wire is constant, which is true as the electric field does not vary along \(z\), \eqref{eq:voltage_def} reduces to $[\mathbf{v}]_i \approx -h\,E_{z}^{\rm tot}(\mathbf{b}_i)$. Hence, the voltage is proportional to the total $z$-directed field at the feed location, including the feed self-field. By stacking the voltages, yields:
\begin{equation}\label{eq:v_tot_vec}
\mathbf{v}\approx -h\,\mathbf{e}^{\rm f}_{\rm tot},\,\,
\mathbf{e}^{\rm f}_{\rm tot}\triangleq
\left[E_{z}^{\rm tot}(\mathbf{p}_1),\cdots,E_{z}^{\rm tot}(\mathbf{p}_{N_{\rm f}})\right]^{\rm T},
\end{equation}
where the total field at the feed locations can be decomposed into a self-contribution and an external contribution: $\mathbf{e}^{\rm f}_{\rm tot}=\mathbf{e}^{\rm f}_{\rm self}+\mathbf{e}^{\rm f}_{\rm ext}$, where
%\begin{equation}\label{eq:feed_field_split}
%\mathbf{e}^{\rm f}_{\rm tot}=\mathbf{e}^{\rm f}_{\rm self}+\mathbf{e}^{\rm f}_{\rm ext},
%\end{equation}
$\mathbf{e}^{\rm f}_{\rm self}$ is induced by each feed on itself and contains a singular term, while $\mathbf{e}^{\rm f}_{\rm ext}$ includes the contribution of all other feeds and the scattering from the metamaterials.

\paragraph{Electric Field at the Feeds}
The external field is the superposition of two field components: \textit{i}) the field radiated by the other feeds; and \textit{ii}) the field scattered by the metamaterials on the top plate, i.e.,
\(\mathbf{e}^{\rm f}_{\rm ext}\triangleq\mathbf{e}^{\rm f}_{\rm f}+\mathbf{e}^{\rm f}_{\rm ms}\). The contribution of \textit{i}) is accounted for via the following term:
\begin{equation}\label{eq:eff_matrix}
\mathbf{e}^{\rm f}_{\rm f}=\mathbf{G}_{\rm ff}\,\mathbf{i},
\end{equation}
where the matrix $\mathbf{G}_{\rm ff}\in\mathbb{C}^{N_{\rm f}\times N_{\rm f}}$ has the following entries:
\begin{equation}\label{eq:Gff_entries}
\left[\mathbf{G}_{\rm ff}\right]_{i,j}\triangleq
\begin{cases}
\displaystyle -\frac{k\eta}{4}H_{0}^{(2)}\!\left(k\lvert \mathbf{b}_i-\mathbf{b}_j\rvert\right), & i\neq j\\[2mm]
0, & i=j
\end{cases}.
\end{equation}
In this definition, the diagonal is set to zero because the self-contribution is handled separately through $\mathbf{e}^{\rm f}_{\rm self}$. In addition, the contribution of \textit{ii}) can be computed 
%Furthermore, the coupling induced by the dipoles is computed via the already established framework. Specifically, the electric field at an arbitrary position in the waveguide is obtained 
by multiplying the magnetic moments in \eqref{eq: Waveguide Green Magnetic to electric zx}, \eqref{eq: Waveguide Green Magnetic to electric zy} and the electric moments in~\eqref{eq: Electric to electric interactions: waveguide}. Combining these expressions with the relation between the dipole moments and the source currents given by~\eqref{eq: dipoles_from_currents_coupled}, the field scattered back to the feeds can be expressed as follows:
\begin{equation}\label{eq:ems_matrix_simplified}
\mathbf{e}_{\rm ms}^{\rm f}
=\mathbf{G}_{\rm f}\mathbf{K}^{-1}\bar{\mathbf{H}}_{f}\,\mathbf{i},
\end{equation}
where $\mathbf{G}_{\rm f}\triangleq\big[\mathbf{G}_{\rm fm},\mathbf{G}_{\rm fp}\big]\in\mathbb{C}^{N_{\rm f}\times 3N}$, including the magnetic-to-feed coupling matrix \(\mathbf{G}_{\rm fm}\in \mathbb{C}^{N_{\rm f} \times 2N}\) and the electric-to-feed coupling matrix \(\mathbf{G}_{\rm fp}\in \mathbb{C}^{N_{\rm f}\times N}\) with the respective entry-wise definitions:
\(\left[\mathbf{G}_{\rm fp}\right]_{i,n} \triangleq G_{\rm WG}^{ee,zz}(\mathbf{b}_i,\mathbf{r}_n)\), \(\left[\mathbf{G}_{\rm fm}\right]_{i,2n-1} \triangleq G_{\rm WG}^{em,zx}(\mathbf{b}_i,\mathbf{r}_n)\), and \(\left[\mathbf{G}_{\rm fm}\right]_{i,2n} \triangleq G_{\rm WG}^{em,zy}(\mathbf{b}_i,\mathbf{r}_n)\). Finally, substituting \eqref{eq:eff_matrix} and \eqref{eq:ems_matrix_simplified} into \eqref{eq:v_tot_vec}, yields the expression:
\begin{equation}\label{eq:eext_final}
\mathbf{e}_{\rm ext}^{\rm f}
=\left(\mathbf{G}_{\rm ff}+\mathbf{G}_{\rm f}\mathbf{K}^{-1}\bar{\mathbf{H}}_{f}\right)\mathbf{i}.
\end{equation}

It follows from \eqref{eq:Ptot_port} that, for the computation of \(P_{\rm tot}\), only the real part of \(\mathbf{e}^{\rm f}_{\rm self}\) is needed. Although \(\mathbf{e}^{\rm f}_{\rm self}\) is singular due to the field being evaluated at the source location, its real part remains finite and is given by \({\rm Re}\{\mathbf{e}^{\rm f}_{\rm self}\} = -0.25\,\eta k \mathbf{i}\). The singularity is confined to the imaginary part, which does not contribute to the supplied power, as will be clarified in the sequel. Consequently, at the level of real quantities, we can write \({\rm Re}\{\mathbf{v}\}={\rm Re}\{\mathbf{Z}_{\rm in}\mathbf{i}\}\), where the following impedance definition has been used:
\begin{equation}\label{eq:Zin_def}
\mathbf{Z}_{\rm in}
\triangleq 0.25\eta k h\mathbf{I}_{N_{\rm f}}
-h\left(\mathbf{G}_{\rm ff}+\mathbf{G}_{\rm f}\mathbf{K}^{-1}\bar{\mathbf{H}}_{f}\right).
\end{equation}
It is noted that, although the equality \(\mathbf{v}=\mathbf{Z}_{\rm in}\mathbf{i}\) does not strictly hold at the complex level due to the imaginary self-term, this term does not contribute to the supplied power; the supplied power depends only on the real part of the quadratic form \(\mathbf{i}^{\rm H}\mathbf{Z}_{\rm in}\mathbf{i}\). Therefore, \(\mathbf{v}=\mathbf{Z}_{\rm in}\mathbf{i}\) can be used in \eqref{eq:Ptot_port} without affecting the computed power. In particular, inserting \eqref{eq:Zin_def} into \eqref{eq:Ptot_port} gives the total power supplied to the PPW-fed antenna:
\begin{equation}\label{eq:Ptot_matrix}
P_{\rm tot}
=\frac{1}{2}\Re\!\left\{\mathbf{i}^{\rm H}\mathbf{Z}_{\rm in}\mathbf{i}\right\}
=\frac{1}{2}\mathbf{i}^{\rm H}\underbrace{\left(0.5\left(\mathbf{Z}_{\rm in}^{\rm H}+\mathbf{Z}_{\rm in}\right)\right)}_{\triangleq \mathbf{R}}\mathbf{i},
\end{equation}
where \(\mathbf{R}\in\mathbb{C}^{N_{\rm f}\times N_{\rm f}}\) is Hermitian by construction and, thus, it has real diagonal entries. Moreover, since the overall system is passive, \(\mathbf{R}\) is positive definite, ensuring that \(P_{\rm tot} > 0\) for any non-zero current vector \(\mathbf{i}\). It is finally noted that the dominant contribution in~\eqref{eq:Zin_def} is $0.25\eta k h\mathbf{I}_{N_{\rm f}}-h\,\mathbf{G}_{\rm ff}$, which captures the direct field coupling among the feeds, as well as the self-fields. The additional term $h\,\mathbf{G}_{\rm f}\mathbf{K}^{-1}\bar{\mathbf{H}}_{f}$ corresponds to a two-step mechanism: fields radiated by the feeds excite the metasurface dipoles, which then reradiate back to the feed locations. This ``bounce-back'' contribution is further weighted by the radiation strength of the metasurface elements and is, therefore, expected to be smaller.

\paragraph{Equivalent Voltage Source Representation}
Up to this point, the feeds have been modeled as ideal current sources. Alternatively, they can be represented as voltage sources, provided that the voltages at the feed terminals are known. To enable this representation, the self-impedance of each feeding wire needs to be properly defined, which requires accounting for their reactive component. To this end, a finite radius \(b\) is introduced for the cylindrical wires, since the reactive self-impedance of an infinitesimally thin wire is not well defined. According to \cite[eq.~(5.165)]{tretyakov2003analytical}, the susceptibility relating the current \(I\) of a thin wire to its self-induced electric field via \(I=\chi e^{\rm f}_{\rm self}\), is given as \(\chi\triangleq\left(-0.25 \eta k \left(1-\jmath \frac{2}{\pi}\ln(0.89\,kb)\right)\right)^{-1}\). Consequently, the corresponding self-impedance, defined through \(IZ^{\rm f}_{\rm self}=-h e^{\rm f}_{\rm self}\), can be expressed as follows:
\begin{equation}\label{eq: self-impedance of thin wire}
    Z^{\rm f}_{\rm self} \triangleq 0.25 \eta k h \left(1-\jmath \frac{2}{\pi}\ln(0.89\,kb)\right).
\end{equation}
The latter derivations allow us to express the feed voltages as \(\mathbf{v}=\mathbf{Z}_{\rm in}\mathbf{i}\), where the input impedance matrix is redefined as:
\begin{equation}\label{eq: Zin_def 2}
\mathbf{Z}_{\rm in}\triangleq Z^{\rm f}_{\rm self}\mathbf{I}_{N_{\rm f}}-h\left(\mathbf{G}_{\rm ff}+\mathbf{G}_{\rm f}\mathbf{K}^{-1}\bar{\mathbf{H}}_{f}\right).
\end{equation}

%\begin{remark}\emph{(Equivalent Voltage Source Representation)}
Overall, if each probe is driven through an external feeding network with overall impedance \(Z_L\) (e.g., a 50~\(\Omega\) coaxial line), the current-source excitation can be equivalently represented by a voltage source in series with the impedance \(Z_L\). In this case, the corresponding source voltage vector \(\mathbf{v}_{\rm src}\) is given by:
\begin{equation}\label{eq: Equivalent Voltage Source Representation}
  \mathbf{v}_{\rm src}\triangleq\mathbf{Z}_{\rm in}\mathbf{i}+Z_L\mathbf{i}.
\end{equation}
%\end{remark}

% Feasible implementations of a thin wire are coaxial probes illuminated via a hole with a waveguide port; the self-impedance of the source for this case can be analytically computed according to \cite[eq.~(C.35)]{landy2013metamaterial}, given the radius and height of the probe as well as the hole's radius. In this work, we do not consider a specific realization for the feeding mechanism but assume equivalent current sources modeled as thin wires. To that end, we will keep consistency with this definition and define the reactive impedane by considering a wire of radius \(b\) and evaluating the self-electric field there, which yields \({\rm Im}\{\mathbf{e}^{\rm f}_{\rm self}\}=0.5k\eta/\pi\,{\rm ln}(0.89\,kb)\mathbf{i}\).

\section{Polarizability Retrieval and FW Validation}\label{sec: Polarizability Retrieval}
\begin{figure}
    \centering
    \includegraphics[width=\linewidth]{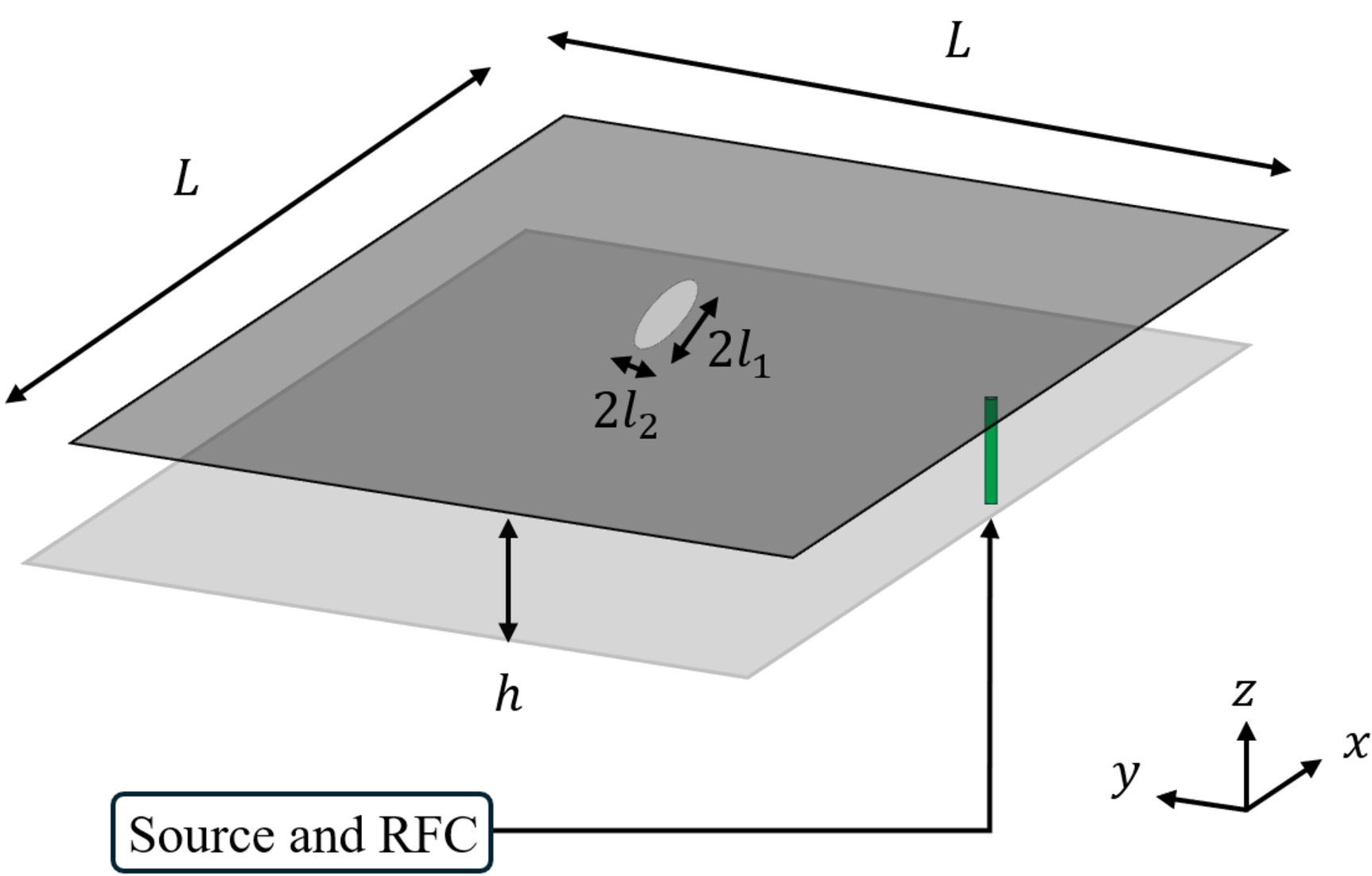}
    \caption{Side view of the PPW-fed metasurface antenna comprising one feed and a single elliptic iris at the center of the structure. The feed is a thin wire, which is assumed to be excited by a digital source along with an RFC.}
    \label{fig:metasurface fig}
\end{figure}

To characterize the response of the metamaterial elements within the considered PPW-fed metasurface antenna array architecture, the effective polarizability parameters need to be retrieved from FW simulations. In this section, we focus on a single element located at the top plate occupying a local surface area $S_a$, which is modeled by its equivalent magnetic dipole moments $m^x$, $m^y$ and its electric dipole moment $p$. To extract these dipole moments, we excite the system using a single thin-wire feed modeled as a discrete port current source connecting the bottom and top plates, as illustrated in Fig.~\ref{fig:metasurface fig}.

\subsection{Dipole Moments Retrieval from Surface Fields}
The dipole moments are calculated by integrating the tangential electric field distribution $\mathbf{E}_t\triangleq[E_x, E_y]$ existing in its surface $S_a$. For the magnetic dipole moments, we invoke the surface equivalence principle \cite{mancera2017polarizability} according to which, the tangential electric field in the aperture corresponds to an equivalent magnetic surface current density $\mathbf{M}_s \triangleq \mathbf{E}_t \times \hat{\mathbf{n}}$, where \(\hat{\mathbf{n}}\) is the unit vector normal to the surface, yielding $\hat{\mathbf{n}} = \hat{\mathbf{z}}$ in our case. Recall that all Green's functions (i.e., $\mathbf{G}_{\text{WG}}$ and $\mathbf{G}_{\text{FS}}$) have been already defined to account for the presence of the metallic boundaries following the image theory. Consequently, to maintain consistency and avoid double counting image sources, the magnetic dipole moments can be retrieved without additional doubling factors, as follows:
\begin{align}
    m^x &= \frac{1}{\jmath \omega \mu_0} \iint_{S_a} E_y(x, y) dx dy, \label{eq:mx_retrieval} \\
    m^y &= -\frac{1}{\jmath \omega \mu_0} \iint_{S_a} E_x(x, y) dx dy. \label{eq:my_retrieval}
\end{align}

We now derive the expression of the normal electric dipole $p$ using the potential theory of double layers~\cite[Sec.~3.16]{stratton1943electromagnetic}. Stratton establishes that a dipole moment density $\tau$ corresponds to a potential discontinuity $\Delta \Phi$ across the surface, i.e.:
\begin{equation}
    \Phi^+ - \Phi^- = \frac{\tau}{\varepsilon_0},
\end{equation}
where $\Phi^+$ and $\Phi^-$ represent the potentials on the upper and lower sides of the aperture. For a normal electric dipole distribution, the potential is inherently antisymmetric w.r.t. the aperture plane (i.e., $\Phi^+ = -\Phi^-$), hence, the total potential jump is $\Delta \Phi = 2\Phi$. Substituting this jump into Stratton's relation, yields the density $\tau = 2\varepsilon_0 \Phi$. The total dipole moment of the equivalent sheet, denoted as the \textit{effective} moment $p_{\text{eff}}$, is obtained by integrating this density, as follows:
\begin{equation}\label{eq: effective electric dipole moment from Phi}
    p_{\text{eff}} \triangleq \iint_{S_a} \tau dS =  2\varepsilon_0 \iint_{S_a} \Phi dS.
\end{equation}
To express this in terms of the tangential electric field ($\mathbf{E}_{\rm tan} \triangleq -\nabla_{xy} \Phi$), we utilize the vector identity $\nabla_{xy} \cdot (\Phi \mathbf{r}) = \mathbf{r} \cdot \nabla_{xy} \Phi + 2\Phi$, which is valid for the 2D position vector $\mathbf{r}\triangleq[x,y]$ centered at the element. Integrating this identity over $S_a$ and applying the divergence theorem (where the line integral vanishes due to the fact that $\Phi=0$ on the PEC rim), yields the relation $2\iint \Phi \, dS = \iint (\mathbf{r} \cdot \mathbf{E}_{\rm tan}) dS$. Consequently, substituting this identity back into \eqref{eq: effective electric dipole moment from Phi}, results in $p_{\text{eff}} = \varepsilon_0 \iint (\mathbf{r} \cdot \mathbf{E}_{\rm tan}) dS$. This $p_{\text{eff}}$ represents the true strength of the equivalent dipole sheet in the aperture, meaning the source plus its image, since its retrieval is based on the scattered field. This differs from the computation of the magnetic dipole moments, which were retrieved via the equivalent current densities. It is noted that, since the Green's functions employed in our formulation already account for the image contributions of the ground plane, using \(p_{\text{eff}}\) directly would lead to double counting. Therefore, the dipole moment used in the model must be half of the effective one, i.e., ($p = p_{\text{eff}}/2$). Thus, the final retrieval formula becomes as follows:
\begin{equation}
    p = \frac{\varepsilon_0}{2} \iint_{S_a} \left(x E_x + y E_y\right) dx dy. \label{eq:p_retrieval}
\end{equation}
It is noted that, the retrieval formula for the electric dipole moment is derived in detail here to clarify its distinction from \cite[eq.~(6)]{mancera2017polarizability}. In particular, in our formulation the dipole moment entering the model must be interpreted consistently with the Green's functions, which already account for the image contribution of the ground plane; this leads to the factor \(1/2\) in \eqref{eq:p_retrieval}.

\subsection{Elements of the Polarizability Matrix}
To determine the specific entries of the polarizability matrix, we place the element at a position $\mathbf{r}$ sharing the same $x$-coordinate as the feed (i.e., $r_{x} = b_{x}$). This geometric alignment ensures that the incident magnetic field is purely $x$-directed ($h_{0}^y = 0$), simplifying the constitutive relations and allowing the direct retrieval of the first column of the magnetic polarizability matrix and the electric polarizability~\cite{mancera2017polarizability}, as:
\begin{equation}\label{eq: retrieval equation 1}
    \alpha^{m}_{xx} = \frac{m^x}{h_{0}^x}, \quad \alpha^{m}_{xy} = \frac{m^y}{h_{0}^x}, \quad \alpha^e = \frac{p}{\varepsilon_0 E_{0}}.
\end{equation}
To retrieve the remaining components, the element is rotated by $90^\circ$ around its normal axis \cite{pulidomancera2018}. This aligns the element's orthogonal response axis with the excitation field $h_{0}^x$. Denoting the moments retrieved from this rotated configuration as $\tilde{m}^x$ and $\tilde{m}^y$, the remaining entries are determined as:
\begin{equation}\label{eq: retrieval equation 2}
    \alpha^{m}_{yy} = \frac{\tilde{m}^x}{h_{0}^x}, \quad \alpha^{m}_{yx} = -\frac{\tilde{m}^y}{h_{0}^x}.
\end{equation}
The latter values constitute the effective polarizabilities, naturally incorporating the effects of the PPW environment.

\subsection{Polarizability Validation with Elliptic Irises}
\subsubsection{Analytical Derivation}
To validate the RR corrections and the aforedescribed polarizability retrieval methodology against analytical benchmarks, we consider the case of elliptic iris slots, for which closed-form expressions of the intrinsic polarizabilities exist. We consider an elliptic aperture with semi-major axis $l_1$ aligned along the $x$-direction and semi-minor axis $l_2$ aligned along the $y$-direction, as depicted in Fig.~\ref{fig:metasurface fig}. The static polarizabilities for a corresponding solid elliptic slot in an unbounded medium are provided in Table 12.1 of \cite{collin1990field}. However, bridging the definition from a scattering obstacle in free space to an aperture connecting two half-spaces (i.e., the waveguide and free space) requires renormalization. Specifically, the intrinsic polarizability of the aperture corresponds to one-fourth of the value for the full elliptic iris, as detailed in \cite[Ch.~7, eq.~(70)]{collin1990field}. Consequently, applying this correction, yields the intrinsic electric (${\alpha}^{{\rm int}.e}$) and magnetic (${\alpha}^{{\rm int},m}_{xx}$ and $\bar{\alpha}^{{\rm int}, m}_{yy}$) polarizability components, as follows:
\begin{align}
    {\alpha}^{{\rm int},e} &= -\frac{\pi l_1^3 (1-e^2)}{3 E(e)}, \label{eq:alpha_e_ellipse}  %\alpha^{e\,\prime}
    \\
    {\alpha}^{{\rm int}, m}_{xx} &= \frac{\pi l_1^3 e^2}{3 (K(e) - E(e))}, \label{eq:alpha_xx_ellipse} %\alpha^{m\,\prime}_{xx}
    \\
    {\alpha}^{{\rm int}, m}_{yy} &= \frac{\pi l_1^3 e^2 (1-e^2)}{3 (E(e) - (1-e^2)K(e))}, \label{eq:alpha_yy_ellipse} %\alpha^{m\,\prime}_{yy}
\end{align}
where $e \triangleq \sqrt{1 - (l_2/l_1)^2}$ is the eccentricity of the ellipse. In addition, the functions $K(e)$ and $E(e)$ denote the complete elliptic integrals of the first and second kind, respectively \cite[Sec.~11.13-1]{polyanin2008handbook}.
% defined as:
% \begin{equation}
%     K(e) = \int_0^{\pi/2} \frac{d\phi}{\sqrt{1 - e^2 \sin^2 \phi}}, \quad 
%     E(e) = \int_0^{\pi/2} \sqrt{1 - e^2 \sin^2 \phi} \, d\phi.
% \end{equation}
Note that these analytical values serve as the frequency-independent intrinsic polarizabilities. To compare them with the effective polarizabilities retrieved from FW simulations at a frequency \(f\), the RR correction formulas in~\eqref{eq: rr_correction} and \eqref{eq: Radiation-reaction electric dipole} need to be first be applied. Finally, due to symmetry across both axes in this case study, the cross-polarization terms vanish.

\subsubsection{Simulation Results}\label{sec:Polarizability_results}
To validate the polarizability retrieval framework, we considered a PPW system operating at the central frequency of $10$~GHz (as in~\cite{mancera2017polarizability}), and comprising two square plates of side length $L=150$~mm, separated by a distance $h=5.21$~mm, as shown in Fig.~\ref{fig:metasurface fig}. The plates had a finite thickness $t=0.1$~mm and the structure was filled with air. The elliptic iris was placed at the center of the top plate with its major radius aligned with the \(x\)-axis, and the feed was placed at the position \(\mathbf{b}=[0,-45]\)~mm, while its current was set to \(I=1\,{\rm A}\). In our FW simulations performed with CST Studio Suite \cite{CST}, the feed was implemented as a discrete port with the thin-wire model and a prescribed current, effectively realizing a current source; in practice, this corresponds to a coaxial cable driven by a digital source with an RFC. In addition, in the setup of the FW solver, the open side walls of the PPW were bounded by Perfectly Matched Layers (PML) conditions; this choice was made to emulate the response of the theoretical model, since it suppresses reflections at the boundaries.

We have examined elliptical iris elements with a fixed semi-major axis $l_1 = 3.6$~mm and varying semi-minor axis $l_2 \leq l_1$. At the considered central frequency of $10$~GHz, the electrical length of the element's major axis corresponds to $k(2l_1) \approx 1.51$ (recall that $k$ is the propagation constant of the guided mode). It is noted that, while the RR theory was rigorously derived for electrically small apertures (i.e., $k(2l_1) \ll 1$), it is demonstrated in~\cite[Figs.~4.14 and~4.15]{tretyakov2003analytical} that the analytic reaction correction terms begin to deviate from the actual ones when $k(2l_1)$ exceeds $1$, but the approximation remains reasonable till \(1.5\); our selected dimension lies at the upper operational limit of this quasi-static approximation.

\begin{figure*}[t]
    \centering
    \includegraphics[scale=0.8]{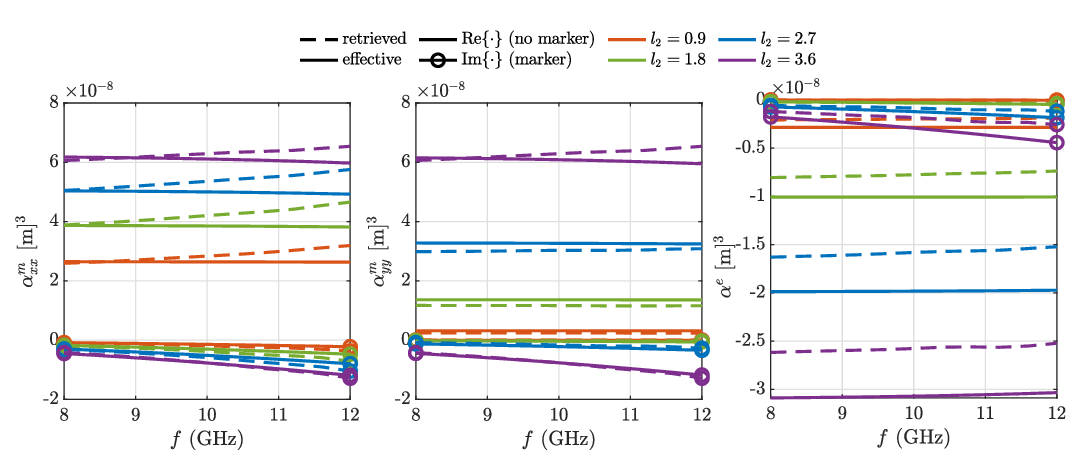}
    \caption{Retrieved (dashed) and analytic (solid) polarizability components versus frequency for different slot lengths $l_2$ in mm. The three subfigures correspond to $\alpha^{m}_{xx}$, $\alpha^{m}_{yy}$, and $\alpha^{e}$ from left to right. For each $l_2$ value, curves share the same color. The real part is shown without markers, while the imaginary part is indicated with circular markers.}
    \label{fig:polarizability_retrieval_l2_v}
\end{figure*}

Figure~\ref{fig:polarizability_retrieval_l2_v} compares the retrieved polarizability components, obtained via \eqref{eq: retrieval equation 1} and \eqref{eq: retrieval equation 2}, with the analytical predictions based on RR theory, given by \eqref{eq: rr_correction} and \eqref{eq: Radiation-reaction electric dipole}. It can be observed that the dominant magnetic polarizability $\alpha_{xx}^{m}$ exhibits better agreement at lower frequencies where the element is electrically smaller, improving the validity of the RR approximation. The agreement also improves systematically for smaller $l_2$ values, as the reduced iris surface is better approximated by a single-point dipole. Overall, it is demonstrated that all polarizability components match the analytical results more closely as $l_2$ decreases, while the electric polarizability $\alpha^{e}_{zz}$ exhibits the largest deviation, most likely due to numerical sensitivity arising from its smaller magnitude~\cite{mancera2017polarizability}.

\begin{figure}[t]
    \centering
    \includegraphics[width=\columnwidth]{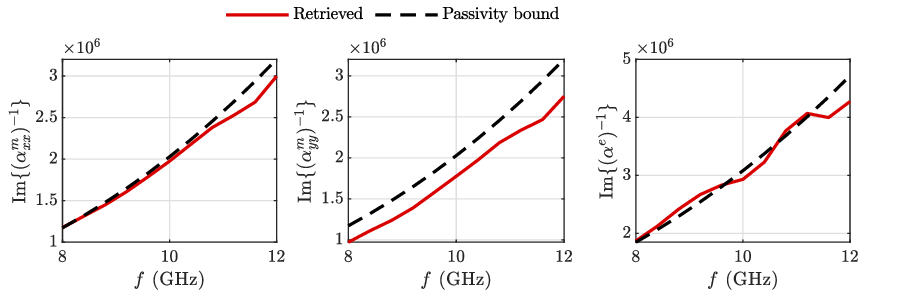}
    \caption{Validation of the passivity constraint using the retrieved polarizabilities for an elliptical iris with semi-minor axis $l_2 = 3$~mm. The imaginary parts of the inverse polarizabilities are compared against the corresponding passivity bounds.}
    \label{fig:passivity_bound}
\end{figure}

In Fig.~\ref{fig:passivity_bound}, the passivity constraint is verified using the retrieved polarizabilities. As established in~\eqref{eq: passivity_constraint} and \eqref{eq: Passivity constraint electric dipole}, the imaginary parts of the inverse polarizabilities must satisfy ${\rm Im}\{(\alpha^{m}_{xx})^{-1}\},{\rm Im}\{(\alpha^{m}_{yy})^{-1}\} \geq k^3/(3\pi)+k^2/(8h)$ for the magnetic case, and ${\rm Im}\{(\alpha^{e})^{-1}\} \geq k^3/(3\pi)+k^2/(4h)$ for the electric case. Since the considered elements are lossless, the retrieved values are expected to attain coincide with these bounds. Overall, good agreement is observed in Fig.~\ref{fig:passivity_bound} for all polarizability components. The largest deviation appears in the magnetic polarizability along the $y$-direction, a trend that is attributed to the finite dimensions of the structure, rendering image theory only approximate, as well as to numerical errors and modeling approximations discussed previously.

\section{FW Validation of the Unified CD Model}\label{sec: Model Validation}
In this section, we validate the proposed magnetic-dipoles-only CD model~\eqref{eq: Dipole Moment from currents} as well as its extension with both electric and magnetic dipoles~\eqref{eq: compact_inverse_solution}, also referred to as the unified CD model, through comparisons between the analytically derived scattered electric field in free space and respective FW simulations via CST. We commence by studying its FF properties via comparing the directivity patterns, and then analyze the NF angular distribution of the electric field for radial distances below the Rayleigh distance \cite{krishnasamy2017Fresnelandfraunhoferlimits}. The considered PPW structure is similar to the one in Section~\ref{sec: Polarizability Retrieval}, with the only differences being that we now consider \(N=10\) randomly placed elliptic irises with varying minor radii, and \(N_{\rm f}=2\) feeds that are modeled as thin-wire current sources with amplitude \(1\)~A and zero phase; the top-view of the structure is illustrated in Fig.~\ref{fig: layout rxy2}. All results presented in this section assume operating frequency at \(f=10\) GHz, unless otherwise specified, while the elliptic irises were approximated as dipoles with polarizability values retrieved numerically via~\eqref{eq: retrieval equation 1}. 

\subsection{FF Characterization}\label{subsec: Far field Characterization}
\begin{figure}
    \centering
    \includegraphics[width=\linewidth]{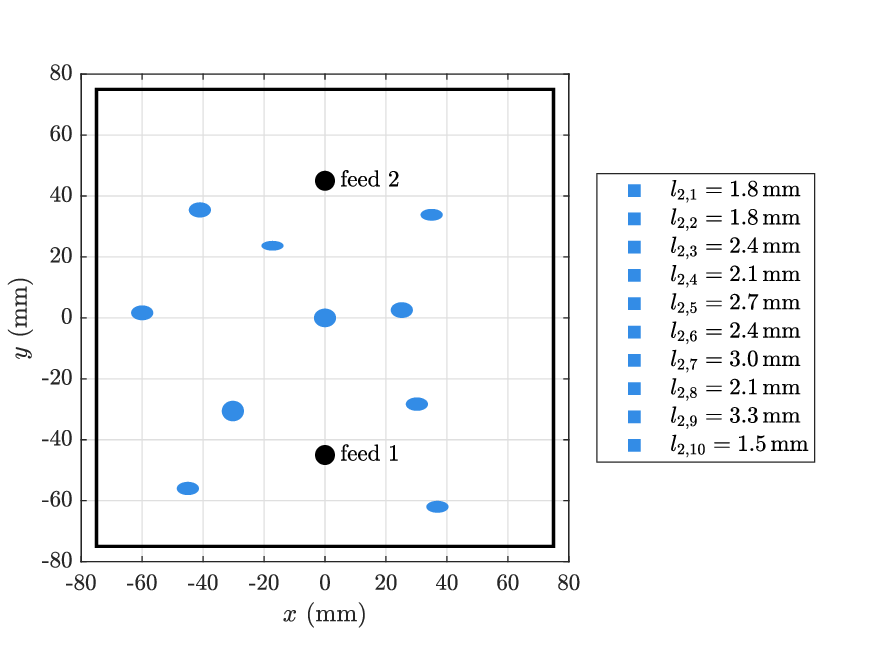}
    \caption{The layout used in the FW simulations (top view). \(N=10\) elliptic iris elements were randomly placed on the top plate, with each $n$-th element having major radius \(l_1=3.6\) mm and varying minor radius \(l_{2,n}\), as shown in the legend. The size of the PPW is the same as in the simulations in Section~\ref{sec:Polarizability_results} ($L=150$~mm and $h=5.21$~mm).}
    \label{fig: layout rxy2}
\end{figure}
To characterize the FF response of the considered PPW-fed metasurface antenna, we first define \(D_{\rm FW}(\phi,\theta)\) and \(D_{\rm model}(\phi,\theta)\) as the directivity functions computed via FW simulations and the proposed unified CD model, respectively. As ``\({\rm model}\)'' we refer either to the case of accounting for both electric and magnetic dipoles (Section~\ref{sec: Both electric and magnetic dipoles} and eq.~\eqref{eq: compact_inverse_solution}) or to the magnetic-dipoles-only (Section~\ref{sec: Modeling Magnetic Dipoles Only} and eq.~\eqref{eq: Dipole Moment from currents}) case; this distinction is specified wherever necessary in the following figures. Recall that, in the relevant state-of-the-art modeling studies~\cite{pulidomancera2018,williams2023EM_DMA,mancera2017polarizability}, the metasurface elements are modeled solely as magnetic
dipoles. Directivity is defined as follows \cite[eq.~(2-21)]{balanis2016antenna}:
\begin{equation}\label{eq: Directivity formula}
    D_{\rm x}(\phi,\theta) \triangleq 4\pi\frac{e^2_{\rm x}(\phi,\theta)}{\int_{\theta=0}^{\pi/2}\int_{\phi=0}^{2\pi} e^2_{\rm x}(\phi,\theta)\sin(\theta)\,d\phi\,d\theta},
\end{equation}
where string \(\rm x\) stands for either ``\(\rm FW\)'' or ``\(\rm model\)'', while \(e_{\rm x}(\phi,\theta)\) is the absolute of the total electric field at a FF distance: \(e_{\rm x}(\phi,\theta) \triangleq({e^{\theta}_{\rm x}(\phi,\theta) +e^{\phi}_{\rm x}(\phi,\theta)})^{1/2}\). Note that the radial distance is irrelevant as long as the observation point is in FF, since its effect cancels out due to the division in~\eqref{eq: Directivity formula}. It is also noted that, since the minor radii \(l_2\) were randomly generated, we have not biased a configuration where the electric response of the elements is pronounced. Finally, to emphasize our modeling generalization w.r.t~\cite{pulidomancera2018}, we herein validate the proposed model for the case of \(N_{\rm f}\geq2\) feeds. The exact layout and the values of \(l_2\) are depicted in Fig.~\ref{fig: layout rxy2}. 

\begin{figure*}[t]
    \centering
    \includegraphics[width=\linewidth]{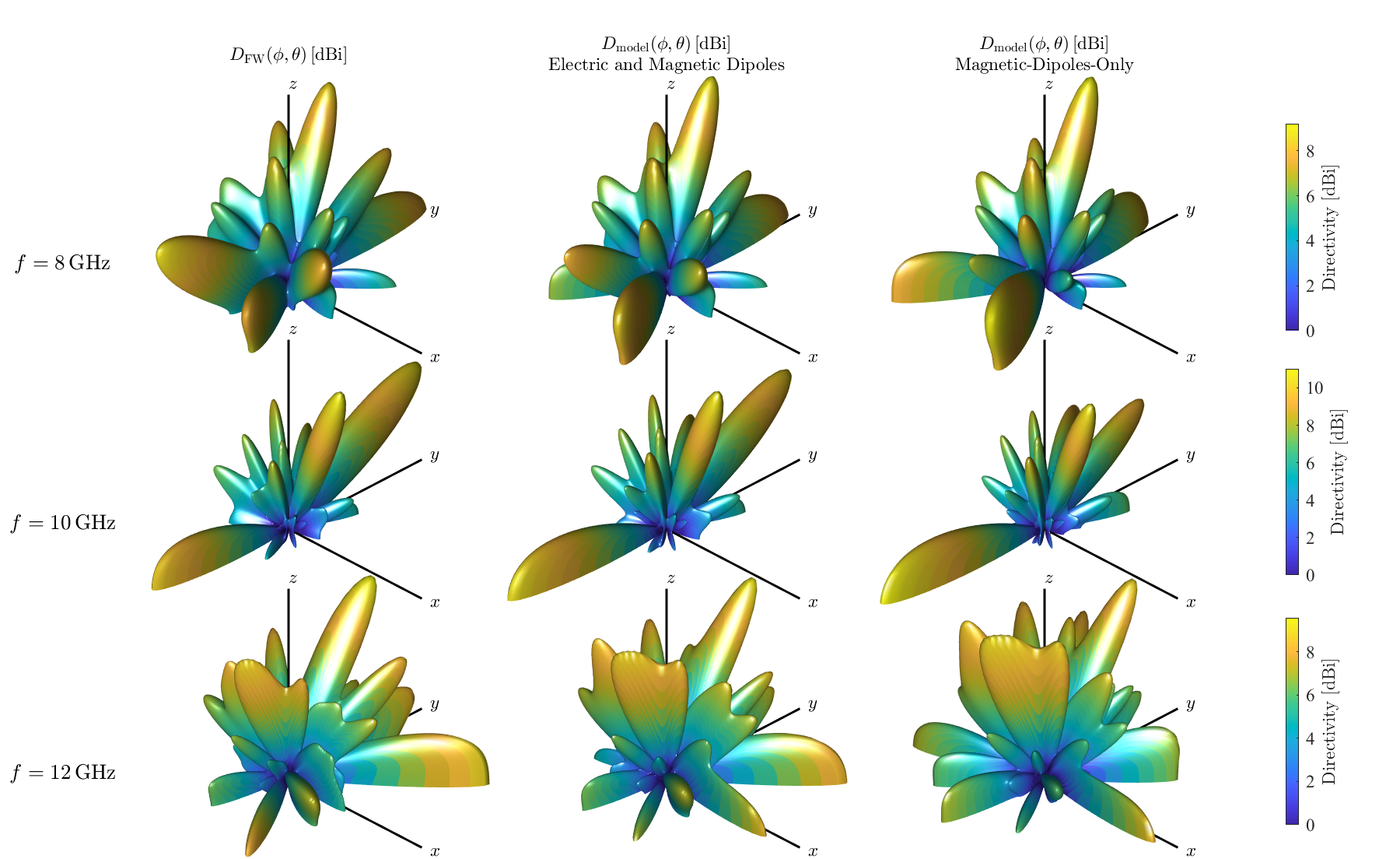}
        \vspace{-5 mm}
   \caption{3D directivity plots where, for each row from left to right, simulations were performed: 1) via the considered FW solver; 2) the unified model considering both electric and magnetic dipoles (Section~\ref{sec: Both electric and magnetic dipoles}); and 3) the magnetic-dipoles-only model (Section~\ref{sec: Modeling Magnetic Dipoles Only}). Each row corresponds to a different frequency, as depicted at the left-hand-side column.}\label{fig: 3D rad pattern frequency}
\end{figure*}

\begin{figure}
    \centering
    \includegraphics[width=\linewidth]{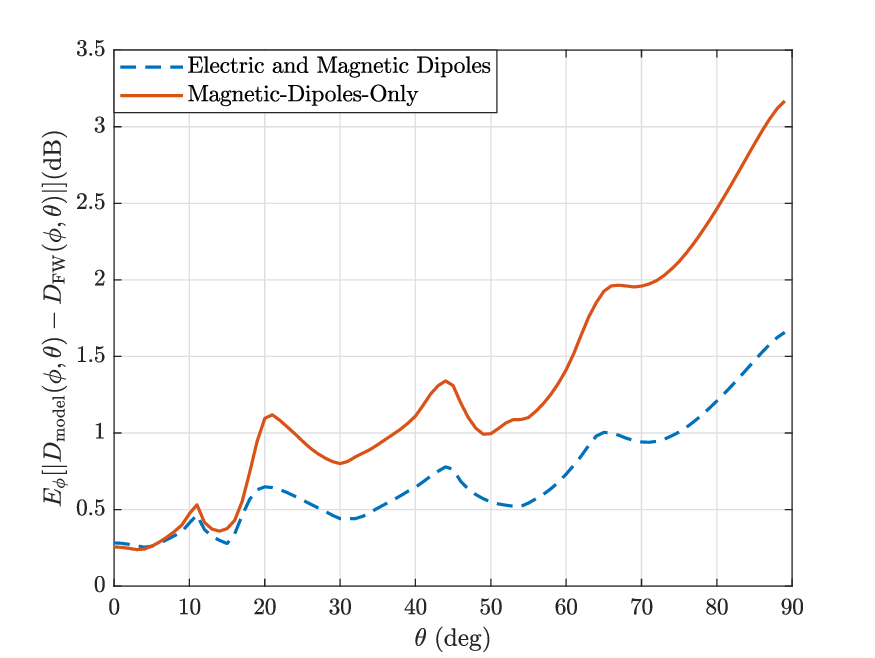}
    \caption{Average error \(\forall \phi \in [0^{\circ},360^{\circ})\) between the FF directivity computed from the proposed unified model (Section~\ref{sec: Both electric and magnetic dipoles}) and FW simulations, as well as between the latter and proposed model considering only magnetic dipoles (Section~\ref{sec: Modeling Magnetic Dipoles Only}).}
    \label{fig:mean error over theta}
\end{figure}
\begin{figure*}[t]
    %  \centering
    % \begin{subfigure}{0.25\linewidth}
    %     \centering
    %     \includegraphics[width=\linewidth]{figures/phi 0 over theta_D.eps}
    %     \caption{}
    %     \label{fig:farfield_phi_0}
    % \end{subfigure}%
    % \hfill
    % \begin{subfigure}{0.25\linewidth}
    %     \centering
    %     \includegraphics[width=\linewidth]{figures/phi 90 over theta_D.eps}
    %     \caption{}
    %     \label{fig:farfield_phi_90}
    % \end{subfigure}%
    % \hfill
    % \begin{subfigure}{0.25\linewidth}
    %     \centering
    %     \includegraphics[width=\linewidth]{figures/phi 180 over theta_D.eps}
    %     \caption{}
    %     \label{fig:farfield_phi_180}
    % \end{subfigure}%
    % \hfill
    % \begin{subfigure}{0.25\linewidth}
    %     \centering
    %     \includegraphics[width=\linewidth]{figures/phi 270 over theta_D.eps}
    %     \caption{}
    %     \label{fig:farfield_phi_270}
    % \end{subfigure}
    \includegraphics[width=\linewidth]{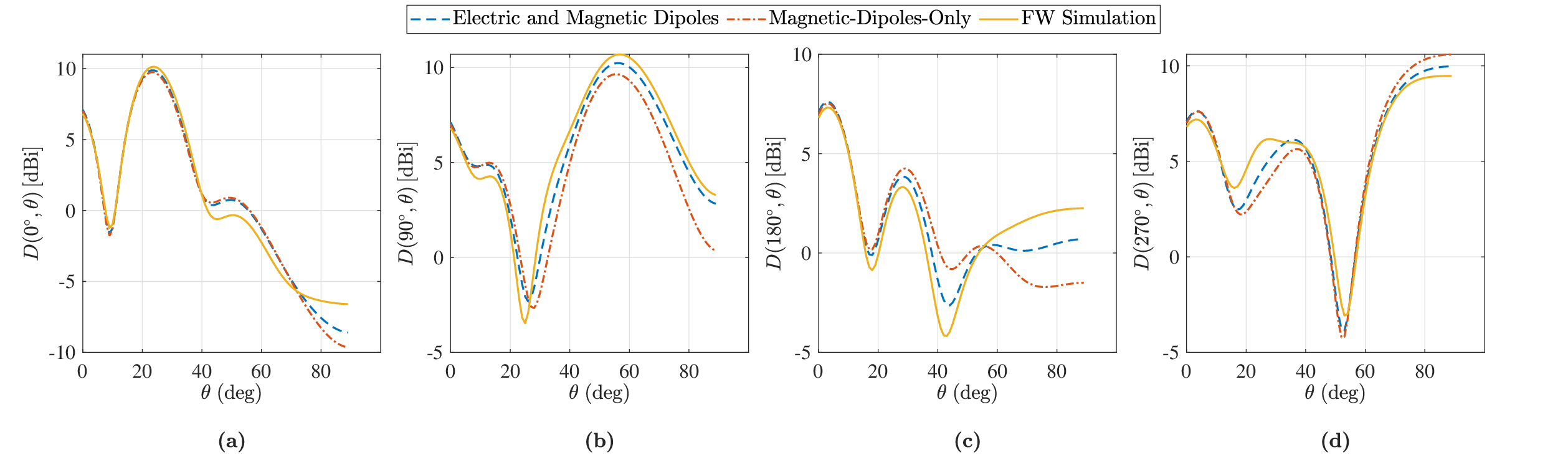}
    \caption{FF directivity patterns versus the elevation angle \(\theta\) for different azimuth planes: (a) \(\phi=0^{\circ}\), (b) \(\phi=90^{\circ}\), (c) \(\phi=180^{\circ}\), and (d) \(\phi=270^{\circ}\). The directivity retrieved from FW simulations is compared with the proposed electric and magnetic dipoles model (Section~\ref{sec: Both electric and magnetic dipoles}) as well as the magnetic-dipoles-only approach (Section~\ref{sec: Modeling Magnetic Dipoles Only}).}
    \label{fig: radiation pattern over theta}
\end{figure*}
Figure~\ref{fig: 3D rad pattern frequency} showcases the \(3\)D radiation patterns for the considered 2D PPW-fed metasurface antenna array at the three different operating frequencies \(f=8,\,10,\) and \(12\) GHz. In the first column, the radiation patterns retrieved from FW simulations are plotted; in the second column, the proposed unified model in Section~\ref{sec: Both electric and magnetic dipoles} is employed; and, in the third column, the magnetic-dipoles-only model presented in Section~\ref{sec: Modeling Magnetic Dipoles Only} is used. It is clear that the electric and magnetic dipoles model is closer to the FW retrieved radiation pattern. In addition, the magnetic-dipoles-only model loses accuracy especially near the elevation angle \(\theta=\pi/2\), where the effect of the electric dipole becomes more prominent due to its \(\sin(\theta)\) dependence, as shown in~\eqref{eq: Elevation Electric Field: electric and magnetic}, while the azimuth coefficient of the electric field, which depends only on the magnetic dipoles, is effectively nulled due to its \(\cos(\theta)\) dependence. The latter observation is validated in Fig.~\ref{fig:mean error over theta}, which demonstrates the average directivity error in the FF regime versus the observation elevation angles \(\theta \in [0^{\circ},90^{\circ}]\), while the averaging was performed over all azimuth angles \(\phi \in [0^{\circ},360^{\circ})\), with both angles sampled at a resolution of \(1^{\circ}\). It can be seen that the proposed unified CD model yields an average error of less than \(1\)~dB up to \(\theta=75^{\circ}\), with a maximum error of \(1.65\)~dB at \(\theta=90^{\circ}\). On the other hand, the magnetic-dipoles-only approach quickly loses accuracy, with the error being consistently over \(1\)~dB after \(\theta=38^{\circ}\) and progressively losing accuracy till a \(3.2\)~dB error at \(\theta=90^{\circ}\). This result validates our proposal that, neglecting the electric dipole contribution in PPW-fed metasurface antenna arrays should not be a universal rule, but rather it should depend on the specific metamaterial implementation. It is noted that, if we had selected to simulate circular slots instead of elliptic ones, the accuracy of the presented magnetic-dipoles-only approach would be even worse, since, at that regime, the magnetic and electric polarizabilities become comparable.

\begin{figure}
    \centering
    \includegraphics[width=\linewidth]{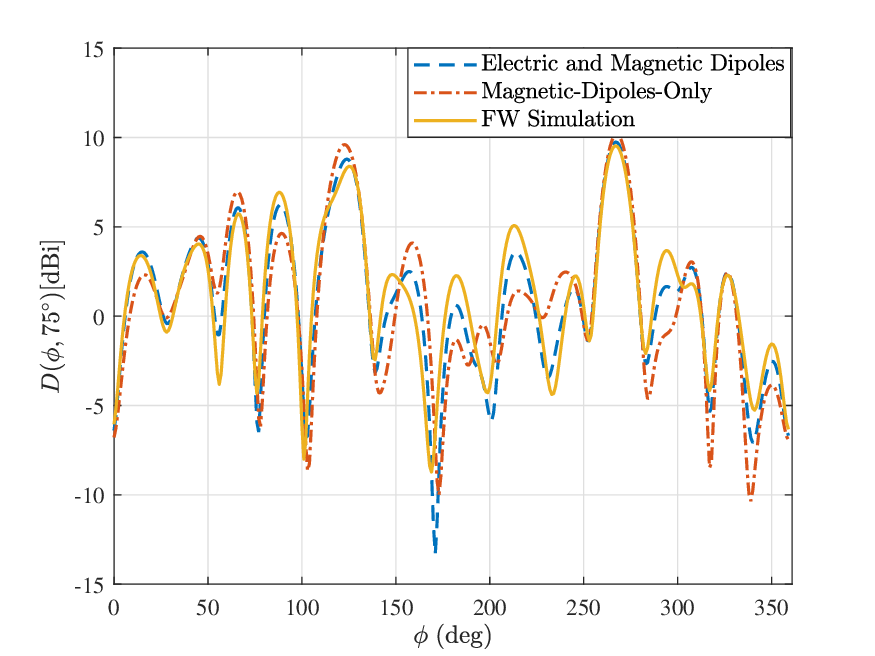}
    \caption{FF directivity pattern versus the azimuth angle \(\phi\), considering the elevation angle \(\theta=75^{\circ}\). The electric and magnetic dipoles model (Section~\ref{sec: Both electric and magnetic dipoles}) and the magnetic-dipoles-only one (Section~\ref{sec: Modeling Magnetic Dipoles Only}) are compared with FW simulations.}
    \label{fig:theta 75 over phi}
\end{figure}

Finally, in Fig.~\ref{fig: radiation pattern over theta}, the FF directivity patterns are plotted versus the elevation angle \(\theta\) for different \(\phi\) planes, specifically for \(\phi=0^{\circ},90^{\circ},180^{\circ},\) and \(270^{\circ}\). As observed, in all planes, the proposed electric and magnetic dipoles model is closer to the FW simulated results. This difference is more apparent in the \(\phi=90^{\circ}\) and \(180^{\circ}\) planes. However, the differences would be even clearer for a case where \(\phi\) is not a multiple of \(90^{\circ}\), so that none of the the contributions of \(\mathbf{m}^{x}\) and \(\mathbf{m}^{y}\) are dominating (please refer to~\eqref{eq: Elevation Electric Field: electric and magnetic}). For this reason, Fig.~\ref{fig:theta 75 over phi} also illustrates the FF directivity pattern as a function of the azimuth angle \(\phi\) for a fixed elevation angle \(\theta=75^{\circ}\), where the discrepancies across \(\phi\) become clearly visible.

\subsection{NF Characterization}
We hereinafter validate our proposed unified CD model in the NF regime. To avoid overcrowding the figures, we have compared between FW simulations and the proposed electric and magnetic dipoles model in Section~\ref{sec: Both electric and magnetic dipoles}, since its superiority over the proposed magnetic-dipoles-only model in Section~\ref{sec: Modeling Magnetic Dipoles Only} has been already showcased. Since directivity is an FF metric, in this section, we focus on the total magnitude of the electric field in free space at an observation point \(\mathbf{s}\), which is defined as follows \cite{balanis2012advanced}: 
\begin{equation}
  e_{\rm x}(\mathbf{s}) \triangleq \left(|e^{\phi}_{\rm x}(\mathbf{s})|^2 + |e^{\theta}_{\rm x}(\mathbf{s})|^2+ |e^{r}_{\rm x}(\mathbf{s})|^2\right)^{1/2},
\end{equation}
where ``\({\rm x}\)'' takes the string values ``\(\rm model\)'' and ``\(\rm FW\),'' depending on whether we refer to the electric field computed via the proposed model or through FW simulations. Moreover, the exponents \(\phi,\theta,\) and \(r\) determine the azimuth, elevation, and longitudinal components of the electric field. As per the proposed NF model, \(e^{\phi}_{\rm model}(\mathbf{s})\) and \( e^{\theta}_{\rm model}(\mathbf{s})\) were computed via \eqref{eq: local_field_vector_form_mp}, while \(e^{r}_{\rm model}(\mathbf{s})\approx 0\) was considered negligible. On the other hand, \(e_{\rm FW}(\mathbf{s})\) is the absolute value of the radiated field acquired from FW simulations and, thus, may include radial field components and reactive effects, which are not analytically modeled, but are accounted for in FW simulations.

Furthermore, to eliminate the different scaling effects when comparing the radiated electric field in varying observation distances and, thus, illustrate how the beampattern changes, we have also evaluated the normalized electric field intensity by defining the following function:
\begin{equation}\label{eq: normalized field intensity function}
F(e_{\rm x}(\mathbf{s}))\triangleq 20\log_{10}\left(e_{\rm x}(\mathbf{s})/\max_{\phi,\theta}e_{\rm x}(\mathbf{s})\right).
\end{equation}
For convenience, the observation point \(\mathbf{s}\) has been defined in spherical coordinates, i.e., \(\mathbf{s}=[r,\phi,\theta]\).

\subsubsection{Total Electric Field}

\begin{figure*}[t]
    \centering
    \includegraphics[width = \linewidth]{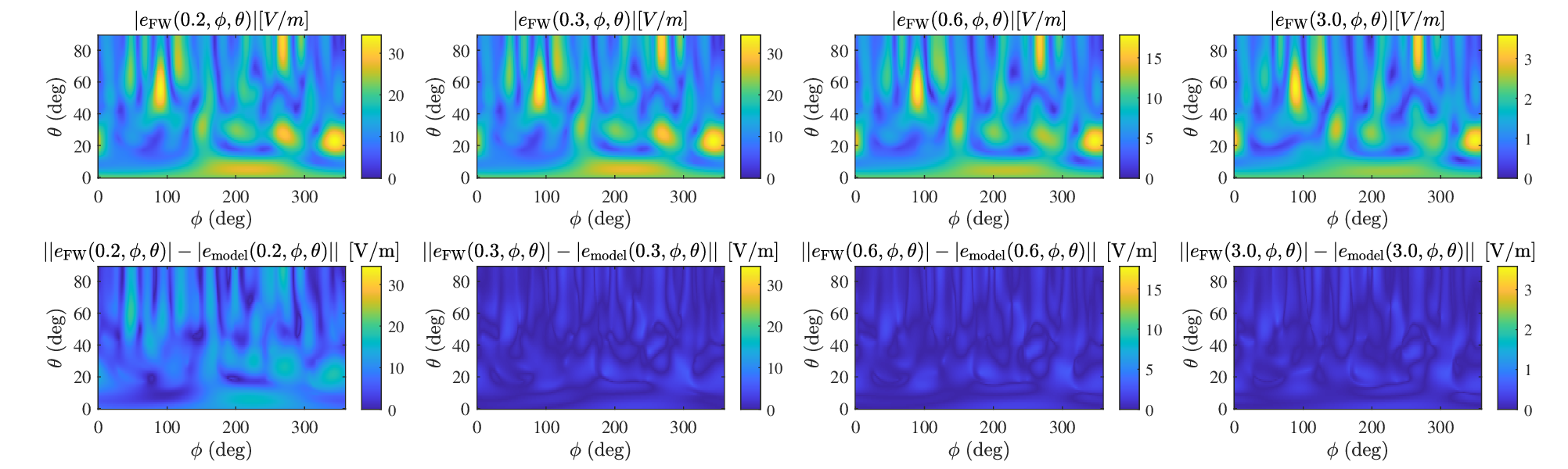}
    \caption{The absolute values of the scattered electric field in free space, retrieved via FW simulations without using the FF approximation, for distances \(r=0.2,0.3,0.6,\) and \(3.0\) m (top row). The absolute difference between the FW simulated electric fields and those acquired using the proposed NF proposed model in Section~\ref{sec: Both electric and magnetic dipoles} (bottom row).}
    \label{fig: 2D colormap nearfield}
\end{figure*}

In Fig.~\ref{fig: 2D colormap nearfield}, the total electric field is illustrated over the whole angular space \(\phi \in [0^{\circ},360^{\circ})\) and \(\theta \in [0,90^{\circ})\), considering the four different distance values: \(r=0.2,0.3,0.6\) and \(3\) m. The top row illustrates the angular distribution of the total electric field from FW simulations, \(e_{\rm FW}(r,\phi,\theta)\) in V/m, while the bottom row shows the absolute difference between the model and the respective FW simulations: \(| e_{\rm FW}(r,\phi,\theta) - e_{\rm model}(r,\phi,\theta) |\) in V/m. 
To provide a measure for the prominence of NF effects in the considered setup, we refer to the NF limits from antenna theory \cite{balanis2016antenna,krishnasamy2017Fresnelandfraunhoferlimits}. In particular, these limits are set regarding the maximum phase error between the two furthest points in an antenna array when approximating phase-progression via the FF approximation. For the considered setup, the distance between the two furthest elements is \(D=124.8\) mm. Hence, the conventional FF limit for this array is given as \(R_{\rm ff}\triangleq2D^2/\lambda\), where, considering operation at \(10\) GHz, i.e, \(\lambda=30\) mm, yields \(R_{\rm ff} = 1.038\)~m \cite{krishnasamy2017Fresnelandfraunhoferlimits}. Furthermore, a definition for the limit of the radiative NF region, below which propagation enters the reactive NF regime, is given by \(R_{\rm nf}\triangleq0.62\sqrt{D^3/\lambda}= 0.158\) m \cite{krishnasamy2017Fresnelandfraunhoferlimits}. Consequently, the selected distance values were chosen to represent different regimes: close to the reactive NF limit at \(r=0.2\)~m, within the radiative NF at \(r=0.3\) and \(0.6\)~m, and FF at \(r=3.0\)~m. As shown in Fig~\ref{fig: 2D colormap nearfield}, there is a sufficient agreement between the FW simulations and the proposed unified model across the entire angular region for distances \(r \geq 0.3\)~m. In contrast, when \(r=0.2\)~m, a noticeably larger mismatch is exhibited. This discrepancy is attributed to several factors, including the absence of the radial polarization component in the proposed model, and the neglect of the free-space dyadic Green's function terms that decay as \(1/r^2\).

To complement the visual comparison of Fig.~\ref{fig: 2D colormap nearfield}, and quantify the agreement between the proposed model and respective FW simulations with a single scalar measure, we have evaluated two solid-angle metrics. First, to isolate discrepancies in the angular distribution of the field's magnitude independently of global scaling effects, we have computed a pattern-normalized solid-angle error. For a fixed observation distance $r$,  we define the following solid-angle integral:
\begin{equation}
\mathcal{I}_{\rm x}(r)
\triangleq
\int_{\Omega} e_{\rm x}(r,\phi,\theta) d\Omega
\,\,\,\text{with}\,\,\,
d\Omega=\sin(\theta)d\theta d\phi.
\end{equation}
Then, we introduce the normalized angular distribution as:
\begin{equation}
u_{\rm x}(r,\phi,\theta)
\triangleq
\frac{e_{\rm x}(r,\phi,\theta)}{\mathcal{I}_{\rm x}(r)} ,
\end{equation}
and define the pattern-normalized error as follows:
\begin{equation}
\varepsilon_{\rm pat}(r)
\triangleq
\frac{1}{2}
\int_{\Omega}
\left|
u_{\rm FW}(r,\phi,\theta)
-
u_{\rm model}(r,\phi,\theta)
\right| d\Omega.
\label{eq:eps_pat_nf}
\end{equation}
This metric takes values in $[0,1]$, with $0$ indicating identical angular distributions and $1$ corresponding to maximally dissimilar (non-overlapping) normalized patterns. Mathematically,
$\varepsilon_{\rm pat}(r)$ is the total variation distance between the normalized solid-angle field distributions, hence, quantifying purely the mismatch in angular structure of the radiated field.

Second, to capture the overall amplitude discrepancy between the two field
maps, we have computed the relative solid-angle integral mismatch via the formula:
\begin{equation}
\varepsilon_{\rm int}(r)
\triangleq
\frac{
\mathcal{I}_{\rm model}(r)
-
\mathcal{I}_{\rm FW}(r)
}{
\mathcal{I}_{\rm FW}(r)
}.
\label{eq:eps_int_nf}
\end{equation}
A positive (negative) value of $\varepsilon_{\rm int}(r)$ indicates that
the proposed model globally overestimates (underestimates) the magnitude of the electric field as compared to respective FW simulations. Equivalently,
$\varepsilon_{\rm int}(r)$ represents the relative error between the
solid-angle averages of $e_{\rm model}$ and $e_{\rm FW}$.
Hence, this metric quantifies a global scaling mismatch between the two field maps:
if the angular patterns are identical but differ by a constant factor,
$\varepsilon_{\rm int}(r)$ captures this offset exactly, whereas purely
angular distortions that preserve the solid-angle integral do not affect it.
\begin{table}[t]
\centering
\begin{tabular}{c|cc|cc}
\hline
& \multicolumn{2}{c|}{NF Model} & \multicolumn{2}{c}{FF Model} \\
$r$ (m) 
& $\varepsilon_{\rm pat}$ (\%) 
& $\varepsilon_{\rm int}$ (\%) 
& $\varepsilon_{\rm pat}$ (\%) 
& $\varepsilon_{\rm int}$ (\%) \\
\hline\hline
0.2 & 5.9 & 59.3 & 10.3 & 58.0 \\
0.3 & 4.6 & 5.8  & 10.3 & 5.35 \\
0.6 & 4.6 & 5.8  & 6.5  & 5.9 \\
3.0 & 4.6 & 5.9  & 4.5  & 6.3 \\
\hline
\end{tabular}
\caption{Angular pattern error ($\varepsilon_{\rm pat}$) and
solid-angle integral mismatch ($\varepsilon_{\rm int}$) for the
NF model and its FF approximation.}\label{tab:nf_metrics}
\end{table}
\begin{figure*}[t]
    % \centering
    % \begin{subfigure}{0.25\linewidth}
    %     \centering
    %     \includegraphics[width=\linewidth]{figures/phi 0.eps}
    %     \caption{}
    %     \label{fig:nearfield_phi_0}
    % \end{subfigure}%
    % \hfill
    % \begin{subfigure}{0.25\linewidth}
    %     \centering
    %     \includegraphics[width=\linewidth]{figures/phi 90.eps}
    %     \caption{}
    %     \label{fig:nearfield_phi_90}
    % \end{subfigure}%
    % \hfill
    % \begin{subfigure}{0.25\linewidth}
    %     \centering
    %     \includegraphics[width=\linewidth]{figures/phi 180.eps}
    %     \caption{}
    %     \label{fig:nearfield_phi_180}
    % \end{subfigure}%
    % \hfill
    % \begin{subfigure}{0.25\linewidth}
    %     \centering
    %     \includegraphics[width=\linewidth]{figures/phi 270.eps}
    %     \caption{}
    %     \label{fig:nearfield_phi_270}
    % \end{subfigure}
\includegraphics[width = \linewidth]{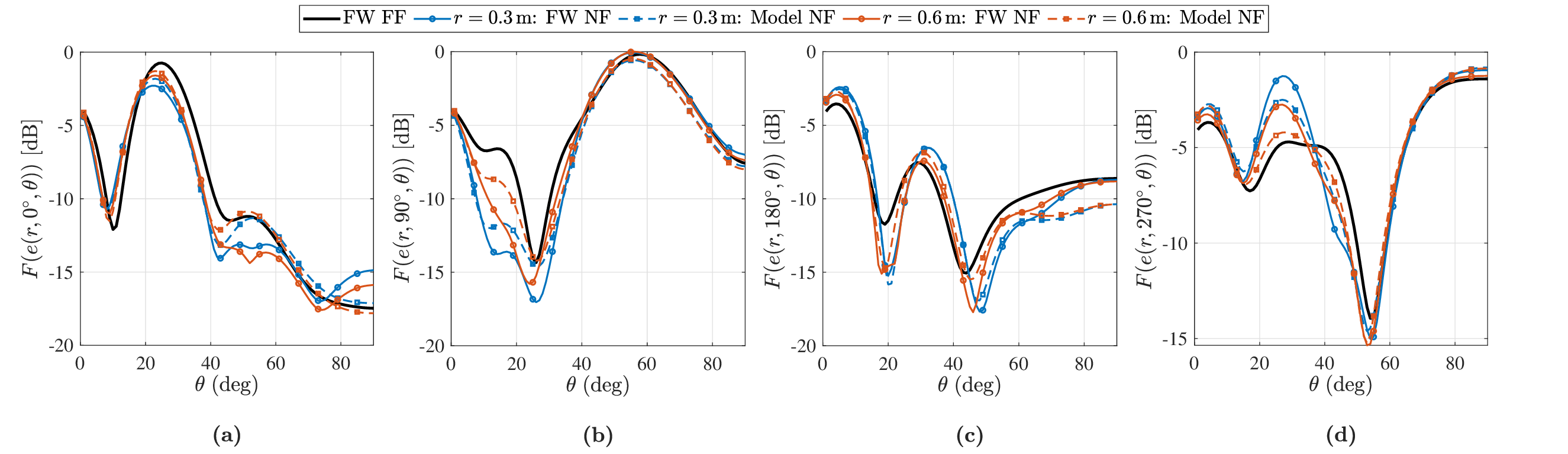}
    \caption{Normalized field intensity patterns versus the elevation angle \(\theta\) for the four azimuth planes: (a) \(\phi=0^{\circ}\), (b) \(\phi=90^{\circ}\), (c) \(\phi=180^{\circ}\), and (d) \(\phi=270^{\circ}\). The proposed NF model in Section~\ref{sec: Both electric and magnetic dipoles} is compared against NF FW simulations, while the pattern retrieved from FF FW simulations is also plotted for reference.}
    \label{fig: nearfield normalized intensity at phi planes}
\end{figure*}
Table~\ref{tab:nf_metrics} reports the quantitative comparison between the
FW simulations and: \textit{i}) the proposed NF model, and \textit{ii}) its FF approximation in Section~\ref{subsec:FF}.
The metrics $\varepsilon_{\rm pat}(r)$ and $\varepsilon_{\rm int}(r)$
are reported in percentage form. This comparison illustrates the benefit of employing the NF formulation at short distances, identifying the regime where the FF approximation becomes sufficient. As shown, at $r=0.2$ m, the proposed NF model significantly improves the angular pattern accuracy compared to the FF approximation (5.9\% versus \ 10.3\%), while both approaches exhibit a large integral mismatch due to strong reactive-field contributions at this short distance. For $r\geq 0.3$ m, the NF model maintains a consistently low
pattern error ($\approx 4.6\%$), whereas the FF approximation gradually approaches similar accuracy as the distance increases. Finally, at $r=3.0$~m, both formulations yield nearly identical performance, as expected. The latter happens because, at this distance, the observation points are in the FF region of the considered PPW-fed metasurface antenna.

\subsubsection{Normalized Electric Field Intensity}
We now compare the normalized field intensities \(F(e_{\rm x}(r,\phi,\theta))\) in dB, as defined in~\eqref{eq: normalized field intensity function}, between the proposed NF model and respective FW simulations. This metric is similar to the directivity comparisons in the FF characterization, since absolute scaling errors cannot be seen; the radiated field intensity \(e^{2}_{\rm x}(\cdot)\) was normalized individually for each case.  Figure~\ref{fig: nearfield normalized intensity at phi planes} depicts the normalized field intensities over the elevation angles \(\theta\in[0,90^{\circ})\) for the four azimuth planes \(\phi=0^{\circ},90^{\circ},180^{\circ},\) and \(270^{\circ}\). We have specifically compared: \textit{i}) FW simulations under the FF approximation (black solid line); \textit{ii}) FW simulations under NF considerations for \(r=0.3\) and \(0.6\)~m (blue solid line with circles and red solid line with squares, respectively); and \textit{iii}) the proposed NF model at \(r=0.3\) and \(0.6\)~m (dashed blue and red lines, respectively). It is shown that the proposed model manages to follow the slight angular shifts of the field pattern in the NF, providing better agreement across all observed angles compared to FF FW simulations. In particular, the maximum error between FF and NF FW simulations, across all four planes depicted in Fig.~\ref{fig: nearfield normalized intensity at phi planes}, is around \(7\)~dB, while the maximum error between the proposed NF model and the respective NF FW-based results is around \(3\)~dB. 

\begin{figure}[t]
    \centering
    \includegraphics[width=\linewidth]{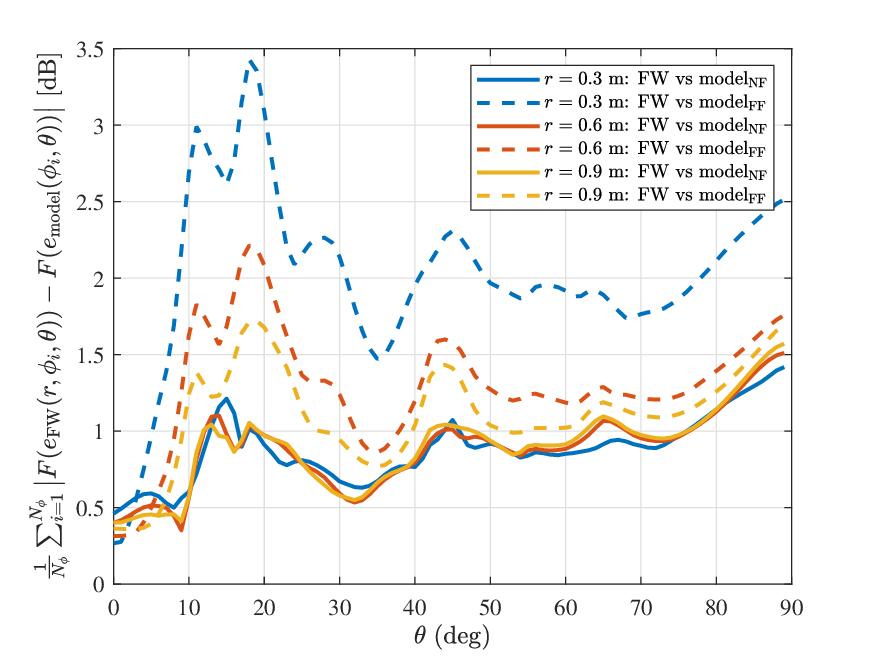}
    \caption{Error in dB between the normalized field intensity pattern computed from NF FW simulations and the proposed NF and FF models in Section~\ref{sec: Both electric and magnetic dipoles}. The error is plotted versus \(\theta\) and averaged over \(\phi \in [0^{\circ},360^{\circ})\) with \(N_{\phi}=360\), considering the distances \(r=0.3,0.6,\) and \(0.9\) m. }
    \label{fig:nearfield error thera}
\end{figure}

Furthermore, to quantify the improved accuracy when the proposed NF model is employed for distances \(R_{\rm nf}\leq r \leq R_{\rm ff}\), we plot, in Fig.~\ref{fig:nearfield error thera}, the absolute error between \(F(e_{\rm FW}(r,\phi,\theta))\) and \(F(e_{\rm model}(r,\phi,\theta))\) as a function of the elevation angle \(\theta\) for the three distance values \(r=0.3,0.6,\) and \(0.9\) m, having performed averaging over \(\phi\)  with \(N_{\phi}=360\) azimuth samples with \(1^{\circ}\) step. As can be seen, for all NF distances, when our NF model is employed, the average error remains the same, while, when our model's FF approximation (Section~\ref{subsec:FF}) is employed, the error increases as the distances decrease. For \(\theta\leq 75^{\circ}\), the average error between the NF model and FW results is near or below \(1\)~dB, achieving similar accuracy as in the FF characterization case (see Fig.~\ref{fig:mean error over theta}). On the other hand, when the FF approximation is applied at \(r=0.3\)~m, the average error is almost consistently over \(2\)~dB, and it reaches a maximum error of approximately \(3.5\)~dB.

\begin{figure*}[t]
    \centering
\includegraphics[width=\linewidth]{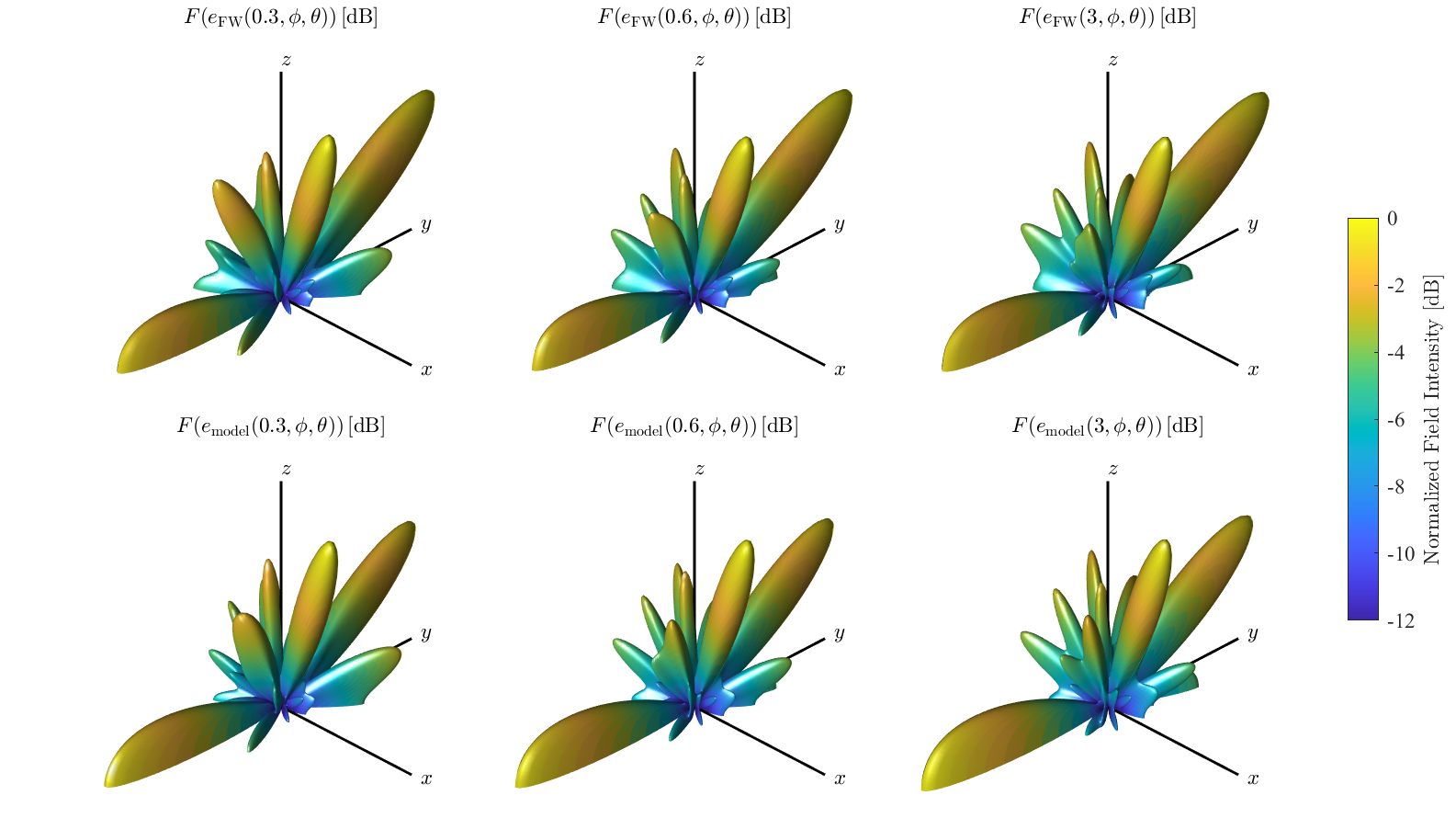}
    \caption{3D normalized field intensity patterns: FW simulations (top row), the proposed NF model in Section~\ref{sec: Both electric and magnetic dipoles} (bottom row).}
    \label{fig:3D_nearfield}
\end{figure*}

Finally, Fig.~\ref{fig:3D_nearfield} illustrates the normalized electric-field intensity as a 3D radiation pattern at distances $r=0.3,0.6,$ and $3$~m. The top row corresponds to the FW simulations, while the bottom row shows the respective results from the proposed NF model. It can be observed that, while the main beams remain largely unchanged across distances, noticeable variations occur in the sidelobe structure, particularly near the $z$-axis (i.e., $\theta=0$). These distance-dependent pattern distortions are accurately reproduced by the proposed model.

\subsection{Voltage at the Sources}
\begin{figure}
\begin{subfigure}{0.5\textwidth}
    \includegraphics[width=\linewidth]{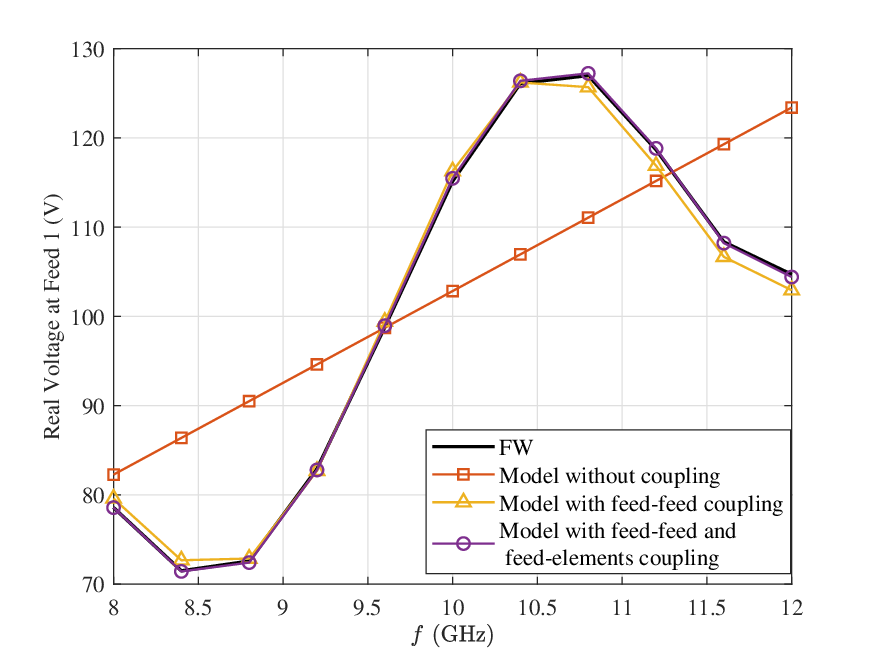}
    \caption{}
    \label{fig: real voltage at feed h 5.21 mm}
\end{subfigure}
\hfill
\begin{subfigure}{0.5 \textwidth}
    \includegraphics[width=\linewidth]{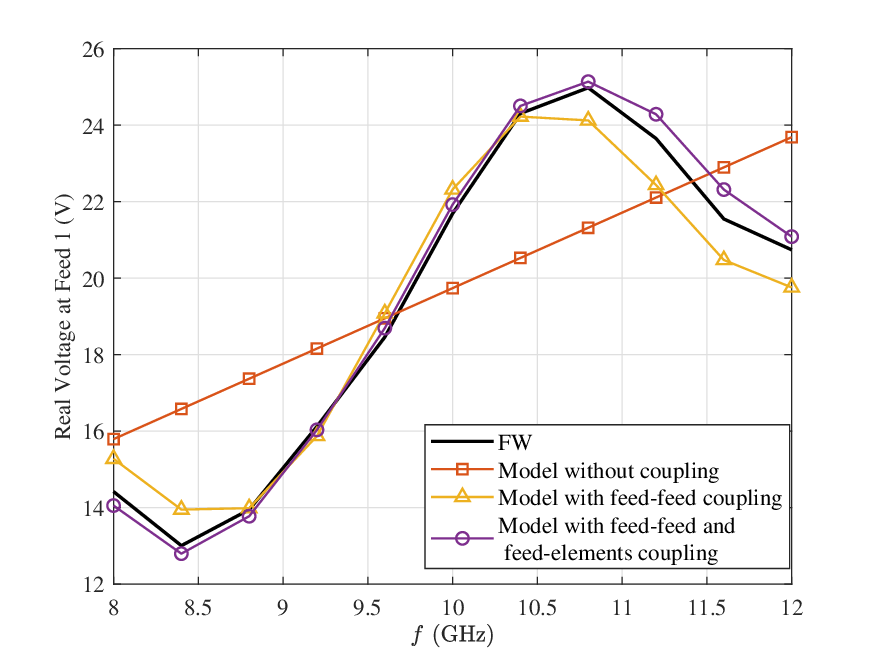}
    \caption{}
    \label{fig: real voltage at feed h 1 mm}
\end{subfigure}

\caption{Real part of the voltage across the terminals of the thin-wire source (at feed $1$ in Fig.~\ref{fig: layout rxy2}). The black line corresponds to the value obtained from FW simulations, where a thin-wire current source was placed at the feed location. The lines with markers represent the predictions with the proposed model: squares account only for the self-interaction term in \eqref{eq:Zin_def}, triangles account for the coupling between the feeds via $\mathbf{G}_{\rm ff}$, and circles account for both coupling mechanisms, i.e., $\mathbf{G}_{\rm ff}+\mathbf{G}_{\rm f}\mathbf{K}^{-1}\bar{\mathbf{H}}_{\rm f}$. The plate separation in the subfigures was set as: (a) \(h=5.21\) mm, and (b) \(h=1\) mm.
%Figure~\ref{fig: real voltage at feed h 5.21 mm} was simulated with plate separation \(h=5.21\) mm, while Fig.~\ref{fig: real voltage at feed h 1 mm} with \(h=1\) mm.
}\label{fig: real voltage at feed}
\end{figure}

To finalize the validation of the proposed unified CD model in Section~\ref{sec: Both electric and magnetic dipoles}, we now validate the analysis in Section~\ref{subsec: Power Delivered to PPW} for the power delivered to the PPW structure. To this end, we have compared the real part of the voltage observed at the source ports, as obtained via voltage monitors in FW simulations, with the analytical impedance expression in~\eqref{eq:Zin_def}. Considering the same setup as in the beginning of this validation section, we have evaluated, in Fig.~\ref{fig: real voltage at feed}, the frequency response of the real voltage at feed~$1$, which is depicted in Fig.~\ref{fig: layout rxy2} and located at the 2D point $[0,-45]$~mm. As demonstrated, the analytically computed voltage at the feed coincides with the voltage retrieved from respective FW simulations. In particular, when the coupling among the two feeds is taken into account, the FW and analytical results are in very close agreement, whereas, when the coupling due to both the feeds and the metamaterial elements is included, the two curves become identical. On the other hand, if the voltage is computed using only the self-interaction term in~\eqref{eq:Zin_def}, this correspondence is lost. The self-interaction term alone would be accurate only if the feed was operating in isolation within a PPW, without the presence of other feeds or metamaterial elements, since their inclusion modifies the effective impedance experienced by the feed. Moreover, due to the fact that all coupling mechanisms are frequency dependent, the resulting PPW input impedance varies with frequency, which, in turn, induces a frequency-dependent voltage response. Intuitively, depending on whether the fields scattered by the other feeds and the metamaterial elements interfere constructively or destructively at the feed location, the effective impedance can increase or decrease, thereby making it respectively more difficult or easier for the source to sustain a prescribed constant current level. Furthermore, since the self-impedance of the feeds is directly related with the plate separation \(h\), as well as all coupling mechanisms within the PPW, the importance of \(h\) becomes non-trivial and needs to be tuned. 

In Fig.~\ref{fig: real voltage at feed}, we also demonstrate the real voltage at feed $1$ for the same layout, but changing \(h\) from \(5.21\), in Fig.~\ref{fig: real voltage at feed h 5.21 mm}, to \(1\)~mm, in Fig.~\ref{fig: real voltage at feed h 1 mm}. For the latter case, the injected power into the PPW is reduced to one fifth, while the corresponding radiated power, as obtained from FW simulations, is reduced only by a factor of two. This indicates that tuning \(h\) can improve radiation efficiency, since a larger fraction of the injected power is radiated. It can be seen, however, from \eqref{eq: rr_correction} and \eqref{eq: Radiation-reaction electric dipole} that the following trade-off is present: when \(h\to 0\), the polarizabilities will also tend to zero, since the denominators therein will tend to infinity. Note that, for the computation of the voltage in the case of a PPW with $h=1$~mm, the effective polarizabilities were not re-retrieved from FW simulations. Instead, the effective polarizabilities obtained at $h=5.21$~mm were first converted to their intrinsic values by removing the RR correction and, subsequently, re-corrected using the RR terms corresponding to $h=1$~mm.

\begin{remark}\label{remark: limit of h} When decreasing $h$, the TEM-only operating region gets enlarged, reinforcing the 2D guide-wave approximation underlying the proposed CD model. However, when $h$ is smaller than the size of the metamaterials, enhanced NF interaction with the bottom plate introduces a strong $h$-dependent reactive self-loading that cannot be captured by RR corrections alone, thereby limiting the applicability of RR-based polarizability estimations. 
This behavior is also reflected in Fig.~\ref{fig: real voltage at feed h 1 mm}, where a slight reduction in accuracy is observed compared to Fig.~\ref{fig: real voltage at feed h 5.21 mm}. 
% Importantly, the dipole model remains accurate provided that the effective polarizability is numerically retrieved at the corresponding value of $h$, thus implicitly accounting for these terms, as can be inferred from \cite{mancera2017polarizability} where \(h=1.27\) mm while the length of the element was \(5\) mm. 
As a practical rule of thumb, the above limitation emerges when the plate separation approaches the reactive NF extent of the metamaterial element, which, for the considered iris-based elements, can be estimated as $d_{\mathrm{nf}} \approx 0.62\sqrt{D^3/\lambda}$ with $D = 2l_1$. For the considered iris dimensions $l_1 = 3.6$~mm at $\lambda = 30$~mm, this yields $d_{\mathrm{nf}} \approx 2.2$~mm, suggesting the PPW height $h$ being around or larger than $2$~mm % $h \gtrsim 2\,\mathrm{mm}$
as an intuitive lower bound.
\end{remark}

\section{Beamforming with PPW-Fed Metasurfaces}\label{sec:sector_optimization}
%\section{Optimization of PPW-Fed Metasurface TXs}\label{sec:sector_optimization}

In this section, capitalizing on the proposed unified CD model for PPW-fed metasurface antenna arrays, we consider a metasurface composed of elliptic iris elements and driven by multiple feeds, and focus on its optimization for transmit FF beamforming. In particular, our design objective is to jointly design the metasurface geometry and the excitations of its feeds in order to achieve directive radiation over a prescribed angular sector. Rather than relying on retrieved values stored in look-up tables or on surrogate fitting models \cite{ZDENEK2014ParallelPlateAntenna,Chou2018ParallelPlateLunsberg}, the electric and magnetic polarizabilities of the elliptic irises are modeled using the analytical expressions in~\eqref{eq: rr_correction},~\eqref{eq: Radiation-reaction electric dipole},~and \eqref{eq:alpha_e_ellipse}--\eqref{eq:alpha_yy_ellipse}. This approach ensures that the dependence of the polarizabilities on the geometric design parameters is captured in a physically consistent, and analytically in closed-form, manner. As a result, their variations remain differentiable w.r.t. the optimization variables, a characteristic that is essential for gradient-based methods and which avoids interpolation artifacts that may arise from retrieved data.

Let \(\mathbf{f}\triangleq[\mathbf{r}_1,\ldots,\mathbf{r}_N,l_{2,1},\ldots,l_{2,N},h]\in \mathbb{R}^{3N+1\times1}\) denote the vector of metasurface geometric variables, where \(\mathbf{r}_n=(x_n,y_n)\) represents the planar location of the \(n\)-th elliptic iris, \(l_{2,n}\) denotes its minor semi-axis, and \(h\) is the plate separation. These parameters can be optimized during the structure's fabrication stage, thus determining the electric and magnetic polarizabilities and, consequently, the CD interaction matrix \(\mathbf{K}\) in~\eqref{eq: compact_inverse_solution}. In contrast, the complex currents \(\mathbf{i}\) in~\eqref{eq: dipoles_from_currents_coupled} at the feeds are reconfigurable, i.e., they can be optimized dynamically during the structure's deployment stage. In the sequel, we assume that the feed positions are predetermined and fixed; thus, they are excluded from the set of metasurface optimization variables.

\subsection{Design Objective}\label{sec:BF_Design_Obj}
%\subsection{Beamforming Optimization}
%The antenna is intended to provide coverage over a prescribed angular sector $\mathcal{S}$, within which users may be located. At any given time instant, a single user is served, and the feed currents are adjusted to form a focused beam toward the corresponding direction. The role of the fabrication design is therefore not to optimize the radiation pattern for a single excitation, but rather to ensure that, for \emph{any} direction within $\mathcal{S}$, an appropriate choice of feed currents can sgenerate a sufficiently directive beam with high radiation intensity.
Let $\mathcal{S}$ denote a desired angular sector that the TX metasurface antenna array needs to serve upon fabrication, within which beams toward specific internal directions need to be dynamically realized. The latter requirement indicates that, for \emph{any} direction within $\mathcal{S}$, there exists a setting for the structure's feed currents that can generate a sufficiently directive beam with high radiation intensity. Let \(\Omega\triangleq\{\phi,\theta\}\) denote a generic direction in the metasurface's FF region (i.e., the elevation and azimuth angles are identical w.r.t. to all elements). For fixed fabrication parameters \(\mathbf{f}\), the radiated field at \(\Omega\) depends linearly on the feed current vector \(\mathbf{i}\) through the channel mapping \(\mathbf{H}(\Omega;\mathbf{f})\triangleq\mathbf{H}_{mp}(\Omega;\mathbf{f})\mathbf{K}^{-1}(\mathbf{f})\bar{\mathbf{H}}_{\rm f}(\mathbf{f}) \in \mathbb{C}^{2 \times N_{\rm f}}\) (see~\eqref{eq: received_signal_mp} for $L=1$ observation point, including the dependencies on $\mathbf{f}$ and $\Omega$).
% In the far field, the electric-field magnitude exhibits the well-known
% $1/r$ decay with the observation distance $r$. Since the design objective concerns
% angular beamforming performance rather than absolute field levels at a specific range,
% we therefore adopt a range-normalized far-field metric by explicitly removing this
% distance dependence.
A natural performance metric for beamforming toward $\Omega$ is the total dual-polarized FF radiation intensity, which can be maximized w.r.t. $\mathbf{i}$ in closed form under a total power constraint $P_{\rm tot}$ on the feeds, i.e., $\mathbf{i}^{\mathrm H}\mathbf{R}\mathbf{i} \leq 2P_{\rm tot}$. Specifically, for a given observation direction $\Omega$ and fixed fabrication parameters $\mathbf{f}$, the maximum achievable radiation intensity is defined as follows under the latter constraint:
\begin{equation}\label{eq:best_beam_gain}
%\begin{split}
g(\Omega;\mathbf{f})
\triangleq
\max_{\mathbf{i}:\,0\leq\mathbf{i}^{\mathrm H}\mathbf{R}(\mathbf{f})\mathbf{i}\le 2P_{\rm tot}}
0.5\frac{r^2}{\eta}\left\|\mathbf{H}(\Omega;\mathbf{f})\mathbf{i}\right\|_2^2,
%\,\,{\rm s.t.}\,\, \mathbf{i}^{\mathrm H}\mathbf{R}(\mathbf{f})\mathbf{i}\le 2P_{\rm tot},
%\end{split}
\end{equation}
where \(\mathbf{R}\) represents the metasurface's input impedance model given by~\eqref{eq:Ptot_matrix}, \(r\) is the observation distance, and \(\eta\) the free-space impedance.
It is noted that $g(\Omega;\mathbf{f})$ is expressed in units of power, specifically $\mathrm{W/sr}$, where $\mathrm{sr}$
denotes the steradian, i.e., the unit of solid angle.

By introducing the positive semidefinite matrix \(\mathbf{Q}(\Omega;\mathbf{f}) \triangleq \mathbf{H}(\Omega;\mathbf{f})^{\mathrm H}\mathbf{H}(\Omega;\mathbf{f}),\) the optimization problem in~\eqref{eq:best_beam_gain} admits a closed-form solution in the form of a vector maximizing a generalized Rayleigh quotient~\cite{7417830}. Specifically, the maximum achievable radiation intensity is given in closed form by:
\begin{equation}\label{eq:best_beam_gain_eig}
g(\Omega;\mathbf{f})
=
\frac{r^2}{\eta}
P_{\rm tot}
\lambda_{\max}\left(
\mathbf{Q}(\Omega;\mathbf{f}),
\mathbf{R}(\mathbf{f})
\right),
\end{equation}
where $\lambda_{\max}(\mathbf{A},\mathbf{B})$ denotes the largest eigenvalue of
$\mathbf{A}^{-1/2}\mathbf{B}\mathbf{A}^{-1/2}$. Consequently, the optimal feed current vector, $\mathbf{i}^{\star}$, is proportional to the dominant eigenvector 
of the matrix $\mathbf{R}(\mathbf{f})^{-1/2}
\mathbf{Q}(\Omega;\mathbf{f})
\mathbf{R}(\mathbf{f})^{-1/2}$, that needs to be appropriately scaled so that 
$(\mathbf{i}^{\star})^{\mathrm H}\mathbf{R}(\mathbf{f})\mathbf{i}^{\star}=2P_{\rm tot}$ holds true.

\subsection{Optimization of the Metasurface Geometric Parameters}\label{subsec:fabrication_design}

We commence by discretizing the desired angular sector $\mathcal{S}$ into the $T$ angular directions: $\{\Omega_t\}_{t=1}^{T}\subset\mathcal{S}$. Similar to~\eqref{eq:best_beam_gain}, let $g(\Omega_t;\mathbf{f})$ denote the maximum achievable radiation intensity toward each direction $\Omega_t$ after optimal excitation of the feeds (i.e., via the current vector $\mathbf{i}^{\star}$), with $\mathbf{f}$ including the geometric parameters of the metasurface that need to be optimized during its fabrication phase. Finally, let $\mathcal{F}$ represent the feasible set for $\mathbf{f}$, encoding the following physical and fabrication constraints for the PPW-fed metasurface TX:
\begin{itemize}
    \item \emph{Geometry Bounds:} Each $n$-th elliptic iris element ($n=1,\ldots,N$) must lie entirely within the finite aperture $\mathcal{A}$ of the top plate. This is enforced by requiring \(x_{\min}+l_{1}+\tfrac{b}{2}\leq x_n\leq x_{\max}-l_{1}-\tfrac{b}{2}\) and \(y_{\min}+l_{1}+\tfrac{b}{2}\leq y_n\leq y_{\max}-l_{1}-\tfrac{b}{2}\), where $(x_{\min},x_{\max})$ and $(y_{\min},y_{\max})$ define the aperture limits and
    \(b\) indicates a physical gap from the aperture edges to ensure that edge effects due to reflection in the aperture edges are reduced.

    \item \emph{Polarizability Bounds:} The minor semi-axis of each $n$-th elliptic iris satisfies the length inequalities: \(l_{2,\min}\leq l_{2,n}\leq l_{1}\), where $l_{2,\min}$ is imposed by fabrication and modeling considerations.

    \item \emph{Element Separation:} To prevent physical overlap between elements, the centers of any two distinct elements $n\neq m$ need to satisfy the axis-aligned separation condition:
\[\big|x_n-x_m\big| \geq 2l_{1}+b_{\rm el}
    \,\,\, \text{or} \,\,\,
    \big|y_n-y_m\big| \geq l_{2,n}+l_{2,m}+b_{\rm el}.\]
    This condition guarantees non-overlapping irises with an additional safety margin \(b_{\rm el}\)
    along both coordinate directions.

    \item \emph{Feed Clearance:} Each $n$-th elliptic iris element must satisfy a minimum clearance from all feeds, which is defined $\forall n=1,\ldots,N$ and $\forall i=1,\ldots,N_{\rm f}$ as follows:
    \[
    \left\|\mathbf{r}_n-\mathbf{b}_i\right\|_2 \geq l_1+b_{\rm{f}},
    \]
    where $b_{\mathrm{f}}>0$ represents a prescribed minimum distance that is used to ensure that higher order modes have been sufficiently suppressed~\cite{mancera2017polarizability}.
    
    \item \emph{Plate Separation:} The distance \(h\) between the two parallel plates satisfies \(h_{\min}\leq h\leq h_{\max}\), where \(h_{\min}\) is selected so that the RR formulas are still valid (see Remark~\ref{remark: limit of h}), while its upper limit $h_{\max}$ is chosen so that only the TEM mode is propagating, i.e., \(h_{\max}\leq 0.5\lambda\), where \(\lambda\) denotes the operation wavelength.
\end{itemize}

To enforce the considered PPW-fed metasurface TX to cover the entire angular sector $\mathcal{S}$, we formulate the following problem for the design of its structural parameters:
\begin{equation*}%\label{eq:fabrication_sector_problem}
\mathcal{OP}:\,\,\max_{\mathbf{f}\in\mathcal{F}}
\min_{t=1,\ldots,T}\ g(\Omega_t;\mathbf{f}).
\end{equation*}

\subsubsection{Positions of the Elliptic Iris Elements}
Direct continuous optimization over the element locations $\{\mathbf{r}_n\}^N_{n=1}$ is rendered
difficult by the highly nonconvex relation between the element positions and the resulting radiation pattern, as well as due to the spacing and boundary constraints. To bypass this difficulty, we adopt a sampling strategy, according to which, candidate element layouts are drawn
from a parameterized spatial distribution that reflects the excitation characteristics of the waveguide. Specifically, we define a power-averaged excitation map over the aperture $\mathcal{A}$ that is based on the electric field on the waveguide due to the feeds\footnote{We have set \(w(\mathbf{r})\) equal to zero within the disks that are centered at the feed positions and have a radius of \(l_1+b_{\rm f}.\)}, as follows:
\begin{equation}\label{eq:w_r}
w(\mathbf{r}) \triangleq \sum_{i=1}^{N_{\rm f}}
\left| H_0^{(2)}\left(k\lvert \mathbf{r}-\mathbf{b}_i\rvert\right) \right|^2,\,\,\mathbf{r}\in\mathcal{A}.
\end{equation}
This quantity captures the average coupling strength between the feeds and a potential element location $\mathbf{r}$, independently of any particular feed excitation (\(E[\mathbf{i}\mathbf{i}^{\rm H}]=\mathbf{I}_{N_{\rm f}}\) has been assumed). Based on this map, candidate element positions can be sampled from the following normalized distribution:
\begin{equation}\label{eq:position_distribution}
p(\mathbf{r};\gamma)
=
\frac{(w(\mathbf{r})+\epsilon)^{\gamma}}
{\int_{\mathcal{A}}(w(\mathbf{u})+\epsilon)^{\gamma}\,d\mathbf{u}},
\end{equation}
where $\gamma\geq 0$ is a shaping parameter and $\epsilon>0$ is a small regularization
constant. This family of distributions interpolates continuously between uniform sampling
($\gamma=0$) and increasingly excitation-focused layouts ($\gamma>0$), i.e., dense element placement around feeds. Note that elements locations can be generated sequentially by sampling from~\eqref{eq:position_distribution} and accepting only those samples that satisfy the spacing, boundary, and feed-clearance constraints included in~$\mathcal{F}$. This procedure yields
feasible layouts by construction, while allowing the spatial distribution of elements to adapt smoothly as $\gamma$ varies.

\subsubsection{Dimensions of the Elliptic Irises and Plates Separation}
For a fixed spatial layout $\{\mathbf{r}_n\}^N_{n=1}$ of the $N$ elements, the fabrication optimization process (i.e., $\mathcal{OP}$) reduces to determining the optimized minor semi-axes $\mathbf{l}_2$ and plate separation \(h\), which are, in general, continuous design variables. 

The objective in $\mathcal{OP}$ is computationally challenging since the inner minimization, $\min_{t=1,\ldots,T} g(\Omega_t; \mathbf{f})$, is non-differentiable at points where the active direction switches. To deal with this, we replace the non-smooth $\max$-$\min$ formulation with a differentiable \textit{Softmin} approximation using the Log-Sum-Exp function \cite[Ch.~3.1.5]{boyd2004convex}, yielding the following revised formulation for the joint design of $\mathbf{l}_2$ and $h$:
\begin{align*}
\mathcal{OP}_1:\,\,&\max_{\mathbf{l}_2,h} J_{\alpha}(\mathbf{l}_2,h) \triangleq -\frac{1}{\alpha} \ln\left( \sum_{t=1}^{T} \exp\left(-\alpha g\left(\Omega_t; \mathbf{l}_2,h\right)\right)\right)\\ 
& \,\,\text{s.t.} \,\,\,\,\, l_{2,\min} \leq l_{2,n} \leq l_{1}\,\forall n,\,h_{\min}\leq h\leq h_{\max},\nonumber
\end{align*}
where $\alpha > 0$ is a smoothing parameter. Now, $\mathcal{OP}_1$'s objective function is smooth w.r.t. $\mathbf{l}_2$ and \(h\), allowing for the use of efficient gradient-based solvers, such as quasi-Newton, rather than slower derivative-free methods. In particular, the parameter $\alpha$ controls the trade-off between smoothness and accuracy: for small $\alpha$, $J_{\alpha}(\cdot,\cdot)$ constitutes a soft average over the angular sector, while, as $\alpha \to \infty$, this function converges to the true worst-case performance $\min_{t=1,\ldots,T} g(\Omega_t; \mathbf{l}_2,h)$. To speed up and improve the performance of the solver, in the Appendix~\ref{app: gradient with z}, we have analytically derived $J_{\alpha}(\cdot,\cdot)$'s gradient w.r.t. \(\mathbf{l_2}\) and \(h\). It is noted that the gradient is given in a general form for any optimization parameter that affects the polarizabilities, and can, thus, be straightforwardly applied to reconfigurable elements (i.e., not only elliptic irises) as well. 

\subsubsection{Overall Design Approach}
%\subsection{Overall Design Approach}
To efficiently compute the optimal parameter $\gamma$, $\gamma^\star$, needed in the distribution for the candidate element positions given by~\eqref{eq:position_distribution}, we adopt a multi-armed bandit strategy based on the successive halving algorithm \cite{jamieson2016multi_armed_bandit}. This approach considers a discrete set of candidate values $\mathcal{C}_{\gamma} \triangleq \{\gamma_k\}_{k=1}^{K}$, and dynamically allocates computational resources to the most promising candidates. The optimization procedure begins with the full set $\mathcal{C}_{\gamma}$. In the first round, a baseline budget of $N_{\text{init}}$ random element layouts is generated for each candidate. For every layout, the minor semi-axes $\mathbf{l}_2$ and \(h\) are optimized using the Softmin objective $J_{\alpha}(\cdot,\cdot)$ of $\mathcal{OP}_1$. To mitigate the high variance associated with the hard-minimum statistic under finite sampling, we rank the candidate \(\gamma_k\)'s based on their average Softmin score across the $N_{\text{init}}$ realizations. At the conclusion of each round, the lower performing half of the candidates in $\mathcal{C}_{\gamma}$ is discarded, and the sampling budget per candidate is doubled for the remaining. This ensures that computational effort is progressively concentrated on the high performing regions of the design space. Once the single best parameter $\gamma^\star$ is identified, a final intensive production run is performed. An ensemble of $N_{\text{final}}\gg N_{\text{init}}$ layouts is generated using $\gamma^\star$, and the final design is selected based on the hard-minimum criterion $\min_{t=1,\ldots,T} g(\Omega_t; \mathbf{f})$ to strictly enforce the coverage specifications. 

The overall approach to compute the plate separation, $h^\star$, as well as the elliptic iris elements positions, $\{\mathbf{r}_n^\star\}_{n=1}^N$, and their dimensions, $\{l_{2,n}^\star\}_{n=1}^N$, solving $\mathcal{OP}$ for the considered PPW-fed metasurface TX is summarized in Algorithm~\ref{alg:fabrication_design}.

\begin{algorithm}[t]
\caption{Metasurface Geometric Parameter Design}
\label{alg:fabrication_design}
\begin{algorithmic}[1]
\Require
Aperture domain $\mathcal{A}$;
sector sample set $\{\Omega_t\}_{t=1}^{T}$;
feed locations $\{\mathbf{b}_i\}_{i=1}^{N_{\rm f}}$;
number of elliptical iris elements~$N$;
bounds $l_{2,\min}$, $l_{1}$, \(h_{\min}\), and \(h_{\max}\);
edge clearence $b$; element clearence $b_{\rm el}$;
feed clearance $b_{\mathrm{f}}$;
smoothing parameter \(\alpha\); set $\mathcal{C}_{\gamma} = \{\gamma_k\}_{k=1}^{K}$;
initial layouts per candidate $N_{\text{init}}$; and
final number of layouts $N_{\text{final}}$.

\State Compute the excitation map via \eqref{eq:w_r} \(\forall \mathbf{r}\in \mathcal{A}\).
%\[
%w(\mathbf{r}) \gets \sum_{i=1}^{N_{\rm f}}
%\left|H_0^{(2)}\!\left(k\lvert \mathbf{r}-\mathbf{b}_i\rvert\right)\right|^2,
%\qquad \mathbf{r}\in\mathcal{A}.
%\]

\State Set current layout budget as $n_{\text{budget}} = N_{\text{init}}$.%$n_{\text{budget}} \gets N_{\text{init}}$.

\While{$|\mathcal{C}_{\gamma}| > 1$} \Comment{\textit{Bandit Search Phase}}
    \State Initialize scores list as $\mathcal{V} = \emptyset$.%$\mathcal{V} \gets \emptyset$.
    \For{each $\gamma_k \in \mathcal{C}_{\gamma}$}
        \State Define distribution $p(\mathbf{r};\gamma_k)$ using Step 1 and \eqref{eq:position_distribution}.
        \For{$\ell = 1,\ldots,n_{\text{budget}}$}
            \State Sample feasible layout $\{\mathbf{r}_n\}_{n=1}^N$ from $p(\mathbf{r};\gamma_k)$.
            \State %Compute $\mathbf{l}_{2}^\star$ and \(h^\star\) maximizing $\mathcal{OP}_1$'s Softmin objective $J_{\alpha}(\cdot,\cdot)$.
            Compute $\mathbf{l}_{2}^\star$ and \(h^\star\) solving $\mathcal{OP}_1$.
            \State Record the score $v_\ell = J_{\alpha}(\mathbf{l}_2^\star,h^{\star})$.%$v_\ell \gets J_{\alpha}(\mathbf{l}_2^\star)$.
        \EndFor
        \State Compute average performance $\bar{v}_{k} = \frac{1}{n_{\text{budget}}}\sum_{\ell} v_\ell$.%$\bar{v}_{k} \gets \frac{1}{n_{\text{budget}}}\sum_{\ell} v_\ell$.
        \State Append $(\gamma_k, \bar{v}_{k})$ to $\mathcal{V}$.
    \EndFor

    \State Sort $\mathcal{C}_{\gamma}$ based on scores $\mathcal{V}$ (descending).
    \State Update $\mathcal{C}_{\gamma}$ to only contain the top $50\%$ of the 
    
    \hspace{-0.28cm} candidates. %$\mathcal{C}_{\gamma} \gets$ Top 50\% of $\mathcal{C}_{\gamma}$ (discard worst half).
    \State Set $n_{\text{budget}} = 2n_{\text{budget}}$ (double budget for survivors).%n_{\text{budget}} \gets 2 \times n_{\text{budget}}$ (Double budget for survivors).
\EndWhile

\State Set the single remaining value in $\mathcal{C}_{\gamma}$ as $\gamma^\star$.%$\gamma^\star \gets$ Single remaining value in $\mathcal{C}_{\gamma}$.
\State Define final distribution $p(\mathbf{r};\gamma^\star)$.
\For{$\ell = 1,\ldots,N_{\text{final}}$}{\Comment{\textit{Final Run}}}
    \State Sample layout $\{\mathbf{r}_n\}_{n=1}^N$ from $p(\mathbf{r};\gamma^\star)$.
    \State Compute $\mathbf{l}_{2}^\star$ and \(h^\star\) solving $\mathcal{OP}_1$.%Optimize $\mathbf{l}_{2}$ and \(h\) maximizing $J_{\alpha}$.
    \State Evaluate strict hard-min: $G_{\text{min}} \triangleq \min_{t=1,\ldots,T} g(\Omega_t; \mathbf{f})$.
    \State Store design if $G_{\text{min}}$ is the best seen so far.
\EndFor

\State \Return Best performing \(h^{\star}\), $\{\mathbf{r}_n^\star\}_{n=1}^N$, and $\{l_{2,n}^\star\}_{n=1}^N$.
\end{algorithmic}
\end{algorithm}

\section{Beamforming Performance Evaluation}\label{sec: Numerical results and discussion}

%\subsection{System Parameters}\label{subsec: System parameters}
In this section, we present performance evaluation results for the proposed transmit beamforming optimization approach in Section~\ref{sec:sector_optimization} w.r.t. the geometric parameters of the considered PPW-fed metasurface antenna array architecture. All metasurface elements were modeled as elliptic irises with common fixed major semi-axis \(l_1=3.6\) mm, while their minor semi-axes were optimized within the range \(l_{2,n} \in [0.2,3.6)\)~mm \(\forall n\). The plate separation was constrained to \(h\in[2,8]\)~mm, and the operating frequency was set to \(f=10\)~GHz. The geometric clearance parameters, namely the aperture edge margin \(b\), the inter-element margin \(b_{\rm el}\), and the feed clearance \(b_{\rm f}\), were all fixed to \(2\)~mm. For the sampling-based layout generation in Algorithm~\ref{alg:fabrication_design}, the candidate set of shaping parameters was set as \(\mathcal{C}_{\gamma}=\{0,0.5,1,1.5,2,2.5\}\). In addition, the successive halving procedure within this algorithm was initialized with \(N_{\rm init}=16\) layouts per candidate, and it was concluded with a final evaluation over \(N_{\rm final}=128\) layouts.

Three metasurface antenna array configurations with increasing number of elements, namely \(N=128\), \(256\), and \(512\), were simulated. In all cases, the total available feed power was fixed to \(P_{\rm tot}=10\)~W, and the number of feeds was chosen as \(N_{\rm f}=25\). To ensure a fair comparison across all size configurations, the physical aperture size was scaled with the number of elements. Specifically, the side length of the square PPW was set to
\(
W = 0.5\sqrt{N}\lambda,
\)
which corresponds to the aperture size of a uniform planar array with inter-element spacing of \(0.5\lambda\). The feeds were placed with uniform spacing within a square region of side length \(W/2\), centered within the PPW. Finally, the desired coverage region was defined over the angular sector
\(\mathcal{S}=\phi\in[0^{\circ},90^{\circ}] \times \theta \in [0^{\circ},30^{\circ}] \),
which was discretized with a resolution of \(2^{\circ}\) in both dimensions, resulting in a total of \(T = 16 \times 46 = 736\) sampled directions.

\subsection{Performance Metrics}

In order to streamline the presentation of the results, we considered the following function taking values in $\mathrm{W/sr}$:
\begin{equation}
U(\Omega;\mathbf{f},\mathbf{i}) \triangleq 
0.5\frac{r^2}{\eta}\left\|\mathbf{H}(\Omega;\mathbf{f})\mathbf{i}\right\|_2^2,
\end{equation}
which represents the radiation intensity toward a direction $\Omega$ for a given metasurface geometric parameter configuration $\mathbf{f}$ and feed excitation $\mathbf{i}$. Recall that the function $g(\Omega;\mathbf{f})$, defined in \eqref{eq:best_beam_gain}, corresponds to the maximum achievable radiation intensity toward $\Omega$, obtained by optimizing $\mathbf{i}$ under a total power constraint (see Section~\ref{sec:BF_Design_Obj}). Consequently, $g(\Omega;\mathbf{f})$ describes the variation of this optimum over $\Omega$, rather than a radiation intensity pattern for a fixed excitation. 

To relate radiation intensity to antenna performance, the antenna gain corresponding to a given \(\mathbf{f}\) and $\mathbf{i}$ is defined as:
\begin{equation}
G(\Omega;\mathbf{f},\mathbf{i}) \triangleq 
10\log_{10}\left(\frac{4\pi U(\Omega;\mathbf{f},\mathbf{i})}{P_{\mathrm{tot}}}\right).
\end{equation}
Note that this gain constitutes a more comprehensive performance metric than directivity, which was was used in the modeling validation in~\eqref{eq: Directivity formula}, since it accounts for radiation efficiency by normalizing w.r.t. the input power rather than the radiated one.

\subsection{Results for Optimized Beamforming Coverage}\label{subsec: results beam coverage}

\begin{figure*}[t]
    \centering
     \begin{subfigure}{0.333\linewidth}
        \centering
        \includegraphics[width=\linewidth]{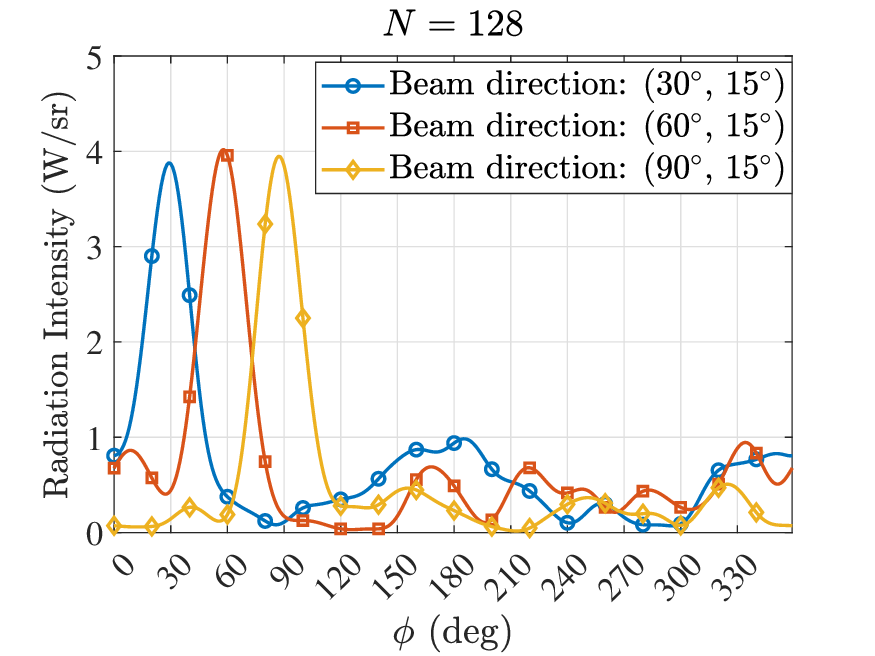}
        \caption{}
        \label{fig:N 128 theta 15}
    \end{subfigure}%
    \hfill
    \begin{subfigure}{0.333\linewidth}
        \centering
        \includegraphics[width=\linewidth]{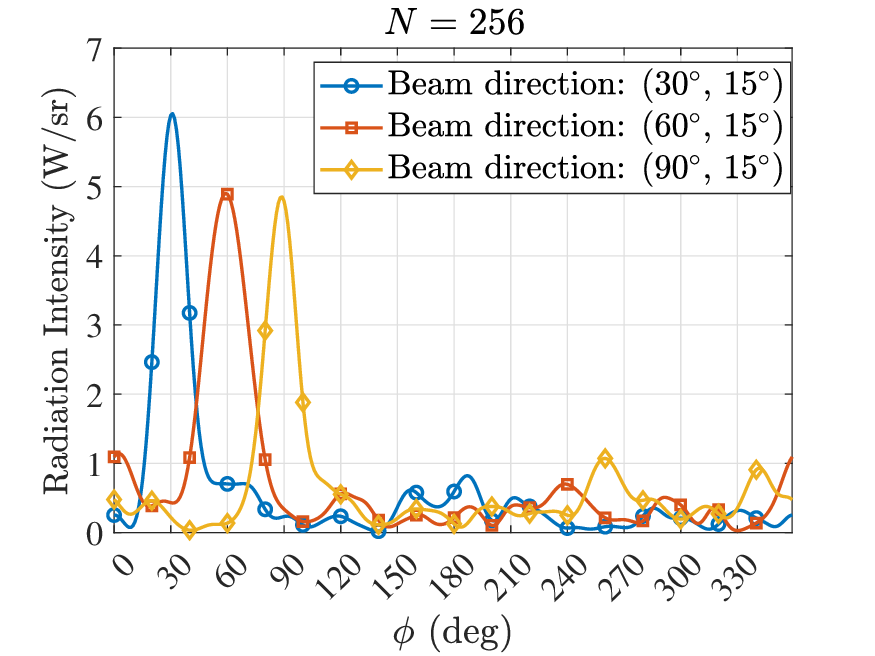}
        \caption{}
        \label{fig:N 256 theta 15}
    \end{subfigure}%
    \hfill
    \begin{subfigure}{0.333\linewidth}
        \centering
        \includegraphics[width=\linewidth]{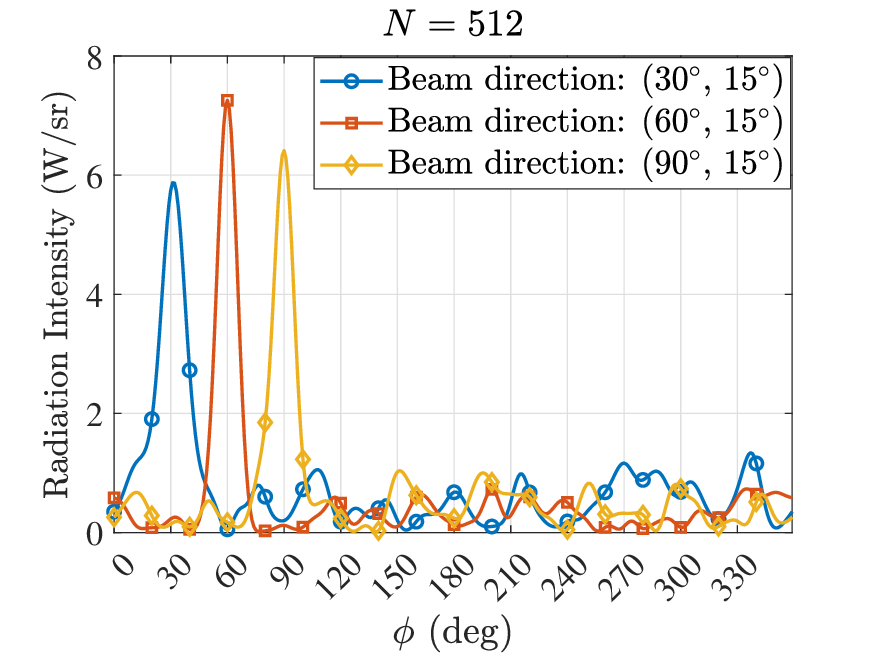}
        \caption{}
        \label{fig:N 512 theta 15}
    \end{subfigure}
    \caption{Radiation intensity (\(\rm W/sr\)) of the optimized 2D PPW-fed metasurface antenna array, given by \(0.5\,{r^2}/{\eta}\big\|\mathbf{H}(\Omega;\mathbf{f})\,\mathbf{i}\big\|_2^2\). The geometric parameters were optimized to provide coverage over the region \(\phi\in[0^{\circ},90^{\circ}] \times \theta \in [0^{\circ},30^{\circ}] \) w.r.t. the antenna center. In each subfigure, the actuating current vector \(\mathbf{i}\) was configured to maximize the radiation intensity at: \textit{i}) \(\phi=30^{\circ}\) and \(\theta =15^{\circ}\), \textit{ii}) \(\phi=60^{\circ}\) and \(\theta = 15^{\circ}\), and \textit{iii}) \(\phi=90^{\circ}\) and \(\theta = 15^{\circ}\), to showcase the beamforming capability over the azimuth angle \(\phi\) in the optimized region. Subfigures (a), (b), and (c) correspond to increasing number of metamaterials: \(N=128\), \(256\), and \(512\), respectively, while the number of feeds \(N_{\rm f}=25\) and the power constraint \(P_{\rm tot}=10\) W remained fixed. }
    \label{fig: radiated power density N 128 256 512}
\end{figure*}

Figure~\ref{fig: radiated power density N 128 256 512} depicts representative radiation intensity curves for the three considered metasurface sizes \(N=128,256\), and $512$, evaluated at the elevation angle \(\theta=15^\circ\), i.e., at the midpoint of the design interval \([0^\circ,30^\circ]\). The plotted quantity is \(U(\Omega;\mathbf{f}^{\star},\mathbf{i})\), where \(\mathbf{f}^{\star}\) denotes the fabrication parameters obtained from the proposed coverage-oriented optimization in Algorithm~\ref{alg:fabrication_design}. For each value of $N$, three different feed current vectors were considered, each designed to focus the beam toward a distinct direction, namely \((\phi,\theta)=(30^\circ,15^\circ)\), \((60^\circ,15^\circ)\), and \((90^\circ,15^\circ)\), by solving \eqref{eq:best_beam_gain} (as described in Section~\ref{sec:BF_Design_Obj}). The corresponding curves illustrate how the optimized metasurface structure enables beam steering across the azimuth sector through reconfiguration of the feed excitation. In fact, two main conclusions can be drawn. First, as $N$ increases, the main beam becomes progressively narrower, indicating improved directivity. Interestingly, this trend is achieved while keeping the number of feeds fixed, which highlights the ability of the PPW-fed metasurface to realize electrically XL hybrid MIMO without increasing the number of active TX RFCs. It is also emphasized that the polarizabilities, which act as the analog weights, remain fixed after fabrication and are optimized only once to improve coverage over the whole sector. Therefore, the reported patterns correspond to a more restrictive setting than a fully reconfigurable metasurface, for which the element responses could also be adapted independently for each beam direction.

In addition, beyond the visual beam sharpening observed in Fig.~\ref{fig: radiated power density N 128 256 512}, the proposed optimized design exhibits also improved coverage metrics when evaluated over the full region of interest. In particular, after computing \(g(\Omega;\mathbf{f}^{\star})\) for all sampled directions \(\Omega\) in the coverage sector, the quantity \(\max_{\Omega\in\mathcal{S}} g(\Omega;\mathbf{f}^{\star})\) was found to be \(5.5\), \(7.6\), and \(8.4\)~\(\mathrm{W/sr}\) for \(N=128,256\), and $512$, respectively; these values are not obtained from the plotted cuts, but from the full evaluation of \(g(\Omega;\mathbf{f}^{\star})\) over \(\mathcal{S}\). Using the following antenna gain relation:
\begin{equation}\label{eq: max antenna gain in sector}
G_{\max,\mathcal{S}}\triangleq10\log_{10}\!\left(\frac{4\pi \max_{\Omega\in\mathcal{S}} g(\Omega;\mathbf{f}^{\star})}{P_{\mathrm{tot}}}\right),
\end{equation}
the corresponding maximum gains within \(\mathcal{S}\) were computed as \(8.39\), \(9.8\), and \(10.23\)~dBi for \(N=128,256\), and $512$, respectively. These results increase with aperture size, but more slowly than would be expected for an ideal uniformly powered array with perfect phase-only beamforming toward each direction. Actually, in that ideal case, doubling the number of elements would approximately double the peak radiation intensity, which, starting from \(8.39\)\, dBi at \(N=128\) would predict \(11.39\) dBi at \(N=256\) and \(14.39\) dBi at \(N=512\). The optimized metasurface therefore exhibits a noticeably weaker scaling. In fact, this was expected for the following three reasons. First, the aperture is not uniformly powered: it is excited through the PPW and, hance, the guided power is distributed non-uniformly across the surface. This implies that metamaterial elements (here, elliptic irises) closer to the feeds are in general more strongly excited than those farther away. Second, the polarizability weights are optimized to satisfy a sector-level performance objective, rather than to maximize radiation toward each individual direction separately. As a result, the field is not perfectly phase-aligned for every angle within the desired sector $\mathcal{S}$, which inevitably reduces the achievable peak gain compared with fully directional beam-specific optimization. Third, phase beamforming is not ideal, since there exists phase-amplitude coupling. 

Most importantly, it can be also seen from Fig.~\ref{fig: radiated power density N 128 256 512} that the worst-case performance over the coverage region $\mathcal{S}$ also improves with \(N\). Specifically, the quantity \(\min_{\Omega\in\mathcal{S}} g(\Omega;\mathbf{f})\), which is the actual fabrication design objective in $\mathcal{OP}$ for the metasurface, increases from \(3.8\) to \(4.3\) and \(5.2\)\,\(\mathrm{W/sr}\) for \(N=128,256\), and $512$, respectively. To relate these values to antenna performance, the minimum sector gain is defined as:
\begin{equation}\label{eq: min antenna gain in sector}
G_{\min,\mathcal{S}} \triangleq 10\log_{10}\!\left(\frac{4\pi \min_{\Omega\in\mathcal{S}} g(\Omega;\mathbf{f}^{\star})}{P_{\mathrm{tot}}}\right),
\end{equation}
which yields \(6.79\), \(7.33\), and \(8.15\) dBi for \(N=128,256\), and $512$, respectively. These values, together with the corresponding maximum radiation densities and gains discussed before, are summarized in Table~\ref{tab:coverage_metrics}.

\begin{table}[t]
\centering
\begin{tabular}{c|cc|cc}
\hline
\multirow{2}{*}{$N$} 
& \multicolumn{2}{c|}{Radiation Intensity (W/sr)} 
& \multicolumn{2}{c}{Gain [dBi]} \\
\cline{2-5}
& $\min_{\Omega\in\mathcal{S}} g(\Omega;\mathbf{f}^{\star})$ 
& $\max_{\Omega\in\mathcal{S}} g(\Omega;\mathbf{f}^{\star})$
& $G_{\min,\mathcal{S}}$ 
& $G_{\max,\mathcal{S}}$ \\
\hline\hline
128 & 3.8 & 5.5 & 6.79 & 8.39 \\
256 & 4.3 & 7.6 & 7.33 & 9.8 \\
512 & 5.2 & 8.4 & 8.15 & 10.23 \\
\hline
\end{tabular}
\caption{Radiation intensity and corresponding gains over the sector $\mathcal{S}$ for different numbers of metasurface elements $N$.}\label{tab:coverage_metrics}
\end{table}

\subsection{Results for Optimized Single-Direction Beamforming}

We herein focus on evaluating the performance of the proposed beamforming optimization framework for a single direction. This corresponds to solving $\mathcal{OP}$ with \(T=1\), i.e., maximizing \(g(\Omega;\mathbf{f})\) for a fixed \(\Omega\), without enforcing sector-wide coverage. Selecting the setting \(\Omega = (\phi,\theta) = (60^\circ, 60^\circ)\), we have applied Algorithm~\ref{alg:fabrication_design}, while all other system parameters remained the same with the previous Section~\ref{subsec: results beam coverage}. This setting serves as a benchmark for the performance of a reconfigurable hybrid XL MIMO architecture, where the response of the considered 2D PPW-fed metasurface antenna array is adapted to each beamforming direction.

\begin{figure*}[t]
    \centering
     \begin{subfigure}{0.333\linewidth}
        \centering
        \includegraphics[width=\linewidth]{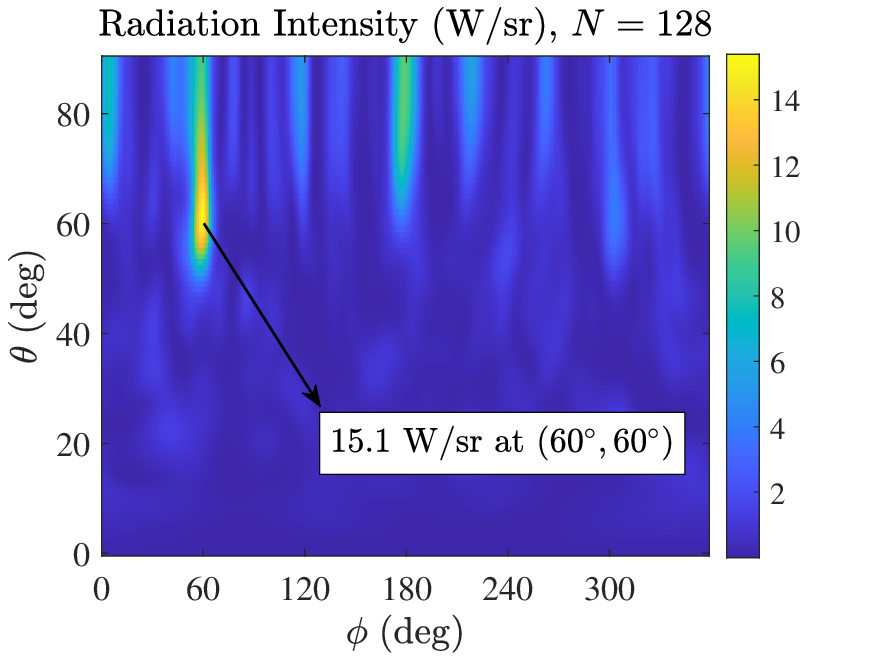}
        \caption{}
        \label{fig:N 128 theta 60}
    \end{subfigure}%
    \hfill
    \begin{subfigure}{0.333\linewidth}
        \centering
        \includegraphics[width=\linewidth]{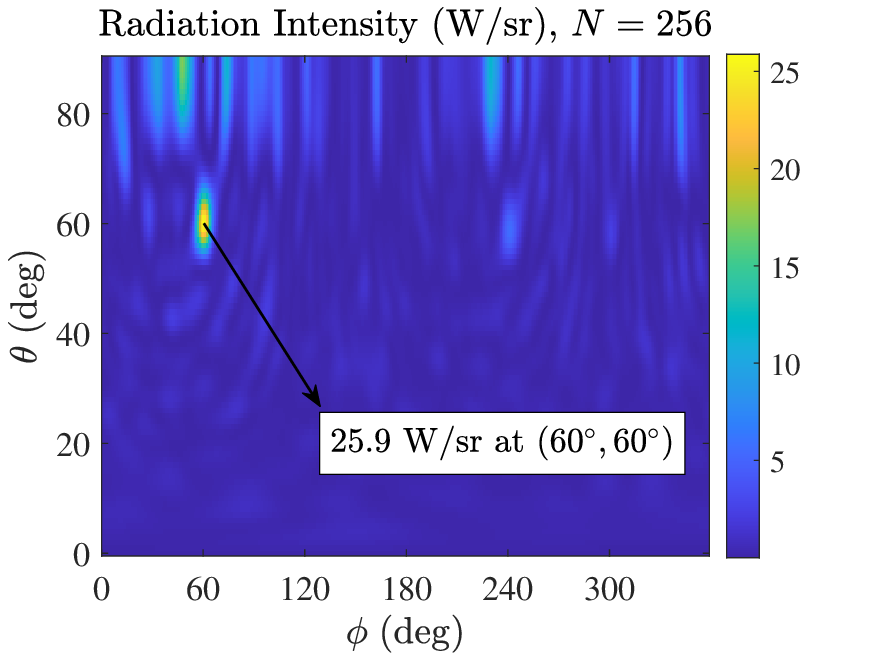}
        \caption{}
        \label{fig:N 256 theta 60}
    \end{subfigure}%
    \hfill
    \begin{subfigure}{0.333\linewidth}
        \centering
        \includegraphics[width=\linewidth]{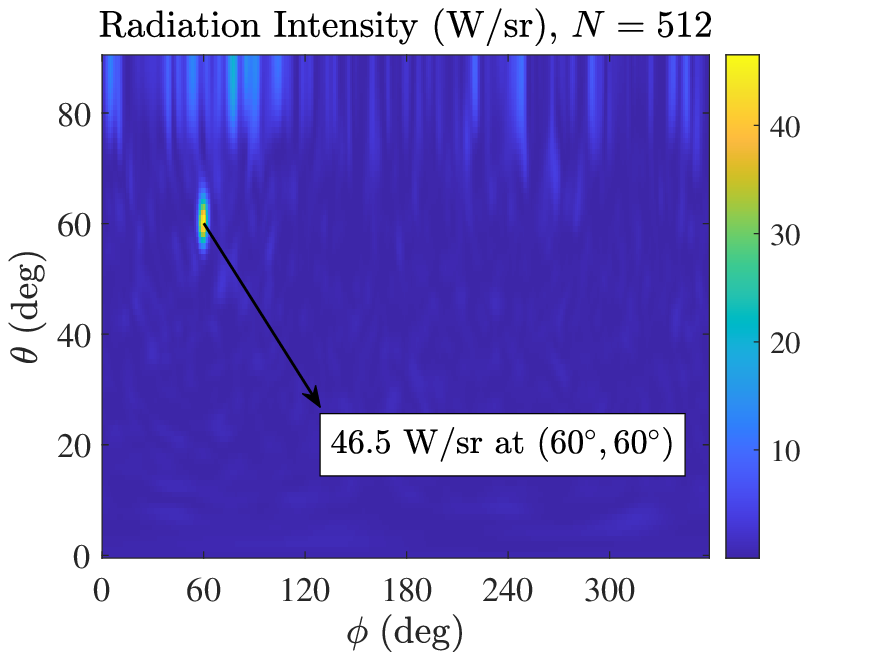}
        \caption{}
        \label{fig:N 512 theta 60}
    \end{subfigure}
    \caption{Radiation intensity patterns of the optimized 2D PPW-fed metasurface antenna for single-direction beamforming toward \(\Omega = (60^{\circ},60^{\circ})\), considering: (a) \(N=128\); (b) \(N=256\); and (c) \(N=512\) elements. In each case, the geometric parameters were optimized for the target direction, while the remaining system parameters were identical to those in Fig.~\ref{fig: radiated power density N 128 256 512}.}
    \label{fig: radiated power density theta phi 60 N 128 256 512}
\end{figure*}

Figure~\ref{fig: radiated power density theta phi 60 N 128 256 512} illustrates the resulting radiation intensity patterns for \(N=128\), \(256\), and \(512\). As expected, in all cases, the maximum is attained at the target direction \(\Omega = (60^{\circ},60^{\circ})\). It is also observed that, as $N$ increases, the beam becomes progressively narrower and more concentrated, indicating improved directivity and higher peak radiation intensity. In terms of antenna gain, the three configurations achieve \(G_{\max}=12.78\), \(15.12\), and \(17.67\)\,dBi, respectively. This corresponds to an increase of approximately \(2.5\)\,dB per doubling of the number of elements, which is close to the ideal \(3\)\,dB scaling expected for uniformly excited arrays with perfect phase-only beamforming. In addition, as compared to the coverage-oriented design in Table~\ref{tab:coverage_metrics}, significantly higher gains were achieved, with improvements of approximately \(4.5\), \(4.5\), and \(7\)~dB for \(N=128\), \(256\), and \(512\), respectively. This performance gap highlights the cost of enforcing sector-wide coverage: when the polarizabilities are optimized for a single direction, the structure can more effectively concentrate the radiated energy, leading to substantially higher peak gains.

It is finally noted that an additional advantage of the proposed geometric parameters design stems from the non-periodic, optimization-driven placement of the metasurface elements. In conventional uniform planar arrays, the spatial periodicity induces a structured angular response, often leading to symmetric sidelobes and, in general, to grating lobes when the inter-element spacing exceeds $\lambda/2$~\cite{balanis2016antenna}. In contrast, the lack of periodicity in the proposed PPW-fed metasurface architecture mitigates these effects, enabling improved suppression of secondary beams. As a result, the radiated energy can be more effectively concentrated into a single dominant direction, further enhancing the achievable beamforming performance.

\section{Conclusions and Future Work} \label{sec: Conclusion}

In this paper, we presented a novel comprehensive EM framework for modeling and optimizing 2D PPW-fed metasurface TX antennas. Building upon prior magnetic-dipoles-only formulations, we first extended the state-of-the-art modeling formulation to account for multiple excitation feeds and to accurately capture both NF and FF behavior. The proposed model was validated with respective FW simulations. Then, we extended the modeling framework to incorporate contributions of both electric and magnetic dipoles, deriving all associated propagation and coupling mechanisms in a unified and physically consistent manner. This novel unified model was also validated against FW simulations and compared with the baseline magnetic-only representation, showcasing that, accounting for both dipole types, leads to improved modeling accuracy. This indicates that neglecting electric dipole contributions should not be adopted as a general modeling assumption, but rather assessed, based on the relative strength of the electric and magnetic polarizabilities of the metamaterial elements. 

In addition, RR corrections tailored to the considered 2D metasurface architecture were derived and validated for both electric and magnetic dipoles, and the passivity of the resulting system was rigorously established. We further derived the input impedance model for the coupled metasurface and feed system, as seen from the feeds, thus, enabling the explicit computation of the accepted power to the PPW-fed antenna array. This allowed us to formulate a physically consistent transmit beamforming optimization problem under a total power constraint, revealing the interplay between metasurface design and power consumption. Furthermore, the model supports equivalent representations of the feeding mechanism in terms of both current and voltage sources, providing flexibility for analysis and implementation. 

Leveraging the analytical structure of the proposed unified CD model, we formulated a joint design optimization problem over the metasurface geometric parameters as well as the excitations of its constituent feeds to achieve sector-wide coverage. To this end, the closed-form polarizability expressions of the proposed model ensure full differentiability w.r.t. the design parameters, enabling efficient gradient-based optimization. In particular, we derived analytical gradients w.r.t. a generic parameter vector, allowing the framework to be readily extended to reconfigurable elements, such as varactor-loaded metamaterials whose circuit behavior can be mapped to equivalent polarizability models~\cite{smith2025circuit_and_polarizability}. The presented numerical investigations demonstrated the capability of the proposed design framework to offer both efficient directive beamforming and wide-area coverage, highlighting the potential of 2D PPW-fed metasurface antenna arrays for next generation hybrid, or even tri-hybrid, XL MIMO architectures.

%Overall, the proposed CD model provides a powerful and flexible tool for the analysis and design of PPW-fed metasurface antennas, combining FW-level accuracy with a fully analytical and differentiable formulation. This makes it particularly suitable for large-scale optimization and for establishing realistic performance benchmarks without relying on oversimplified models.

Future work will focus on extending the presented TX model to the reception mode and validating the corresponding model against FW simulations, as well as developing a relevant complete communication  framework encompassing both channel estimation and beamforming. Additional directions include the study of the NF beam focusing capabilities and their fundamental limitations, the incorporation of reactive NF terms in the model for improved modeling accuracy, as well as the exploration of polarization-aware beamforming strategies. Finally, developing accurate models for both electric and magnetic polarizabilities of reconfigurable metasurface elements, based on circuit representations, constitutes an important step toward enabling their direct integration into the proposed modeling and design framework, as well as the investigation of practical reconfigurable metasurface implementations.

\begin{appendices}
\section{Power Conservation}
\subsection{Power Conservation with Magnetic Dipoles}\label{app: power magnetic}
Starting from the relation \(P_{\rm rad}\geq P_{\rm sup}\) using expressions~\eqref{eq: total supplied power magnetic} and~\eqref{eq: total radiated power magnetic}, it is apparent that it suffices to show that:
\begin{align}\label{eq: Appendix Prad less than Psup magnetic}
    %P_{\rm sup}\geq P_{\rm rad} \leftrightarrow 
    {\rm Im_{\rm sym}}\left(\bar{\mathbf{A}}^{-1}-\bar{\mathbf{G}}\right)\geq -{\rm Im_{\rm sym}}\left(\bar{\mathbf{G}}_{\rm full}\right),
\end{align}
where we have used the definition: \(\rm Im_{\rm sym}(\mathbf{A})\triangleq-0.5\jmath(\mathbf{A}-\mathbf{A}^{\rm H})\). Furthermore, using the RR correction from~\eqref{eq: rr_correction}, yields:
\begin{equation}\label{eq: Appendix prad less than psup magnetic 2}
    \rm Im_{\rm sym}\left(\bar{\mathbf{A}}^{-1}-\bar{\mathbf{G}}\right)={\rm Im}\{\left(\bar{\mathbf{A}}^{\rm int}\right)^{-1}\} - {\rm Im_{\rm sym}\left(\bar{\mathbf{G}}_{\rm full}\right)},
\end{equation}
where \(\bar{\mathbf{A}}^{\rm int}\in \mathbb{C}^{2N \times 2N}\) is the block diagonal matrix containing the intrinsic polarizabilities of the elements. Importantly, for passive elements, it holds \({\rm Im}\{(\bar{\mathbf{A}}^{\rm int})^{-1}\}\geq \mathbf{0}\), i.e., the imaginary part of the inverse is a positive semidefinite matrix, with the equation holding for the lossless case. To this end, by substituting \eqref{eq: Appendix prad less than psup magnetic 2} to \eqref{eq: Appendix Prad less than Psup magnetic}, it can be seen that the condition \(P_{\rm sup}\geq P_{\rm rad}\) holds, leading to \({\rm Im}\{(\bar{\mathbf{A}}^{\rm int})^{-1}\}\geq \mathbf{0}\), which holds true. Finally, it is worth noting that, for \(P_{\rm rad}\) to be non-negative, \(-{\rm Im_{\rm sym}}(\bar{\mathbf{G}}_{\rm full})\) needs to be a positive semidefinite matrix, which should always hold for passive environments. 
\subsection{Power Conservation with Magnetic and Electric Dipoles}\label{app: power magnetic and electric}

Once more, starting from the relation \(P_{\rm rad}\geq P_{\rm sup}\) using expressions~\eqref{eq: total supplied power magnetic and electric dipoles} and~\eqref{eq: total radiated power electric and magnetic dipoles}, it suffices to show the following:
\begin{align}\label{eq: Appendix Prad less than Psup magnetic and electric dipoles}
    %P_{\rm sup}\geq P_{\rm rad} \leftrightarrow 
    {\rm Im_{\rm sym}}\left(\mathbf{S}_{\mu_0}\mathbf{K}\right)\geq -{\rm Im_{\rm sym}}\left(\mathbf{S}_{\mu_0}{\mathbf{G}}_{\rm full}\right).
\end{align}
Similarly to Appendix~\ref{app: power magnetic}, employing the RR corrections from~\eqref{eq: rr_correction} and~\eqref{eq: Radiation-reaction electric dipole}, leads to:
\begin{equation}\label{eq: Appendix prad less than psup electric and magnetic dipoles 2}
    \rm Im_{\rm sym}\left(\mathbf{S}_{\mu_0}\mathbf{K}\right)={\rm Im}\{\mathbf{S}_{\mu_0}\left({\mathbf{A}}^{\rm int}\right)^{-1}\} - {\rm Im_{\rm sym}\left(\mathbf{S}_{\mu_0}{\mathbf{G}}_{\rm full}\right)},
\end{equation}
where \({\mathbf{A}}^{\rm int}\in \mathbb{C}^{3N \times 3N}\) is a block diagonal matrix containing the intrinsic polarizabilities of the magnetic and electric dipoles. Hence, it can be seen that \(P_{\rm sup}\geq P_{\rm rad}\) leads to \({\rm Im}\{({\mathbf{A}}^{\rm int})^{-1}\}\geq \mathbf{0}\), which holds true due to metamaterial passivity. Finally, for \(P_{\rm rad}\) to be non-negative, it suffices that \(-{\rm Im_{\rm sym}}(\mathbf{S}_{\mu_0}{\mathbf{G}}_{\rm full})\)  is a positive semidefinite matrix. However, this does not always hold, and is a stricter than necessary requirement, since the vector \(\mathbf{x}=[\mathbf{m};\mathbf{p}]\) cannot take any values, since \(\mathbf{m}\) and \(\mathbf{p}\) are coupled variables. Hence, \(-{\rm Im_{\rm sym}}(\mathbf{S}_{\mu_0}{\mathbf{G}}_{\rm full})\) should be positive semidefinite only on the subspace of feasible \(\mathbf{x}\) values. To this end, since \(\mathbf{x}=\mathbf{K}^{-1}\bar{\mathbf{H}}_{\rm f}\mathbf{i}\), the correct condition to ensure passivity of the system is that \(-{\rm Im_{\rm sym}}((\mathbf{K}^{-1}\bar{\mathbf{H}}_{\rm f})^{\rm H}\mathbf{S}_{\mu_0}{\mathbf{G}}_{\rm full}(\mathbf{K}^{-1}\bar{\mathbf{H}}_{\rm f}))\) is a positive semidefinite matrix. This should always hold, and if violated, it indicates either an inconsistent sign/scaling convention or a breakdown/numerical instability of the approximation.

\section{Magnetic and Electric Dipoles in the Waveguide}\label{app: Magnetic and electric dipoles propagation}

When both electric and magnetic dipoles are present inside the 2D waveguide (i.e., PPW), the fields cannot be computed using only the magnetic vector potential \(\mathbf{F}\), as done in \cite{pulidomancera2018}. Instead, both the electric vector potential \(\mathbf{A}\) and \(\mathbf{F}\) must be considered. Note that, although the vector potentials are vectors, capital bold letters are used for consistency with EM literature. Due to the two metallic plates of the waveguide, each dipole can be modeled as an infinitely long line source along its principal axis. Consequently, each element is represented by two transverse magnetic dipoles (\(x\)- and \(y\)-directed) and one normal electric dipole (\(z\)-directed). The corresponding magnetic and electric vector potentials are therefore given by:
\begin{align}
\mathbf{F} &= -0.25\jmath\varepsilon_0 H_0^{(2)}(k\rho)
\left( M_x \hat{\mathbf{x}} + M_y \hat{\mathbf{y}} \right),
\label{eq:vector_potential_F}\\
\mathbf{A} &= -0.25\jmath\mu_0 H_0^{(2)}(k\rho) J_z \hat{\mathbf{z}},
\label{eq:vector_potential_A}
\end{align}
where $\hat{\mathbf{x}}$, $\hat{\mathbf{y}}$, and $\hat{\mathbf{z}}$ denote the 3D Cartesian unit vectors. The quantities \(M_x\), \(M_y\), and \(J_z\) represent the line current densities associated with the dipoles, which are related to the dipole moments through: $M_x = \jmath\omega\mu_0 \frac{m_x}{h}$, $M_y = \jmath\omega\mu_0 \frac{m_y}{h}$, and $J_z = \jmath\omega \frac{p_z}{h}$, where \(m_x\) and \(m_y\) are the magnetic dipole moments, \(p_z\) is the electric dipole moment, and \(h\) is the waveguide height.

To compute the electric field component \(E_z\) and the transverse magnetic field \(\mathbf{h}=[H_x;H_y]\) inside the waveguide (i.e., away from the source location), the following relations can be used \cite[eq.~(10--16)]{balanis2012advanced}:
\begin{align}
E_z &= -\frac{1}{\varepsilon_0}\left(\nabla\times\mathbf{F}\right)_z 
- \frac{\jmath}{\omega\mu_0\varepsilon_0}\left(\nabla\times\nabla\times\mathbf{A}\right)_z,
\label{eq:E_from_potentials}\\
\mathbf{h} &= \frac{1}{\mu_0}\nabla\times\mathbf{A}
- \frac{\jmath}{\omega\mu_0\varepsilon_0}\nabla\times\nabla\times\mathbf{F},
\label{eq:H_from_potentials}
\end{align}
while the vector potentials satisfy the following equations: \(\nabla \times \nabla \times \mathbf{A} = \omega^2 \mu_0 \varepsilon_0 \mathbf{A}\) and \(\nabla \times \nabla \times \mathbf{F} = \omega^2 \mu_0 \varepsilon_0 \mathbf{F} + \nabla(\nabla\cdot\mathbf{F})\). Note that, for a vector \(\mathbf{x}\) in Cartesian coordinates, the \(\nabla \times \mathbf{x}\) and \(\nabla (\nabla \cdot \mathbf{x})\) operators are given in \cite[eqs.~(II-14)--(II-18)]{balanis2012advanced}. 

After expanding the differential operators and substituting the current densities $M_x$, $M_y$, and $J_z$ with the dipole moments $m_x$, $m_y$, and $p_z$, respectively, the electric field induced at an observation point \(\mathbf{r}\) from a source positioned at \(\mathbf{r}_{\rm s}\) inside the waveguide is given by \eqref{eq: appendix Ez_final} (top of next page). Similarly, the transverse magnetic field components are given in \eqref{eq: appendix H_x final} and \eqref{eq: Appendix H_y final} (top of next page).
\begin{figure*}[t]
\begin{equation}
E_z(\mathbf{r})
=
\underbrace{\frac{-\jmath k^2 }{4\varepsilon_0 h}H_0^{(2)}(k\rho)}_{=G^{ee,zz}_{\rm WG}}\,p_z
+
\underbrace{\frac{-k^2 \eta}{4h}H_1^{(2)}(k\rho)
\sin(\psi)}_{=G^{em,zx}_{\rm WG}} m_x
+
\underbrace{\frac{k^2 \eta}{4h}H_1^{(2)}(k\rho) \cos(\psi)}_{=G^{em,zy}_{\rm WG}} m_y,
\label{eq: appendix Ez_final}
\end{equation}

\begin{equation}
H_x(\mathbf{r}) =
\underbrace{\frac{-k^2}{4h\eta\varepsilon_0}H_1^{(2)}(k\rho)\sin(\psi)}_{= G^{me,xz}_{\rm WG}}\, p_z
+
\underbrace{\frac{-\jmath k^2}{8h}
\left(H_0^{(2)}(k\rho) \! + \! \cos(2\psi)\,H_2^{(2)}(k\rho)\right)}_{=G^{mm,xx}_{\rm WG}}m_x
+
\underbrace{\frac{-\jmath k^2}{8h}\sin(\!2\psi\!)\,H_2^{(2)}(k\rho)}_{=G^{mm,xy}_{\rm WG}}\,m_y,
\label{eq: appendix H_x final}
\end{equation}

\begin{equation}
H_y(\mathbf{r}) =
\underbrace{\frac{k^2}{4h\eta\varepsilon_0}H_1^{(2)}(k\rho)\cos(\!\psi\!)}_{=G^{me,yz}_{\rm WG}}\, p_z
+
\underbrace{\frac{-\jmath k^2}{8h}\sin(\!2\psi\!)\,H_2^{(2)}(k\rho)}_{=G^{mm,yx}_{\rm WG}}\,m_x
+
\underbrace{\frac{-\jmath k^2}{8h}\left(H_0^{(2)}(k\rho)-\cos(\!2\psi\!)\,H_2^{(2)}(k\rho)\right)}_{=G^{mm,yy}_{\rm WG}}m_y .
\label{eq: Appendix H_y final}
\end{equation}
\end{figure*}
These expressions describe the electric and magnetic fields produced inside the 2D waveguide by an electric dipole $p_z$ and transverse magnetic dipoles $m_x$ and $m_y$. Therein, $\rho$ denotes the distance in the \(x\text{--}y\) plane between the observation point $\mathbf{r}$ and the source point $\mathbf{r}_s$, while $\psi$ represents the angle formed by the vector $\mathbf{r}-\mathbf{r}_s$ w.r.t. the $x$-axis: \(\psi={\rm atan}((\mathbf{r}_y-\mathbf{r}_{{\rm s}_y})/(\mathbf{r}_x-\mathbf{r}_{{\rm s}_x}))\). It is noted that expressions \eqref{eq: appendix Ez_final}--\eqref{eq: Appendix H_y final} form the basis for the Green function operators used in the main text.

\section{Gradient w.r.t. Metasurface Parameters}\label{app: gradient with z} 
%\(\mathbf{l_2}\) Vector
%a generic metamaterial optimization vector
Let $\mathbf{z}\in\mathbb{R}^{N_z\times 1}$ include the collection of the optimization parameters for all metamaterials. In our case, this corresponds to the \(\mathbf{l}_2\) vector, but it can be extended to reconfigurable metamaterials by considering, for example, the tunable capacitances of varactor diodes.
The (effective) magnetic and electric polarizabilities can be described through the following block-diagonal matrix:
\begin{equation}\label{eq:Adef}
\mathbf{A}(\mathbf{z})
\triangleq
\mathrm{diag}\big(\mathbf{A}^{m}(\mathbf{z}),\,\mathbf{A}^{e}(\mathbf{z})\big).
\end{equation}
The coupled interactions are captured by the block matrix:
\begin{equation}\label{eq:Cdef}
\mathbf{G}_{\rm mut}
=
\begin{bmatrix}
\bar{\mathbf{G}}^{mm} & \bar{\mathbf{G}}^{me}\\
\bar{\mathbf{G}}^{em} & \bar{\mathbf{G}}^{ee}
\end{bmatrix},
\end{equation}
which is independent of $\mathbf{z}$ for fixed element locations. Consequently, substituting the latter expressions in \eqref{eq: compact_inverse_solution}, the coupled-dipole
interaction matrix can be compactly written as
\(\mathbf{K}(\mathbf{z})
=
\mathbf{A}^{-1}(\mathbf{z})
-
\mathbf{G}_{\rm mut}\). It is noted that this formulation is general enough. Actually, for a given metasurface implementation, $\mathbf{A}(\mathbf{z})$ is known explicitly. For instance, in the case of elliptic irises, $\mathbf{A}^{m}(\mathbf{z})$ and $\mathbf{A}^{e}(\mathbf{z})$ follow from the effective polarizabilities (see \eqref{eq: rr_correction} and \eqref{eq: Radiation-reaction electric dipole} in conjunction with the intrinsic polarizability definitions in expressions~\eqref{eq:alpha_e_ellipse}-\eqref{eq:alpha_yy_ellipse}).

Under the power constraint $\mathbf{i}^{\mathrm H}\mathbf{R}\mathbf{i}\le 2 P_{\rm tot}$,
with $\mathbf{R}=0.5(\mathbf{Z}_{\rm in}^{\rm H} + \mathbf{Z}_{\rm in})$, the maximum achievable intensity is:
\begin{equation}\label{eq:gOmega_def}
\!\!\! g(\Omega;\mathbf{z})
=
\frac{r^2}{\eta}\,P_{\rm tot}\,
\lambda_{\max}\!\big(\mathbf{R}(\mathbf{z})^{-1/2}\mathbf{Q}(\Omega;\mathbf{z})\mathbf{R}(\mathbf{z})^{-1/2}\big).
\end{equation}
Using this definition, the sector objective can be defined as a soft-min exponential sum, as follows:
\begin{equation}\label{eq:J_softmin}
J_{\alpha}(\mathbf{z})
=
-\frac{1}{\alpha}
\log\left(
\sum_{t=1}^{T}
\exp\!\left(-\alpha g(\Omega_t;\mathbf{z})\right)\right),
\end{equation}
whose gradient is computed as follows:
\begin{align}\label{eq:softmin_weights}
\nabla_{\mathbf{z}}J_{\alpha}(\mathbf{z})
&=
\sum_{t=1}^{T}
w(\Omega_t;\mathbf{z})\,
\nabla_{\mathbf{z}} g(\Omega_t;\mathbf{z}),\\
w(\Omega_t;\mathbf{z})
&\triangleq
\frac{\exp\!\big(-\alpha g(\Omega_t;\mathbf{z})\big)}
{\sum_{i=1}^{T}
\exp\!\big(-\alpha g(\Omega_i;\mathbf{z})\big)}.\nonumber
\end{align}
To avoid differentiating the matrix square root $\mathbf{R}(\mathbf{z})^{-1/2}$ w.r.t. $\mathbf{z}$,
we equivalently express the problem in generalized eigenvalue form. Specifically, for each direction $\Omega$, we consider the
dominant generalized eigenpair
\(
\mathbf{Q}(\Omega;\mathbf{z})\,{\mathbf{u}}_{\Omega}(\mathbf{z})
=
\lambda_{\Omega}(\mathbf{z})\,
\mathbf{R}(\mathbf{z})\,
{\mathbf{u}}_{\Omega}(\mathbf{z})
\),
with
\({\mathbf{u}}_{\Omega}^{\mathrm H}(\mathbf{z})\,
\mathbf{R}(\mathbf{z})\,
{\mathbf{u}}_{\Omega}(\mathbf{z})=1.
\)
The eigenvalue \(\lambda_{\Omega}(\mathbf{z})\) is equivalent with \(\lambda_{\max}(\mathbf{R}(\mathbf{z})^{-1/2}\mathbf{Q}(\Omega;\mathbf{z})\mathbf{R}(\mathbf{z})^{-1/2})\) in \eqref{eq:gOmega_def} for each \(\Omega\) direction, while the principal eigenvector \(\mathbf{u}_{\Omega}(\mathbf{z})\) is related with the principal eigenvector, \(\tilde{\mathbf{u}}_{\Omega}\), of \(\mathbf{R}(\mathbf{z})^{-1/2}\mathbf{Q}(\Omega;\mathbf{z})\mathbf{R}(\mathbf{z})^{-1/2}\) via the expression \(\mathbf{u}_{\Omega}(\mathbf{z}) = \mathbf{R}(\mathbf{z})^{-1/2}\tilde{\mathbf{u}}_{\Omega}(\mathbf{z})\).
For compactness, the dependence of $\mathbf{z}$ on \(\lambda_{\Omega}\) and \(\mathbf{u}_{\Omega}\) is omitted in the following expressions.

Using the generalized eigenvalue sensitivity theory, the derivative of $\lambda_\Omega$ satisfies the following~\cite[eq.~(7.2)]{andrew1993derivatives}:
\begin{equation}
\nabla_{\mathbf{z}}\lambda_{\Omega}
={\mathbf{u}}_\Omega^{\mathrm H}
\left( \nabla_{\mathbf{z}}\mathbf{Q}(\Omega)
-
\lambda_\Omega \nabla_{\mathbf{z}}\mathbf{R}
\right) {\mathbf{u}}_\Omega,
\end{equation}
which, in turn, yields the derivative of \(g(\Omega;\mathbf{z})\) as follows: 
\begin{equation}\label{eq:dgOmega_dz}
\!\!\!\nabla_{\mathbf{z}} g(\Omega;\mathbf{z})
\!=\!
\frac{r^2}{\eta}P_{\rm tot}
{\mathbf{u}}_\Omega^{\mathrm H}
\left(
\nabla_{\mathbf{z}}\mathbf{Q}(\Omega;\mathbf{z})
\!-
\!\lambda_\Omega
\nabla_{\mathbf{z}}\mathbf{R}(\mathbf{z})
\right){\mathbf{u}}_\Omega.
\end{equation}
While obtaining a closed-form expression for 
${\mathbf{u}}_\Omega$ as a function of $\mathbf{z}$ would
be analytically intractable, this is not necessary here.
Instead, the presented analysis provides a closed-form gradient expression for use in gradient-based solvers, avoiding finite-difference approximations.

Let us define the matrix $\mathbf{X}(\mathbf{z})\in\mathbb{C}^{3N\times N_{\rm f}}$ that satisfies \(\mathbf{X}(\mathbf{z})\triangleq\mathbf{K}^{-1}(\mathbf{z})\,\bar{\mathbf{H}}_{\rm f}\). Then, using \( \mathbf{Q}(\Omega;\mathbf{z}) = \mathbf{X}(\mathbf{z})^{\mathrm H}\mathbf{M}(\Omega)\mathbf{X}(\mathbf{z}) \) with \( \mathbf{M}(\Omega) = \mathbf{H}_{mp}(\Omega)^{\mathrm H}\mathbf{H}_{mp}(\Omega),\)
we obtain the following gradient:
\begin{align}
\nabla_{\mathbf{z}}\mathbf{Q}(\Omega;\mathbf{z})
=&
\big(\nabla_{\mathbf{z}}\mathbf{X}(\mathbf{z})\big)^{\mathrm H}
\mathbf{M}(\Omega)
\mathbf{X}(\mathbf{z})
+ \label{eq:dQ_dz} \\
&\mathbf{X}(\mathbf{z})^{\mathrm H}
\mathbf{M}(\Omega)
\big(\nabla_{\mathbf{z}}\mathbf{X}(\mathbf{z})\big).\nonumber
\end{align}
To this end, differentiating \(\mathbf{X}(\mathbf{z})\) yields:
\begin{align}
\nabla_{\mathbf{z}}\mathbf{X}(\mathbf{z})
=&
-\mathbf{K}^{-1}(\mathbf{z})
\big(\nabla_{\mathbf{z}}\mathbf{K}(\mathbf{z})\big)
\mathbf{X}(\mathbf{z}), \label{eq:dX_dz}\\
\nabla_{\mathbf{z}}\mathbf{K}(\mathbf{z})
=&
\nabla_{\mathbf{z}}\mathbf{A}^{-1}(\mathbf{z}),\label{eq:dK_dz}
\end{align}
where the last identity holds since
$\mathbf{C}$ is independent of $\mathbf{z}$. Finally, using the relation
\(
\mathbf{Z}_{\rm in}
=
Z^{\rm f}_{\rm self}\,\mathbf{I}_{N_{\rm f}}
-
h\left(\mathbf{G}_{\rm ff}
+
\mathbf{G}_{\rm f}\mathbf{X}(\mathbf{z})\right)
\)
and the fact that $\mathbf{G}_{\rm ff}$ and $\mathbf{G}_{\rm f}$ are independent of $\mathbf{z}$ for fixed geometry,
the derivative of \(\mathbf{R}\) becomes as follows using ${\rm Re_{sym}}(\mathbf{A})\triangleq 0.5(\mathbf{A}+\mathbf{A}^{\mathrm H})$:
\begin{equation}
\nabla_{\mathbf{z}}\mathbf{R}
=
-{\rm Re_{sym}}\!\left(
h\,\mathbf{G}_{\rm f}\,
\nabla_{\mathbf{z}}\mathbf{X}(\mathbf{z})
\right).
\end{equation}

All in all, the gradient $\nabla_{\mathbf{z}}J_{\alpha}(\mathbf{z})$ can be obtained in closed form once the polarizability gradients $\partial\mathbf{A}(\mathbf{z})/\partial\mathbf{z}$ are available, thus, providing an exact analytic gradient for utilization in gradient-based solvers.

\subsubsection{Extension to Gradient over \(h\)}\label{app: grad with h}
Let us now consider the plate separation $h$ as an additional design variable. In this case, the
polarizability model becomes $\mathbf{A}(\mathbf{z},h)$, since the effective polarizabilities depend on
$h$ through the RR corrections (cf. \eqref{eq: rr_correction} and \eqref{eq: Radiation-reaction electric dipole}). Moreover, the PPW
interaction blocks in \eqref{eq:Cdef} inherit an explicit dependence on $h$ (through $1/h$
scalings), and we now write $\mathbf{G}_{\rm mut}(h)$ to make this dependence explicit. To this end, the interaction matrix is expressed as follows:
\begin{equation}\label{eq:Kzh_def}
\mathbf{K}(\mathbf{z},h)=\mathbf{A}^{-1}(\mathbf{z},h)-\mathbf{G}_{\rm mut}(h),
\end{equation}
and the partial derivative of $\mathbf{K}$ w.r.t. $h$ is given by:
\begin{equation}\label{eq:dK_dh}
\frac{\partial \mathbf{K}(\mathbf{z},h)}{\partial h}
=
\frac{\partial \mathbf{A}^{-1}(\mathbf{z},h)}{\partial h}
-
\frac{\partial \mathbf{G}_{\rm mut}(h)}{\partial h},
\end{equation}
where the dependence of the polarizabilities w.r.t. \(h\) is given via the RR formulas. Accordingly, \eqref{eq:dX_dz} extends to:
\begin{equation}\label{eq:dX_dh}
\frac{\partial \mathbf{X}(\mathbf{z},h)}{\partial h}
=
-\mathbf{K}^{-1}(\mathbf{z},h)
\left(\frac{\partial \mathbf{K}(\mathbf{z},h)}{\partial h}\right)
\mathbf{X}(\mathbf{z},h).
\end{equation}

In the following expressions the dependence on
$(\mathbf{z},h)$ is omitted for compactness. Similarly to \eqref{eq:dgOmega_dz}, the derivative of the generalized eigenvalue is computed as:
\begin{equation}
\frac{\partial g(\Omega)}{\partial h}
=
\frac{r^2}{\eta}P_{\rm tot}\,
\tilde{\mathbf{u}}_\Omega^{\mathrm H}
\left(
\frac{\partial \mathbf{Q}(\Omega)}{\partial h}
-
\lambda_\Omega
\frac{\partial \mathbf{R}}{\partial h}
\right)
\tilde{\mathbf{u}}_\Omega.
\end{equation}
The derivative of $\mathbf{Q}(\Omega)$ w.r.t. $h$ follows the same structure
as in \eqref{eq:dQ_dz}, while $\partial\mathbf{X}(\Omega)/\partial h$ is given by \eqref{eq:dX_dh}. Finally, the derivative of $\mathbf{R}$
w.r.t. $h$ can be evaluated by differentiating the input impedance model in \eqref{eq:Zin_def}. In particular, it holds that:
\begin{equation}\label{eq:dZin_dh_app}
\!\!\!\!\frac{\partial \mathbf{R}}{\partial h}
\!\!=
0.25\eta k\,\mathbf{I}_{N_{\rm f}}
\!\!-\!{\rm Re_{sym}}\left(\mathbf{G}_{\rm ff}
\!- \!h\mathbf{G}_{\rm f}\,\frac{\partial (\mathbf{K}^{-1})}{\partial h}\,\bar{\mathbf{H}}_{f}\right)\!\!.
\end{equation}
 Moreover,
\begin{equation}\label{eq:dKinv_dh_app}
\frac{\partial (\mathbf{K}^{-1})}{\partial h}
=
-\mathbf{K}^{-1}\left(\frac{\partial \mathbf{K}}{\partial h}\right)\mathbf{K}^{-1},
\end{equation}
and, therefore, $\partial \mathbf{R}/\partial h$ can be evaluated. 
\end{appendices}

\bibliographystyle{IEEEtran}
\bibliography{references}

\begin{thebibliography}{10}
\providecommand{\url}[1]{#1}
\csname url@samestyle\endcsname
\providecommand{\newblock}{\relax}
\providecommand{\bibinfo}[2]{#2}
\providecommand{\BIBentrySTDinterwordspacing}{\spaceskip=0pt\relax}
\providecommand{\BIBentryALTinterwordstretchfactor}{4}
\providecommand{\BIBentryALTinterwordspacing}{\spaceskip=\fontdimen2\font plus
\BIBentryALTinterwordstretchfactor\fontdimen3\font minus \fontdimen4\font\relax}
\providecommand{\BIBforeignlanguage}[2]{{%
\expandafter\ifx\csname l@#1\endcsname\relax
\typeout{** WARNING: IEEEtran.bst: No hyphenation pattern has been}%
\typeout{** loaded for the language `#1'. Using the pattern for}%
\typeout{** the default language instead.}%
\else
\language=\csname l@#1\endcsname
\fi
#2}}
\providecommand{\BIBdecl}{\relax}
\BIBdecl

\bibitem{8264743}
S.~Hu, F.~Rusek, and O.~Edfors, ``Beyond massive {MIMO}: The potential of positioning with large intelligent surfaces,'' \emph{IEEE Trans. Signal Process.}, vol.~66, no.~7, pp. 1761--1774, 2018.

\bibitem{8030501}
A.~F. Molisch, V.~V. Ratnam, S.~Han, Z.~Li, S.~L.~H. Nguyen, L.~Li, and K.~Haneda, ``Hybrid beamforming for massive {MIMO}: A survey,'' \emph{IEEE Commun. Mag.}, vol.~55, no.~9, pp. 134--141, 2017.

\bibitem{9933358}
G.~C. Alexandropoulos, M.~A. Islam, and B.~Smida, ``Full-duplex massive multiple-input, multiple-output architectures: Recent advances, applications, and future directions,'' \emph{IEEE Veh. Technol. Mag.}, vol.~17, no.~4, pp. 83--91, 2022.

\bibitem{AlexandropoulosRIS}
G.~C. Alexandropoulos, A.~Zappone, N.~Shlezinger, M.~Di~Renzo, and Y.~C. Eldar, \emph{Reconfigurable Intelligent Surfaces for Wireless Communications: Modeling, Architectures, and Applications}.\hskip 1em plus 0.5em minus 0.4em\relax Singapore: Springer Nature, 2026.

\bibitem{10352433}
G.~C. Alexandropoulos, N.~Shlezinger, I.~Alamzadeh, M.~F. Imani, H.~Zhang, and Y.~C. Eldar, ``Hybrid reconfigurable intelligent metasurfaces: Enabling simultaneous tunable reflections and sensing for 6g wireless communications,'' \emph{IEEE Veh. Technol. Mag.}, vol.~19, no.~1, pp. 75--84, 2024.

\bibitem{10124713}
J.~He, A.~Fakhreddine, C.~Vanwynsberghe, H.~Wymeersch, and G.~C. Alexandropoulos, ``{3D} localization with a single partially-connected receiving {RIS}: Positioning error analysis and algorithmic design,'' \emph{IEEE Trans. Veh. Technol.}, vol.~72, no.~10, pp. 13\,190--13\,202, 2023.

\bibitem{Receiving_RISs}
G.~C. Alexandropoulos, K.~D. Katsanos, and E.~Vlachos, ``Receiving {RISs} for channel estimation and autonomous configuration,'' \emph{arXiv preprint arXiv:2506.10662}, 2025.

\bibitem{Shlezinger2021Dynamic}
N.~Shlezinger, G.~C. Alexandropoulos, M.~F. Imani, Y.~C. Eldar, and D.~R. Smith, ``Dynamic metasurface antennas for {6G} extreme massive {MIMO} communications,'' \emph{IEEE Wireless Comm.}, vol.~28, no.~2, pp. 106--113, 2021.

\bibitem{Gong2024HMIMO}
T.~Gong \emph{et~al.}, ``Holographic {MIMO} communications: Theoretical foundations, enabling technologies, and future directions,'' \emph{IEEE Commun. Surveys \& Tutorials}, vol.~26, no.~1, pp. 196--257, 2024.

\bibitem{Shlezinger2019_DMA_uplink_MIMO}
N.~Shlezinger, O.~Dicker, Y.~C. Eldar, I.~Yoo, M.~F. Imani, and D.~R. Smith, ``Dynamic metasurface antennas for uplink massive {MIMO} systems,'' \emph{IEEE Trans. Commun.}, vol.~67, no.~10, pp. 6829--6843, 2019.

\bibitem{Li2023_DMA_energy_efficiency}
L.~You, J.~Xu, G.~C. Alexandropoulos, J.~Wang, W.~Wang, and X.~Gao, ``Energy efficiency maximization of massive {MIMO} communications with dynamic metasurface antennas,'' \emph{IEEE Trans. Wireless Commun.}, vol.~22, no.~1, pp. 393--407, 2023.

\bibitem{10505154}
J.~Xu, L.~You, G.~C. Alexandropoulos, X.~Yi, W.~Wang, and X.~Gao, ``Near-field wideband extremely large-scale {MIMO} transmissions with holographic metasurface-based antenna arrays,'' \emph{IEEE Trans. Wireless Commun.}, vol.~23, no.~9, pp. 12\,054--12\,067, 2024.

\bibitem{DMA1bit_comms}
P.~Gavriilidis and I.~A. G.~C. Alexandropoulos, ``Metasurface-based receivers with 1-bit {ADCs} for multi-user uplink communications,'' in \emph{Proc. IEEE ICASSP}, Seoul, South Korea, 2024.

\bibitem{10938788}
P.~Gavriilidis and G.~C. Alexandropoulos, ``Near-field beam tracking with extremely large dynamic metasurface antennas,'' \emph{IEEE Trans. Wireless Commun.}, vol.~24, no.~7, pp. 6257--6272, 2025.

\bibitem{guohao2020_DMAsensingMag}
G.~Lan, M.~F. Imani, P.~d. Hougne, W.~Hu, D.~R. Smith, and M.~Gorlatova, ``Wireless sensing using dynamic metasurface antennas: Challenges and opportunities,'' \emph{IEEE Commun. Mag.}, vol.~58, no.~6, pp. 66--71, 2020.

\bibitem{9500663}
Z.~Abu-Shaban, K.~Keykhosravi, M.~F. Keskin, G.~C. Alexandropoulos, G.~Seco-Granados, and H.~Wymeersch, ``Near-field localization with a reconfigurable intelligent surface acting as lens,'' in \emph{Proc. IEEE ICC}, Montreal, Canada, 2021.

\bibitem{DMA_CRB}
I.~Gavras and G.~C. Alexandropoulos, ``Near-field localization with dynamic metasurface antennas at {THz}: A {CRB} minimizing approach,'' \emph{IEEE Wireless Commun. Lett.}, vol.~14, no.~7, p. 1854–1858, 2025.

\bibitem{DMA1bit_Loc}
I.~Gavras, I.~Atzeni, and G.~C. Alexandropoulos, ``Near-field localization with 1-bit quantized hybrid {A/D} reception,'' in \emph{Proc. IEEE ICASSP}, Seoul, South Korea, 2024.

\bibitem{Gavras2025DMA_bistatic_Optimization}
I.~Gavras and G.~C. Alexandropoulos, ``Electromagnetics-compliant optimization of dynamic metasurface antennas for bistatic sensing,'' \emph{arXiv preprint arXiv:2509.19801}, 2025.

\bibitem{10403512}
K.~Stylianopoulos, M.~Bayraktar, N.~González-Prelcic, and G.~C. Alexandropoulos, ``Autoregressive attention neural networks for non-line-of-sight user tracking with dynamic metasurface antennas,'' in \emph{Proc. IEEE CAMSAP}, Herradura, Costa Rica, 2023, pp. 391--395.

\bibitem{11161718}
I.~Gavras and G.~C. Alexandropoulos, ``Circuit-compliant optimization of dynamic metasurface antennas for near-field localization,'' in \emph{Proc. IEEE ICC}, Montreal, Canada, 2025, pp. 2485--2490.

\bibitem{Rezvani2024_DMA_channelestimation}
M.~Rezvani and R.~Adve, ``Channel estimation for dynamic metasurface antennas,'' \emph{IEEE Trans. Wireless Commun.}, vol.~23, no.~6, pp. 5832--5846, 2024.

\bibitem{10694467}
I.~Gavras and G.~C. Alexandropoulos, ``Simultaneous near-field thz communications and sensing with full duplex metasurface transceivers,'' in \emph{Proc. IEEE SPAWC}, Lucca, Italy, 2024, pp. 126--130.

\bibitem{Gollub2017LargeMetasurfaceImaging}
J.~N. Gollub \emph{et~al.}, ``Large metasurface aperture for millimeter wave computational imaging at the human-scale,'' \emph{Scientific Rep.}, vol.~7, no.~1, p. 42650, 2017.

\bibitem{Boyarsky2018SingleFrequency3D}
M.~Boyarsky, T.~Sleasman, L.~Pulido-Mancera, A.~V. Diebold, M.~F. Imani, and D.~R. Smith, ``Single-frequency {3D} synthetic aperture imaging with dynamic metasurface antennas,'' \emph{Appl. Opt.}, vol.~57, no.~15, pp. 4123--4134, 2018.

\bibitem{Diebold2018PhaselessGhostImaging}
A.~V. Diebold, M.~F. Imani, T.~Sleasman, and D.~R. Smith, ``Phaseless computational ghost imaging at microwave frequencies using a dynamic metasurface aperture,'' \emph{Appl. Opt.}, vol.~57, no.~9, pp. 2142--2149, 2018.

\bibitem{9769592}
T.~V. Hoang, V.~Fusco, T.~Fromenteze, and O.~Yurduseven, ``Dynamic metasurface antenna for computational polarimetric imaging,'' in \emph{Proc. EuCAP}, Madrid, Spain, 2022.

\bibitem{11362106}
O.~Yurduseven, M.~Garc{\'\i}a-Fern{\'a}ndez, G.~{\'A}lvarez-Narciandi, M.~Khalily, and A.~M. Molaei, ``Programmable metasurfaces for computational {DoA} estimation: Experimental validation,'' in \emph{Proc. ISAP}, Fukuoka, Japan, 2025.

\bibitem{Heath_2026_trihybrid}
R.~W. Heath, J.~Carlson, N.~V. Deshpande, M.~R. Castellanos, M.~Akrout, and C.-B. Chae, ``The tri-hybrid {MIMO} architecture,'' \emph{IEEE Wireless Commun.}, vol.~33, no.~1, pp. 199--206, 2026.

\bibitem{pulidomancera2018}
L.~Pulido-Mancera, M.~F. Imani, P.~T. Bowen, N.~Kundtz, and D.~R. Smith, ``Analytical modeling of a two-dimensional waveguide-fed metasurface,'' \emph{arXiv preprint arXiv:1807.11592}, 2018.

\bibitem{williams2023EM_DMA}
R.~J. Williams, P.~Ramirez-Espinosa, J.~Yuan, and E.~de~Carvalho, ``Electromagnetic based communication model for dynamic metasurface antennas,'' \emph{IEEE Trans. Wireless Commun.}, vol.~22, no.~2, pp. 1464--1464, 2023.

\bibitem{gavras20262D_DMA}
I.~Gavras, P.~Gavriilidis, and G.~C. Alexandropoulos, ``\(2\){D} waveguide-fed metasurface antenna arrays: {M}odeling and optimization for bistatic sensing,'' in \emph{Proc. IEEE EuCAP}, Dublin, Irelend, 2026.

\bibitem{davidsmith2017}
D.~R. Smith, O.~Yurduseven, L.~Pulido-Mancera, P.~Bowen, and N.~B. Kundtz, ``Analysis of a waveguide-fed metasurface antenna,'' \emph{Physical Review Applied}, 2017.

\bibitem{gavriilidis2025microstrip}
P.~Gavriilidis and G.~C. Alexandropoulos, ``How do microstrip losses impact near-field beam depth in dynamic metasurface antennas?'' in \emph{Proc. EUSIPCO}, Palermo, Italy, 2025, pp. 1005--1009.

\bibitem{Carlson2026WidebandDMABeamforming}
J.~M. Carlson, N.~V. Deshpande, M.~R. Castellanos, and R.~W. Heath, ``Wideband dynamic metasurface antenna performance with practical design characteristics,'' \emph{IEEE Trans. Wireless Commun.}, vol.~25, pp. 4674--4690, 2026.

\bibitem{tretyakov2020ppwdoublepenetrable}
X.~Ma, M.~S. Mirmoosa, and S.~A. Tretyakov, ``Parallel-plate waveguides formed by penetrable metasurfaces,'' \emph{IEEE Trans. Antennas Propag.}, vol.~68, no.~3, pp. 1773--1785, 2020.

\bibitem{hosseini2023ppw_huygens}
K.~Hosseini, H.~Younesiraad, and M.~Dehmollaian, ``Study of parallel-plate waveguides bordered by reactive {H}uygens metasurfaces,'' \emph{IEEE Trans. Antennas Propag.}, vol.~71, no.~4, pp. 3371--3381, 2023.

\bibitem{balanis2012advanced}
C.~A. Balanis, \emph{Advanced engineering electromagnetics}.\hskip 1em plus 0.5em minus 0.4em\relax John Wiley \& Sons, 2012.

\bibitem{CDF_tacit}
A.~Rabault, L.~Le~Magoarou, J.~Sol, G.~C. Alexandropoulos, N.~Shlezinger, H.~V. Poor, and P.~del Hougne, ``On the tacit linearity assumption in common cascaded models of {RIS}-parametrized wireless channels,'' \emph{IEEE Trans. Wireless Commun.}, vol.~23, no.~8, pp. 10\,001--10\,014, 2024.

\bibitem{rashidfaqiri2023physfad}
R.~Faqiri, C.~Saigre-Tardif, G.~C. Alexandropoulos, N.~Shlezinger, M.~F. Imani, and P.~del Hougne, ``{PhysFad}: {P}hysics-based end-to-end channel modeling of {RIS}-parametrized environments with adjustable fading,'' \emph{IEEE Trans. Wireless Commun.}, vol.~22, no.~1, pp. 580--595, 2023.

\bibitem{Landy2014DiscreteDipoleMetamaterial}
N.~I. Landy and D.~R. Smith, ``Two-dimensional metamaterial device design in the discrete dipole approximation,'' \emph{J. Appl. Phys.}, vol. 116, no.~4, p. 044906, 2014.

\bibitem{YURKIN2007DiscreteDipoleApproximation}
M.~Yurkin and A.~Hoekstra, ``The discrete dipole approximation: An overview and recent developments,'' \emph{J. Quantitative Spectrosc. Radiative Transfer}, vol. 106, no.~1, pp. 558--589, 2007.

\bibitem{liu2023near}
Y.~Liu, Z.~Wang, J.~Xu, C.~Ouyang, X.~Mu, and R.~Schober, ``Near-field communications: A tutorial review,'' \emph{IEEE Open J. Commun. Society}, vol.~4, pp. 1999--2049, 2023.

\bibitem{Novotny_Hecht_2006}
L.~Novotny and B.~Hecht, \emph{Principles of nano-optics}.\hskip 1em plus 0.5em minus 0.4em\relax Cambridge University Press, 2006.

\bibitem{bohren2008absorption}
C.~F. Bohren and D.~R. Huffman, \emph{Absorption and scattering of light by small particles}.\hskip 1em plus 0.5em minus 0.4em\relax John Wiley \& Sons, 2008.

\bibitem{balanis2016antenna}
C.~A. Balanis, \emph{Antenna theory: analysis and design}.\hskip 1em plus 0.5em minus 0.4em\relax John wiley \& sons, 2016.

\bibitem{Tretyakov2000}
S.~A. Tretyakov and A.~J. Viitanen, ``Line of periodically arranged passive dipole scatterers,'' \emph{Electrical Engineering}, vol.~82, no.~6, pp. 353--361, Nov. 2000.

\bibitem{tretyakov2020magneticdipoles}
M.~S. Mirmoosa, G.~A. Ptitcyn, R.~Fleury, and S.~A. Tretyakov, ``Instantaneous radiation from time-varying electric and magnetic dipoles,'' \emph{Phys. Rev. A}, vol. 102, p. 013503, 2020.

\bibitem{petersen2008matrix}
K.~B. Petersen, M.~S. Pedersen \emph{et~al.}, ``The matrix cookbook,'' \emph{Technical University of Denmark}, vol.~7, no.~15, p. 510, 2008.

\bibitem{tretyakov2003analytical}
S.~Tretyakov, \emph{Analytical modeling in applied electromagnetics}.\hskip 1em plus 0.5em minus 0.4em\relax Artech House, 2003.

\bibitem{mancera2017polarizability}
L.~Pulido-Mancera, P.~T. Bowen, M.~F. Imani, N.~Kundtz, and D.~R. Smith, ``Polarizability extraction of complementary metamaterial elements in waveguides for aperture modeling,'' \emph{Phys. Rev. B}, Dec 2017.

\bibitem{7037416}
G.~C. Alexandropoulos, V.~I. Barousis, and C.~B. Papadias, ``Precoding for multiuser {MIMO} systems with single-fed parasitic antenna arrays,'' in \emph{Proc. IEEE GLOBECOM}, Austin, USA, 2014, pp. 3897--3902.

\bibitem{Nossek2010CircuitTheoryCommunication}
M.~T. Ivrla{\v{c}} and J.~A. Nossek, ``Toward a circuit theory of communication,'' \emph{IEEE Trans. Circuits Systems I: Regular Papers}, vol.~57, no.~7, pp. 1663--1683, 2010.

\bibitem{yang2019surface}
F.~Yang and Y.~Rahmat-Samii, \emph{Surface electromagnetics: with applications in antenna, microwave, and optical engineering}.\hskip 1em plus 0.5em minus 0.4em\relax Cambridge University Press, 2019.

\bibitem{yatsenko2003plane}
V.~V. Yatsenko, S.~I. Maslovski, S.~A. Tretyakov, S.~L. Prosvirnin, and S.~Zouhdi, ``Plane-wave reflection from double arrays of small magnetoelectric scatterers,'' \emph{IEEE Trans. Antennas Propag.}, vol.~51, no.~1, pp. 2--11, 2003.

\bibitem{stratton1943electromagnetic}
J.~A. Stratton, \emph{Electromagnetic Theory}.\hskip 1em plus 0.5em minus 0.4em\relax McGraw-Hill, 1943.

\bibitem{collin1990field}
R.~E. Collin, \emph{Field theory of guided waves}.\hskip 1em plus 0.5em minus 0.4em\relax John Wiley \& Sons, 1990.

\bibitem{polyanin2008handbook}
P.~Polyanin and A.~V. Manzhirov, \emph{Handbook of integral equations}.\hskip 1em plus 0.5em minus 0.4em\relax Chapman and Hall/CRC, 2008.

\bibitem{CST}
\emph{CST Studio Suite}, Dassault Systèmes, 2022.

\bibitem{krishnasamy2017Fresnelandfraunhoferlimits}
K.~T. Selvan and R.~Janaswamy, ``Fraunhofer and {F}resnel distances: Unified derivation for aperture antennas,'' \emph{IEEE Antennas Propag. Mag.}, vol.~59, no.~4, pp. 12--15, 2017.

\bibitem{ZDENEK2014ParallelPlateAntenna}
Z.~Kubík, D.~Nikolayev, P.~Karban, J.~Skála, and M.~Hromádka, ``Optimization of electrical properties of parallel plate antenna for {EMC} testing,'' \emph{J. Comput. Appl. Math.}, vol. 270, pp. 283--293, 2014.

\bibitem{Chou2018ParallelPlateLunsberg}
H.-T. Chou and Z.-D. Yan, ``Parallel-plate {Luneburg} lens antenna for broadband multibeam radiation at millimeter-wave frequencies with design optimization,'' \emph{IEEE Trans. Ant. Propag.}, vol.~66, no.~11, pp. 5794--5804, 2018.

\bibitem{7417830}
G.~C. Alexandropoulos and M.~Kountouris, ``Maximal ratio transmission in wireless {P}oisson networks under spatially correlated fading channels,'' in \emph{Proc. IEEE GLOBECOM}, San Diego, USA, 2015, pp. 1--6.

\bibitem{boyd2004convex}
S.~Boyd and L.~Vandenberghe, \emph{Convex optimization}.\hskip 1em plus 0.5em minus 0.4em\relax Cambridge university press, 2004.

\bibitem{jamieson2016multi_armed_bandit}
K.~Jamieson and A.~Talwalkar, ``Non-stochastic best arm identification and hyperparameter optimization,'' in \emph{Proc. Int. Conf. Artif. Intell. Statistics (AISTATS)}, vol.~51, Tangier, Morocco, 2016, pp. 240--248.

\bibitem{smith2025circuit_and_polarizability}
D.~R. Smith, M.~Sazegar, and I.~Yoo, ``Equivalence of polarizability and circuit models for waveguide-fed metamaterial elements,'' \emph{IEEE Trans. Antennas Propag.}, vol.~73, no.~1, pp. 7--21, 2025.

\bibitem{andrew1993derivatives}
A.~L. Andrew, K.-W.~E. Chu, and P.~Lancaster, ``Derivatives of eigenvalues and eigenvectors of matrix functions,'' \emph{SIAM J. Matrix Anal. Appl.}, vol.~14, no.~4, pp. 903--926, 1993.

\end{thebibliography}
\end{document}